%% file: main.tex
\documentclass[11pt]{amsart}
\usepackage[margin=1in]{geometry}
\usepackage{amsmath,amssymb,amsthm,mathtools}
\usepackage[dvipsnames]{xcolor}
\usepackage{graphicx,booktabs,enumitem,makecell}
\usepackage{placeins}

\usepackage{algorithm,algpseudocode}
\usepackage[colorlinks=true,linkcolor=blue!50!black,citecolor=green!40!black]{hyperref}
\usepackage[capitalise,nameinlink]{cleveref}
\hypersetup{pdftitle={Certified folding-free deformable image registration: a whole-cell Jacobian certificate and a
  sequential variational solver}, pdfauthor={Baihan Zhao, Zicong Zhou, Guojun Liao}}
\usepackage{tikz}
\usetikzlibrary{positioning,arrows.meta,decorations.pathreplacing,calc}

\input{macros}

\newtheorem{theorem}{Theorem}[section]
\newtheorem{lemma}[theorem]{Lemma}
\newtheorem{proposition}[theorem]{Proposition}
\newtheorem{corollary}[theorem]{Corollary}
\theoremstyle{definition}

\theoremstyle{remark}
\newtheorem{remark}[theorem]{Remark}
\usepackage{etoolbox}
\AtBeginEnvironment{lemma}{\crefalias{theorem}{lemma}}
\AtBeginEnvironment{proposition}{\crefalias{theorem}{proposition}}
\AtBeginEnvironment{corollary}{\crefalias{theorem}{corollary}}
\AtBeginEnvironment{definition}{\crefalias{theorem}{definition}}
\AtBeginEnvironment{assumption}{\crefalias{theorem}{assumption}}
\AtBeginEnvironment{remark}{\crefalias{theorem}{remark}}
\crefname{lemma}{Lemma}{Lemmas}\crefname{proposition}{Proposition}{Propositions}
\crefname{corollary}{Corollary}{Corollaries}\crefname{definition}{Definition}{Definitions}
\crefname{assumption}{Assumption}{Assumptions}\crefname{remark}{Remark}{Remarks}

\title[Certified folding-free registration]{Certified folding-free deformable image registration: a whole-cell Jacobian certificate and a sequential variational solver}
\author{Baihan Zhao}
\address{Department of Mathematics, University of Texas at Arlington, Arlington, TX 76019, USA}
\email{bxz6705@mavs.uta.edu}
\author{Zicong Zhou}
\address{Department of Pharmacology and Neuroscience, University of North Texas Health Science Center, Fort Worth,
TX 76107, USA}
\email{Zicong.Zhou@unthsc.edu}
\author{Guojun Liao}
\address{Department of Mathematics, University of Texas at Arlington, Arlington, TX 76019, USA}
\email{liao@uta.edu}
\thanks{Corresponding author: Baihan Zhao.}
\date{\today}

\begin{document}

\begin{abstract}
\input{sections/00_abstract}
\end{abstract}

\maketitle

\input{sections/01_introduction}
\input{sections/02_related}
\input{sections/03_model}
\input{sections/04_certificate}
\input{sections/05_algorithm}
\input{sections/06_analysis}
\input{sections/07_implementation}
\input{sections/08_experiments}
\FloatBarrier
\input{sections/09_discussion}

\appendix
\input{sections/A_proofs}
\input{sections/B_parameters}
\input{sections/C_prescreen}
\FloatBarrier

\section*{Acknowledgment}
Generative AI (Claude, Anthropic) was used to polish the text of this article and to carry out part of the
experimental work. The authors take full responsibility for the content of this article.

\bibliographystyle{amsplain}
\bibliography{refs}

\end{document}

%% file: macros.tex
\newcommand{\R}{\mathbb{R}}

\newcommand{\Om}{\Omega}
\newcommand{\Omb}{\overline{\Omega}}
\newcommand{\cells}{\mathcal{K}}
\newcommand{\nodes}{\mathcal{V}}
\newcommand{\nodesint}{\mathcal{V}^{\circ}}
\newcommand{\adm}{\mathcal{A}}

\newcommand{\Id}{\mathrm{Id}}
\newcommand{\cof}{\operatorname{cof}}
\newcommand{\curl}{\operatorname{curl}}

\newcommand{\diag}{\operatorname{diag}}
\newcommand{\hull}{\operatorname{hull}}

\newcommand{\Dat}{\mathcal{D}}
\newcommand{\Reg}{\mathcal{R}}
\newcommand{\Bpot}{\mathcal{B}}
\newcommand{\Fout}{\mathcal{F}}
\newcommand{\Ein}{\mathcal{E}}
\newcommand{\Pen}{\mathcal{P}}

\newcommand{\Gpts}{\mathsf{G}}   %
\newcommand{\Spts}{\mathsf{S}}   %

\newcommand{\fl}{\operatorname{fl}}
\newcommand{\down}{\operatorname{\downarrow}}
\newcommand{\up}{\operatorname{\uparrow}}
\newcommand{\GF}{\mathrm{GF}}
\newcommand{\Tb}{T_{\mathrm b}}          %
\newcommand{\Gh}{\mathbf G_K}            %
\newcommand{\uflow}{\varepsilon_{\mathrm u}}  %
\newcommand{\Jmaxh}{\hat J_{\max}}
\newcommand{\eps}{\varepsilon}

\newcommand{\norm}[1]{\lVert #1\rVert}

%% file: sections/00_abstract.tex
Registration transformations are used to propagate labels, to measure local volume change and to map results
back through their inverse; all three uses require an invertible transformation. Registration methods usually
guarantee invertibility only for an idealised continuous transformation. Folding of the computed transformation is
then reported with a voxelwise finite-difference check, which can miss foldings between grid points. We consider the
canonical continuous extension of a nodal displacement field, its trilinear interpolant, and give a test that either
proves that its Jacobian determinant is positive at every point of every grid cell or rejects the map. The test is
rigorous under the IEEE~754 binary64 model (correctly rounded operations without contraction), and with a fixed
boundary a proof implies that the transformation is a bi-Lipschitz homeomorphism of the image domain. We build a
registration method, SeqVP, around this certificate. A proximal formulation of the Jacobian-determinant--curl variational
principle, with the controls eliminated, gives a sequence of inner problems whose penalties on the change of the map,
its Jacobian determinant and its curl vanish to first order at the current map. The scheme decreases a fixed outer
energy (verified up to a floating-point resolution guard), and every committed iterate is certified, so stopping at
any time after the certified initialisation returns a folding-free map. A Poisson-preconditioned gradient method with Jacobian-aware step
prediction and affine-scaling damping, started from a certified handoff of an uncertified binary32
pre-optimiser, makes this affordable for 1\,mm brain MRI with about $2\times10^7$ unknowns in $90$\,s on one GPU.
On 49 held-out OASIS pairs the method reaches a mean Dice of $0.821$ over 35 structures and $0.889$ over 4 tissue
classes with certified zero folding, $+0.022$ and $+0.025$ over ANTs SyN and $+0.022$ / $+0.023$ over uniGradICON
with instance optimisation at equal time (paired, Holm-adjusted $p<0.001$), at the cost of rougher maps; on LPBA40 it
is level with ANTs, on IBSR18 behind it, and on 20 large-deformation pairs ahead of every label-free baseline.
Applied to the fields of nine established methods, the certificate proves non-positive cells on every ANTs SyN map
(a median of ten cells, invisible to the voxelwise check) and on every map of NiftyReg, ConvexAdam, uniGradICON and
TransMorph.

%% file: sections/01_introduction.tex
\section{Introduction}\label{sec:intro}

Deformable image registration estimates a spatial transformation that aligns the anatomy in a moving image with
that in a fixed image. In computational anatomy the transformation does more than resample the moving image:
its Jacobian determinant is analysed voxel by voxel in deformation- and tensor-based morphometry
\cite{Chung2001,Leow2006,Hua2008}, label maps are carried from an atlas to a subject, and the inverse transformation
is needed to map results back \cite{VPreg}. All of these uses assume that the
transformation is one-to-one and orientation preserving. A transformation that folds produces negative volume
changes, propagates labels inconsistently and has no inverse.

\begin{figure}[t]
\centering
\includegraphics[width=\textwidth]{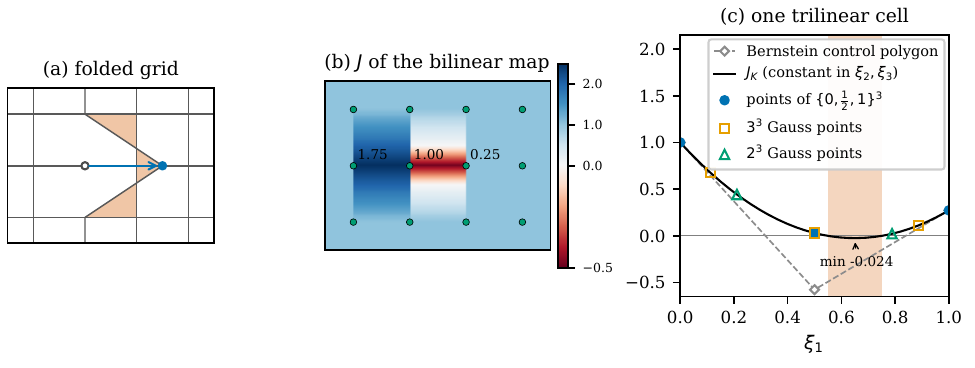}
\caption{Positivity at sample points does not imply positivity of the interpolated map. (a)--(b) Two-dimensional toy
example with unit grid spacing: one interior node is displaced by $1.5$ spacings (arrow) and overtakes its neighbour,
so the bilinear interpolant folds two cells (shaded in (a); negative $J$ in (b)). The central-difference determinant,
as used for voxelwise folding counts, is positive at every node (dots; smallest value $0.25$), because the central
difference at a node does not involve the node itself. (c) One trilinear cell,
$\Phi(\xi)=(\xi_1,(1-a\xi_1)\xi_2,(1-b\xi_1)\xi_3)$ with $a=20/11$, $b=4/3$, so that $J_K=(1-a\xi_1)(1-b\xi_1)$. It is
positive at all points of $\{0,\frac12,1\}^3$ (the barrier points, including the corners and the cell centre) and at
the $2^3$ and $3^3$ Gauss points, but negative for $\xi_1\in(11/20,3/4)$, with minimum $-4/165$ at $\xi_1=13/20$
(exact rational arithmetic). The middle Bernstein coefficient, $-19/33$, exposes the possible sign change, and
subdivision locates it. Script in the ancillary files.}\label{fig:toy}
\end{figure}

Most registration methods address invertibility through the representation of the transformation. Large
deformation diffeomorphic metric mapping and its geodesic-shooting variants, stationary-velocity-field methods
such as diffeomorphic demons and DARTEL, and the symmetric normalisation of ANTs construct the transformation as the
flow of a smooth velocity field, which is a diffeomorphism in the continuous setting \cite{Beg2005,Ashburner2007,
Vercauteren2009,Avants2008,MangBiros2015}. Other methods add
penalties or constraints on the Jacobian determinant to a variational model \cite{ChoiLee1999,Sdika2008,
ChunFessler2009,Burger2013}, and learning-based methods typically add a
smoothness or folding penalty to a trained network \cite{Balakrishnan2019,Chen2022transmorph,
Hoffmann2022,Tian2024}.
What is computed in every case, however, is a discrete object: a displacement field on the image grid, which is
interpolated between grid points, most often trilinearly, whenever the transformation is applied to points, to
surfaces or to images at another resolution, composed or inverted. The continuous guarantee does not transfer
automatically to this object. Folding is then assessed after the fact by evaluating a finite-difference Jacobian
determinant at the voxels, as in the Learn2Reg challenge \cite{Hering2023}. This check samples a
different discrete map and can miss foldings between grid points (\cref{fig:toy}) \cite{Liu2024fdj}, as VPreg also notes
\cite[\S2.1.1]{VPreg}; a reported value of zero folded voxels means that no folding was detected, not that the map is
invertible.

The canonical continuous extension of a nodal displacement field, its trilinear interpolant $\Phi_u$, is a
piecewise polynomial map, and on each grid cell its Jacobian determinant is a polynomial of degree two in each
variable. Whether it is positive everywhere is therefore a finite question that interval arithmetic and Bernstein
expansions can answer rigorously, and, as we show, a positive answer together with a fixed boundary implies that
$\Phi_u$ is a homeomorphism of the image domain. Controlling the Jacobian on the continuous domain rather than
at grid points has precedents. For hierarchical parametric models built from first-degree B-splines, Musse
et al.\ showed that in two dimensions the Jacobian is linear on each cell, so that positivity at the cell corners,
imposed as linear constraints, guarantees a globally one-to-one map \cite{Musse2001}. Noblet et al.\ extended the
approach to three dimensions, where the map on each box is trilinear and the constraints become nonlinear: before each
update of a node, interval branch-and-bound over the boxes of its support brackets the largest step that keeps the
Jacobian within prescribed bounds everywhere \cite{Noblet2005}. That work handled about $9\times10^4$ parameters for
$128^3$ images, did not address floating-point rounding, and assessed its real-data results visually. These methods restrict
the deformation to a coarse parametric hierarchy and enforce the constraints inside the optimisation of each block
of parameters. Here the map is a dense nodal field at the full image
resolution, and the difficulty is not the check itself but its use inside an optimiser: a 1\,mm brain MRI has
about seven million cells, the optimiser evaluates many trial maps, and if it may only accept certified maps it
must still make progress when parts of the map approach the boundary of the feasible set.

In this paper we develop a registration method, SeqVP (sequential variational-principle registration), in which
every committed map, and every map it returns, is certified. It builds on the
variational principle (VP) approach to grid generation and registration, which controls the Jacobian determinant
and the curl of the transformation \cite{ZhouLiao2021,VPreg}. Our contributions are:
\begin{enumerate}[label=(\arabic*)]
  \item \emph{A certificate of whole-cell positivity at full resolution and its global consequence.} We give a test
  that decides, rigorously under the IEEE~754 binary64 model, whether the Jacobian determinant of a dense trilinear
  displacement map at the image resolution is positive at every point of every cell. It combines a constant-cost
  edge-hull bound with the Bernstein validity test for linear hexahedral elements of finite-element meshing
  \cite{Johnen2017}, made rigorous by outward rounding and an a priori error bound
  (\cref{sec:certificate}). It is fast enough to be applied to every trial map of the optimiser: its CPU stages
  overlap with GPU work and, in a pre-freeze measurement, add a median of $14$\,ms of waiting per trial map of
  $6.8\times10^6$ cells (\cref{sec:implementation}). We prove that certified maps with zero boundary displacement are bi-Lipschitz
  homeomorphisms of the image domain (\cref{thm:invertibility}).
  \item \emph{A sequential Jacobian--curl model with an anytime guarantee.} Eliminating the controls from a
  proximal VP formulation yields a sequence of inner problems whose penalties on the change of the map, of its
  Jacobian determinant and of its curl vanish to first order at the current map (\cref{sec:sequential-model}).
  The scheme is monotone on a fixed outer energy, and every committed iterate is certified, so stopping at any time
  after the certified initialisation returns a folding-free map (\cref{sec:analysis}).
  \item \emph{A solver that makes the guarantee affordable.} We combine a Poisson-type preconditioner with
  diagonal equilibration and affine-scaling damping near compressed cells, a step prediction based on the exact
  cubic dependence of the Jacobian determinant on the step length, and a certified handoff from an uncertified
  binary32 pre-optimiser (\cref{sec:algorithm}). On a single GPU a 1\,mm brain registration with about
  $2\times10^7$ unknowns runs in $90$\,s end to end ($88$\,s on average on the held-out pairs) including
  certification of every trial (\cref{sec:implementation}).
  \item \emph{Evaluation.} On 49 held-out OASIS pairs, 20 large-deformation OASIS pairs, 10 LPBA40 pairs and 36
  IBSR18 pairs we compare, under pre-registered protocols, with ANTs, NiftyReg, deedsBCV, ConvexAdam, FireANTs,
  uniGradICON, VoxelMorph and TransMorph. On OASIS the method is ahead of every label-free baseline in seg35, seg4
  and HD95 (paired, Holm-adjusted), with rougher maps; on LPBA40 it is level with ANTs and on IBSR18 behind it. The
  certificate applied to the baselines' own fields proves folds that the voxelwise check reports as zero
  (\cref{sec:experiments}).
\end{enumerate}

The closest predecessor is VPreg \cite{VPreg}, which controls the divergence and the curl through a div--curl
system, combines a penalty stage with a hard-constrained stage, and reports folding with central differences, noting
explicitly that this is a diagnostic rather than a certificate (VPreg, \S2.1.1). Here we work with the Jacobian
determinant itself, use the sequential self-referenced formulation of \cref{sec:sequential-model}, and certify every
accepted map.

The paper is organised as follows. \Cref{sec:related} reviews related work. \Cref{sec:model} introduces the
discrete model and \cref{sec:certificate} the certificate. \Cref{sec:algorithm} describes the solver and
\cref{sec:analysis} its properties. \Cref{sec:implementation} summarises the implementation and
\cref{sec:experiments} reports experiments. \Cref{sec:discussion} discusses limitations.

%% file: sections/02_related.tex
\section{Related work}\label{sec:related}

\paragraph{Diffeomorphic parameterisations} Most registration methods obtain invertibility from the
representation of the transformation. Large deformation diffeomorphic metric mapping computes the map as the end
point of the flow of a time-dependent velocity field in a sufficiently smooth space, which makes the continuous
flow a diffeomorphism \cite{Beg2005}. Stationary velocity fields, exponentiated by scaling and squaring, are used in
DARTEL \cite{Ashburner2007} and in diffeomorphic demons \cite{Vercauteren2009}, and symmetric normalisation (SyN)
integrates two time-dependent flows that meet at a midpoint \cite{Avants2008}. In all of these the computed
transformation is a field on the image grid obtained with finitely many time steps or squarings and trilinear
interpolation. The authors are explicit about the consequences: DARTEL treats positivity of the Jacobian of the
exponentiated flow as a heuristic and allows small violations from the discrete composition \cite{Ashburner2007};
diffeomorphic demons notes that the discretisation and the smoothing of the map are not necessarily consistent with
the continuous framework and reports rare, slightly negative finite-difference Jacobians \cite{Vercauteren2009}; SyN
controls invertibility through the residual of a numerically computed inverse \cite{Avants2008}. High-performance
solvers follow the same principle. CLAIRE optimises a stationary velocity field with an inexact Newton--Krylov method
and spectral preconditioning \cite{MangBiros2015,Mang2019claire,Brunn2021}; it controls the Jacobian only indirectly,
by choosing the regularisation weight so that the grid Jacobian stays within prescribed bounds on a tuning pair, and
reports that the bound is not always met on other pairs \cite{Mang2019claire}. FireANTs composes small updates on a
dense grid with adaptive Riemannian optimisation; its invertibility argument concerns a single continuous step, and it
reports voxelwise singularity fractions \cite{Jena2024fireants}. Variational
models with a relaxed Jacobian constraint check signed areas at grid points and repair the map locally where needed
\cite{Li2024bivariant}, and learning-based post-processing reduces, but does not eliminate, voxels with a
non-positive Jacobian \cite{Pal2022}.

\paragraph{Jacobian constraints and penalties in registration} Several methods restrict the Jacobian of a
parametric transformation. For uniform cubic B-spline free-form deformations, Choi and Lee gave a sufficient
condition for local injectivity in terms of a bound on the control-point displacements \cite{ChoiLee1999}, and
Rueckert et al.\ composed many such bounded deformations \cite{Rueckert2006}; folding was then checked at the
voxels. Chun and Fessler derived a sufficient condition on B-spline coefficients and used it as a penalty that
encourages local invertibility \cite{ChunFessler2009}. Sdika imposed positivity of the Jacobian as constraints at the
voxels in a large-scale constrained optimisation; between voxels positivity is not guaranteed \cite{Sdika2008}. Burger,
Modersitzki and Ruthotto designed a hyperelastic regulariser whose volume term grows without bound as the Jacobian
tends to zero, discretised on a partition of each voxel into 24 tetrahedra, which keeps the piecewise-linear map
orientation-preserving during the optimisation \cite{Burger2013}. The closest precursors of our certificate are the
constrained hierarchical models of Musse et al.\ \cite{Musse2001} and Noblet et al.\ \cite{Noblet2005}, discussed in
\cref{sec:intro}: with first-degree B-splines they control the Jacobian on the continuous domain, in two and three
dimensions respectively, for coarse parametric hierarchies and without treatment of rounding.

\paragraph{Jacobian determinant and curl as controls} The variational principle and its use in registration are
reviewed in \cref{sec:sequential-model} \cite{Chen2016,ZhouLiao2021,ZhouLiao2022,VPreg}. In this family positivity
of the Jacobian of the computed map has so far been observed numerically, not certified; VPreg states explicitly that
its central-difference check is a diagnostic \cite[\S2.1.1]{VPreg}.

\paragraph{Validity of trilinear and curved elements} Mesh generation faces the same question for finite elements:
is the Jacobian of an element map positive on the whole reference element? Johnen, Remacle and Geuzaine expand the
Jacobian determinant of polynomial elements in a B\'ezier basis and refine the bounds by adaptive subdivision
\cite{Johnen2013}; Johnen, Weill and Remacle specialise this to the linear (trilinear) hexahedron, whose Jacobian
determinant is triquadratic, and show that 20 samples of the Jacobian determinant suffice to obtain its 27
B\'ezier coefficients \cite{Johnen2017}. A
recent preprint shows that strict positivity on the boundary of a trilinear hexahedron implies positivity inside
\cite{Zhang2026hex}. These works target meshes and either assume exact arithmetic or evaluate in floating point without
rounding control. Our certificate uses the Bernstein test of
\cite{Johnen2017} as its last stage, adds a cheaper sufficient test and rigorous rounding, and is applied to every
trial map of an optimiser.

\paragraph{Folding in evaluation} Registration benchmarks report folding by counting voxels with a non-positive
finite-difference Jacobian determinant and report the standard deviation of its logarithm as a smoothness measure
\cite{Hering2023}. Liu et al.\ show that a single finite-difference stencil can miss folding and derive conditions
for piecewise-linear interpolation on simplices that require several stencils to be positive \cite{Liu2024fdj}. In
the public Learn2Reg evaluation code, the folding count is taken after the shifted determinant has been clipped to a
positive value and is therefore zero for every input \cite{L2Rcode}. None of these checks concerns the trilinear map
$\Phi_u$ (\cref{rem:other-checks}).

\paragraph{Learning-based registration} Networks predict a displacement or a stationary velocity field in one
forward pass. VoxelMorph learns a displacement field with a smoothness penalty \cite{Balakrishnan2019}; TransMorph
replaces the encoder by a transformer \cite{Chen2022transmorph}; LapIRN predicts velocity fields on a Laplacian pyramid
and integrates them by scaling and squaring \cite{Mok2020}; SynthMorph is trained only on images synthesised from label
maps, so that it is contrast-invariant \cite{Hoffmann2022}; uniGradICON is trained on many anatomies with an
inverse-consistency loss and can be refined by test-time optimisation \cite{Tian2024}; ConvexAdam combines a discrete
convex optimisation with Adam-based instance optimisation, optionally on learned segmentation features
\cite{Siebert2025}. These methods run in seconds or less. Several of them use anatomical labels, either as features or
in the loss, or labelled validation data to select hyperparameters, and the best Learn2Reg OASIS entries were trained
with the evaluated labels \cite{Hering2023}. Where regularity is reported, it is measured on the voxel grid, by the fraction of voxels with a
non-positive finite-difference Jacobian determinant or by SDlogJ; none of these methods certifies positivity of the
interpolated map.

%% file: sections/03_model.tex
\section{Discrete model}\label{sec:model}

\subsection{Grid, images and the trilinear map}\label{sec:grid}
\Cref{tab:notation} collects the main notation.
\begin{table}[t]
\centering\small\setlength{\tabcolsep}{4pt}
\caption{Main notation.}\label{tab:notation}
\begin{tabular}{@{}p{0.27\textwidth}p{0.66\textwidth}@{}}
\toprule
$\Om$, $h_k$, $n_k$, $\omega$ & image box, grid spacing and number of nodes along axis $k$, cell volume $h_1h_2h_3$\\
$\nodes$, $\nodesint$, $\cells$ & nodes, interior nodes, closed cells\\
$u=(u_a)$, $\Phi_u$ & nodal displacement (zero on boundary nodes), its trilinear (Q1) map \eqref{eq:q1map}\\
$\xi\in[0,1]^3$, $J_K(u;\xi)$, $\curl_K u$ & reference coordinates on cell $K$, Jacobian determinant and curl of $\Phi_u$ on $K$\\
$\Gpts$, $\Spts$; $q^G$, $q_a$ & $3^3$ Gauss points and the closed point set $\{0,\frac12,1\}^3$ of a cell; Gauss and nodal weights\\
$\Dat$, $\Reg$, $\Bpot$, $\Fout=\Dat+\Reg+\Bpot$ & data term, curvature term, local barrier, outer energy \eqref{eq:outer}\\
$\Pen_i$, $\Ein_i=\Fout+\Pen_i$ & sequential penalties and inner energy of outer iteration $i$ \eqref{eq:inner}--\eqref{eq:penalty}\\
$\lambda,\mu,\nu,\alpha,\kappa,J_b$ & weights of proximity, Jacobian change, curl change, curvature and barrier; barrier activation level\\
$\adm$ & admissible set \eqref{eq:admissible}\\
$\beta$, $T$, $\Gh$ & Bernstein coefficients of $J_K$, the $3\times3$ map from point values to them, interval enclosure of $\nabla\Phi_u$ on $K$ (\cref{sec:certificate})\\
$\hat\beta$, $\hat{\mathsf B}$ & binary64 Bernstein coefficients and their a priori error bound, test (ii)\\
$\fl$, $\down$, $\up$ & rounding to nearest; neighbouring binary64 numbers towards $\mp\infty$\\
$u_i$, $u_0$, $\GF$ & committed iterate, certified initialiser, mass-scaled gradient norm of $\Fout$ (\cref{sec:algorithm})\\
$P_i$; $M$, $S$, $R$ & metric of outer iteration $i$; Poisson matrix, equilibration and damping diagonals \eqref{eq:metric}\\
$k_{\max}$, $\eta_i$, $s_E$, $s_F$ & accepted inner steps per outer iteration, forcing factor, resolution guards\\
$\rho_k$, $\Tb$ & handoff scales, wall-clock budget\\
\bottomrule
\end{tabular}
\end{table}
Let $\Om=\prod_{k=1}^{3}(0,(n_k-1)h_k)\subset\R^3$ be a box carrying a regular grid with $n_k$ nodes and
spacing $h_k>0$ along axis $k$; all lengths are in millimetres. We write $\nodes$ for the set of nodes $X_a$,
$\nodesint$ for the interior nodes, $\cells$ for the set of closed grid cells and $\omega=h_1h_2h_3$ for the cell
volume. The fixed image $I_f$ and the moving image $I_m$ are given at the nodes; they have been affinely aligned
and resampled to this common grid beforehand (\cref{sec:data}). The moving image is extended to $\Omb$ by its
tensor-product cubic B-spline interpolant $\widetilde I_m$ with not-a-knot end conditions.

The unknown is a nodal displacement $u=(u_a)_{a\in\nodes}$, $u_a\in\R^3$, with $u_a=0$ on boundary nodes. On a
cell $K$ with lowest corner $X_K$ we use reference coordinates $\xi\in[0,1]^3$, $X=X_K+\diag(h)\,\xi$, and define
\begin{equation}\label{eq:q1map}
  \Phi_u(X)=X+\sum_{v\in\{0,1\}^3}N_v(\xi)\,u_{K,v},\qquad
  N_v(\xi)=\prod_{k=1}^{3}\ell_{v_k}(\xi_k),\quad \ell_0(s)=1-s,\ \ell_1(s)=s,
\end{equation}
where $u_{K,v}$ is the displacement at the corner of $K$ with offset $v$; the $N_v$ are the trilinear (Q1) shape
functions. The map $\Phi_u$ is continuous on $\Omb$, trilinear on every cell, and equal to the identity on
$\partial\Om$. It is the canonical continuous (trilinear)
extension of the nodal field: it coincides with $X_a+u_a$ at the nodes, and it is the map evaluated whenever a
displacement field stored on the image grid is interpolated trilinearly between nodes. All geometric statements in
this paper, including the certificate of \cref{sec:certificate}, refer to $\Phi_u$.

\subsection{Jacobian determinant, cofactor and curl}\label{sec:geometry}
On each cell the physical deformation gradient is $\nabla\Phi_u=\Id+\nabla u$, with $\partial_{X_j}=h_j^{-1}\partial_{\xi_j}$.
We write
\[
  J_K(u;\xi)=\det\nabla\Phi_u(\xi),\qquad A=\cof\nabla\Phi_u,\qquad
  \curl u=(\partial_2u_3-\partial_3u_2,\ \partial_3u_1-\partial_1u_3,\ \partial_1u_2-\partial_2u_1).
\]
These quantities are defined cellwise; they may jump across cell faces, and values from neighbouring cells are
never averaged. Two algebraic facts are used throughout: the first variation of the determinant is $\delta J=A:\nabla v$,
and along a direction $V$
\begin{equation}\label{eq:detcubic}
  \det(F+tV)=\det F+t\,\cof(F):V+t^2\,F:\cof(V)+t^3\det V ,
\end{equation}
so the Jacobian determinant along any straight search line is a cubic polynomial in the step length. Because the
$j$-th column of $\nabla\Phi_u$ does not depend on $\xi_j$ and is bilinear in the other two reference variables,
$J_K(u;\cdot)$ is a polynomial of degree at most two in each reference variable (\cref{lem:triquadratic}).

\paragraph{Quadrature} For the geometric penalties we use the $27$-point tensor Gauss--Legendre rule
$\Gpts$ on each cell with weights $q^{G}=\omega w_iw_jw_k$, $w=(5/18,4/9,5/18)$. The squared change of $J_K$
between two Q1 maps has degree at most four in each variable and the squared change of the curl has degree at
most two, so this rule integrates both exactly. A $2\times2\times2$ rule would suffice for the curl term but not for
the Jacobian term. For the barrier we use the
closed uniform point set $\Spts=\{0,\tfrac12,1\}^3$ on each cell with equal weights $\omega/27$; this is a declared
discrete penalty, not a quadrature of a continuous barrier. Nodal quantities use trapezoidal weights
$q_a=\omega\prod_k t_{a,k}$ with $t_{a,k}=\tfrac12$ at the two ends of axis $k$ and $1$ otherwise.

\subsection{From the Jacobian--curl variational principle to a sequential model}\label{sec:sequential-model}
Constructing transformations with a prescribed Jacobian determinant $f$ and curl $g$ was formulated as a
variational problem by Chen and Liao \cite{ChenLiao2015,Chen2016} and
recast in composition form, the variational principle (VP), by Zhou and Liao \cite{ZhouLiao2021}. Registration
methods in this family use the Jacobian determinant (or a divergence) and the curl as controls of an optimal
control problem \cite{Liao2008,Hsiao2014,ZhouLiao2022,VPreg}. The curl
is needed in addition to the Jacobian determinant: post-composing a map $\Phi$ with a volume-preserving
diffeomorphism $\psi$ that is the identity near $\partial\Om$ changes the map but neither its boundary values nor its
Jacobian determinant, since $\det\nabla(\psi\circ\Phi)=(\det\nabla\psi)\circ\Phi\cdot\det\nabla\Phi=\det\nabla\Phi$.
Earlier uniqueness arguments for these data concern small perturbations of the identity
\cite{Zhou2017uniq,Chen2016}; boundary values, Jacobian determinant and curl determine the map under structural
conditions, but not in general \cite{Zhao2026uniq}.

In the VP approach the controls $f$ and $g$ are unknowns with their own regularisation or prescribed targets
\cite{ZhouLiao2021,VPreg}. Here we consider instead a sequence of problems in which the
map and the controls are penalised for changing from the last committed state $(\Phi_i,f_i,g_i)$, not for
deviating from the identity. With soft coupling of the map to the controls, the $i$-th problem is
\begin{multline}\label{eq:joint}
  L_i(\Phi,f,g)=\Dat(\Phi)+\tfrac{\lambda}{2}\norm{\Phi-\Phi_i}^2
   +\tfrac{a_J}{2}\norm{J(\Phi)-f}^2+\tfrac{a_C}{2}\norm{\curl\Phi-g}^2\\
   +\tfrac{b_J}{2}\norm{f-f_i}^2+\tfrac{b_C}{2}\norm{g-g_i}^2 .
\end{multline}

\begin{proposition}[Elimination of the controls]\label{prop:elimination}
For fixed $\Phi$, \eqref{eq:joint} is minimised by $f^\ast=(a_JJ(\Phi)+b_Jf_i)/(a_J+b_J)$ and
$g^\ast=(a_C\curl\Phi+b_Cg_i)/(a_C+b_C)$, and
\[
  \min_{f,g}L_i(\Phi,f,g)=\Dat(\Phi)+\tfrac{\lambda}{2}\norm{\Phi-\Phi_i}^2+\tfrac{\mu}{2}\norm{J(\Phi)-f_i}^2
  +\tfrac{\nu}{2}\norm{\curl\Phi-g_i}^2,\qquad \mu=\tfrac{a_Jb_J}{a_J+b_J},\ \nu=\tfrac{a_Cb_C}{a_C+b_C}.
\]
In the exact-control limit $a_J,a_C\to\infty$ with $b_J,b_C$ fixed, $\mu\to b_J$, $\nu\to b_C$, and the recovered
controls are the geometry of the committed map, $f_{i+1}=J(\Phi_{i+1})$ and $g_{i+1}=\curl\Phi_{i+1}$.
\end{proposition}
The proof is a completion of squares (\cref{app:elimination}). We use the exact-control limit: the references of
the $i$-th problem are the Jacobian determinant and curl of the last committed map. The closest precedent
of this sequential formulation in the VP family is the fixed-point iteration of VPreg, which
resets its controls to the Jacobian determinant and curl of the current map \cite[Alg.~A.2]{VPreg}. In the discrete setting the
$i$-th inner energy is
\begin{equation}\label{eq:inner}
  \Ein_i(u)=\Fout(u)+\Pen_i(u),
\end{equation}
\begin{multline}\label{eq:penalty}
  \Pen_i(u)=\frac{\lambda}{2}\sum_{a\in\nodes}q_a\,|u_a-u_{i,a}|^2
   +\frac{\mu}{2}\sum_{K\in\cells}\sum_{\xi\in\Gpts}q^{G}\big(J_K(u;\xi)-J_K(u_i;\xi)\big)^2\\
   +\frac{\nu}{2}\sum_{K\in\cells}\sum_{\xi\in\Gpts}q^{G}\big|\curl_K u(\xi)-\curl_K u_i(\xi)\big|^2 ,
\end{multline}
where $u_i$ is the last committed displacement (the reference) and $\Fout$ is the outer energy of \cref{sec:outer}.
Throughout, \emph{committed} refers to outer iterates that passed the commit test of \cref{sec:algorithm}, and
\emph{accepted} to trial steps of the inner method. The three
penalty terms constrain how much the map, its local volume change and its local rotation may change in one
outer step; they do not pull the map back to the identity. At the reference they vanish to first order
(\cref{prop:consistency}): $\Pen_i(u_i)=0$ and $\nabla\Pen_i(u_i)=0$.

\subsection{Outer energy}\label{sec:outer}
The outer energy contains only absolute terms, which do not depend on the sequential references:
\begin{equation}\label{eq:outer}
  \Fout(u)=\Dat(u)+\Reg(u)+\Bpot(u),
\end{equation}
\begin{align}
  \Dat(u)&=\tfrac12\sum_{a\in\nodes}q_a\big(\widetilde I_m(X_a+u_a)-I_f(X_a)\big)^2,\label{eq:data}\\
  \Reg(u)&=\tfrac{\alpha}{2}\sum_{a\in\nodesint}q_a\,|(Lu)_a|^2,\label{eq:curvature}\\
  \Bpot(u)&=\kappa\sum_{K\in\cells}\frac{\omega}{27}\sum_{\xi\in\Spts}b\big(J_K(u;\xi)\big),\qquad
  b(J)=\begin{cases}-(1-J/J_b)^2\log(J/J_b),&0<J<J_b,\\ 0,&J\ge J_b,\end{cases}\label{eq:barrier}
\end{align}
with $b(J)=+\infty$ for $J\le0$. Here $L$ is the seven-point Laplacian with physical spacings applied
componentwise to the interior nodes (boundary values are zero), $\alpha\ge0$ is a curvature weight, $\kappa>0$ the
barrier weight and $J_b\in(0,1)$ the barrier activation level. The local barrier $b$ is nonnegative,
nonincreasing, twice continuously differentiable on $(0,\infty)$ (value, first and second derivative vanish at
$J=J_b$), strictly convex on $(0,J_b)$ and blows up as $J\downarrow0$. It acts only on cells that are already
strongly compressed at the sampled points, and it does not penalise expansion.

Two remarks on this design. First, because $\Reg$ and $\Bpot$ are absolute, the outer energy $\Fout$ is one fixed
function along the whole sequence of inner problems on a given grid; this is what makes the scheme monotone
(\cref{prop:monotone}). Second, the barrier is evaluated at finitely many points and cannot by itself exclude
non-positive Jacobians between those points (\cref{fig:toy}(c)); whole-cell positivity is enforced separately by the
certificate.

\subsection{Admissible set}\label{sec:admissible}
The admissible set is
\begin{equation}\label{eq:admissible}
  \adm=\Big\{u:\ u_a=0\ (a\in\nodes\setminus\nodesint),\quad X_a+u_a\in\Om\ (a\in\nodesint),\quad
  \min_{\xi\in[0,1]^3}J_K(u;\xi)>0\ \ \forall K\in\cells\Big\}.
\end{equation}
Every trial step accepted by the inner method, every committed iterate and every map the method returns is
certified to lie in $\adm$ (\cref{sec:certificate,sec:algorithm}); the only uncertified stage is the pre-optimiser of
variant P, whose output is certified at the handoff before it is used. By \cref{thm:invertibility}, the first and third conditions already imply
that $\Phi_u$ is a homeomorphism of $\Omb$ onto itself, so the second (sampling-domain) condition is implied;
the implementation keeps it as an inexpensive fail-safe.

%% file: sections/04_certificate.tex
\section{A floating-point rigorous certificate of whole-cell positivity}\label{sec:certificate}

The certificate decides, for a given nodal displacement $u$, whether
\begin{equation}\label{eq:certified-statement}
  J_K(u;\xi)>0\qquad\text{for every }K\in\cells\text{ and every }\xi\in[0,1]^3 ,
\end{equation}
where $J_K$ is computed in exact real arithmetic from the stored binary64 nodal displacements and spacings, and
$J_K$ on the closed cell is understood as the value of the cell polynomial (one-sided on faces). It returns
\emph{certified} (\eqref{eq:certified-statement} is proven), \emph{infeasible} (a reason for rejection was found)
or \emph{uncertain} (neither was established within the work budget or the arithmetic preconditions). Only
\emph{certified} is ever accepted; \emph{uncertain} is treated exactly like \emph{infeasible}. The certificate
concerns the map that is stored and returned, not an idealised unrounded iterate. Before any cell is examined it
checks that all inputs are finite, that $u_a=0$ on boundary nodes, and that every node is mapped into the image
domain (a floating-point test, \cref{rem:domain-test}).

\begin{figure}[t]
\centering
\resizebox{\textwidth}{!}{\input{figures/tikz/certificate.tex}}
\caption{The certificate for one cell. A test is applied only to cells left open by the previous one, and only test
(iii) can produce a Jacobian witness. Tests (i) and (iii) use interval arithmetic with outward rounding; test (ii)
uses the a priori bound of \cref{lem:prescreen}. The map is certified only if every cell is certified, and
\emph{uncertain} is treated as \emph{infeasible}. The post-hoc audit re-certifies every committed iterate on the CPU
with tests (i) and (iii) only.}\label{fig:certificate}
\end{figure}
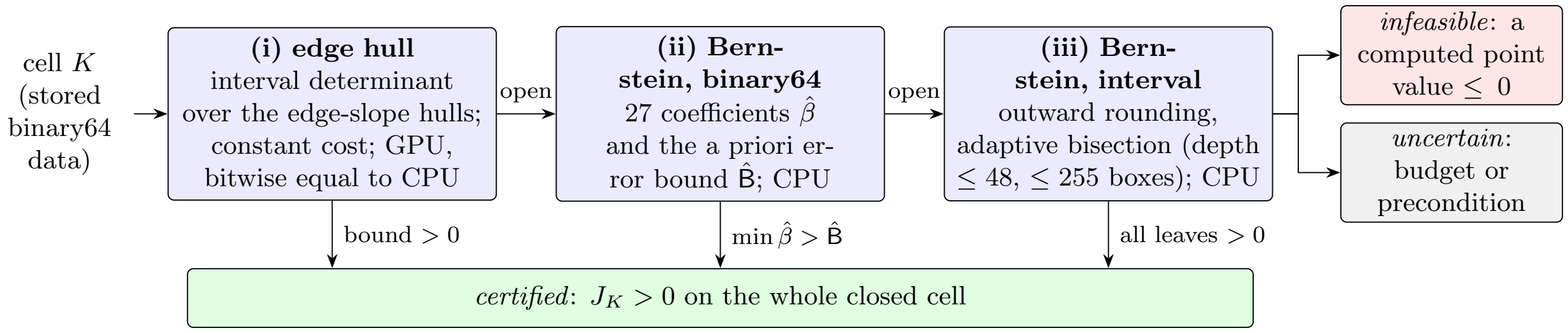

Each cell passes through up to three tests of increasing cost; a cell is certified as soon as one test certifies
it, and the next test is used only for cells the previous one left open:
(i) an edge-hull interval bound of constant cost (\cref{sec:edgehull}); (ii) floating-point Bernstein coefficients
with an a priori rounding-error bound (\cref{sec:prescreen}); (iii) interval Bernstein coefficients with adaptive
subdivision (\cref{sec:bernstein}). Only test (iii) can declare a cell infeasible on the basis of its Jacobian; the
pre-checks above reject the whole map (\cref{fig:certificate}). We describe test (iii) first, because tests (i) and (ii)
are shortcuts to it.

\subsection{Bernstein form of the Jacobian determinant on a cell}\label{sec:bernstein}
\begin{lemma}\label{lem:triquadratic}
For every cell $K$, $\xi\mapsto J_K(u;\xi)$ is a polynomial of degree at most two in each of $\xi_1,\xi_2,\xi_3$.
\end{lemma}
\begin{proof}
By \eqref{eq:q1map}, the $j$-th column of $\nabla\Phi_u$ is $e_j+h_j^{-1}\partial_{\xi_j}u$, which does not depend on
$\xi_j$ and is affine in each of the other two variables. Every term of the Leibniz expansion of the determinant
is a product of one entry from each column, hence has degree at most $0+1+1=2$ in each variable.
\end{proof}
Consequently $J_K$ is determined by its $27$ values on $\{0,\frac12,1\}^3$. Let
$B_0(s)=(1-s)^2$, $B_1(s)=2s(1-s)$, $B_2(s)=s^2$ and
\[
  T=\begin{pmatrix}1&0&0\\-\tfrac12&2&-\tfrac12\\0&0&1\end{pmatrix}.
\]
If $y_{pqr}$ denotes the value of $J_K$ at $(p/2,q/2,r/2)$, the Bernstein coefficients
$\beta=(T\otimes T\otimes T)y$ satisfy
\begin{equation}\label{eq:bernstein}
  J_K(u;\xi)=\sum_{i,j,k=0}^{2}\beta_{ijk}\,B_i(\xi_1)B_j(\xi_2)B_k(\xi_3).
\end{equation}
Since the Bernstein basis functions are nonnegative and sum to one on $[0,1]^3$,
\begin{equation}\label{eq:bernstein-bound}
  \min_{\xi\in[0,1]^3}J_K(u;\xi)\ \ge\ \min_{ijk}\beta_{ijk},
\end{equation}
and the eight corner coefficients equal the corner values of $J_K$. Restricting \eqref{eq:bernstein} to a
sub-box obtained by halving one reference axis gives a polynomial of the same degree whose Bernstein coefficients
follow exactly from two rounds of averaging of neighbouring coefficients (de Casteljau subdivision); the bound
\eqref{eq:bernstein-bound} then applies to each sub-box, and the sub-box bounds converge quadratically to the true
minimum as the boxes shrink \cite{CohenSchumaker1985}. Bernstein expansions with adaptive subdivision are the
standard tool for proving the validity of curvilinear finite elements \cite{Johnen2013}, and exactly this test---27 Bernstein
coefficients of the triquadratic $J$ obtained from samples by a fixed linear map, a corner value $\le0$ rejecting,
adaptive subdivision otherwise---was proposed for the linear hexahedral element by Johnen, Weill and Remacle, who
also showed that 20 of the 27 samples suffice \cite{Johnen2017}. Those works compute in floating point without rounding control;
the earlier one caps the subdivision depth and treats undetermined elements as invalid, as we do. What is new here
is (a) a treatment of rounding that makes the test rigorous for the stored binary64 data (\cref{sec:rounding,sec:prescreen}), (b) a constant-cost sufficient pre-test (\cref{sec:edgehull}), and (c) the use of the test as the
acceptance condition for every trial map of an optimiser, at $\sim7\times10^6$ cells per test. A recent preprint
proves that strict positivity on the boundary of a trilinear hexahedron implies positivity inside
\cite{Zhang2026hex}; the certificate is independent of that result.

\paragraph{Test (iii)} The 27 values are computed as intervals, mapped to interval Bernstein coefficients, and the
cell is certified if all lower endpoints are strictly positive. If some actually computed point value has an upper
endpoint $\le0$, the cell is \emph{infeasible} with a Jacobian witness. Otherwise the box is bisected along its
longest reference edge and the children are examined, lowest lower bound first; a negative Bernstein coefficient by
itself is \emph{not} a proof of infeasibility. A cell is certified only when all leaves are certified. Each cell has
a budget of subdivision depth $48$ and $255$ visited boxes; a cell not resolved within the budget is
\emph{uncertain}.

\subsection{Test (i): the edge-hull bound}\label{sec:edgehull}
\begin{figure}[t]
\centering
\includegraphics[width=\textwidth]{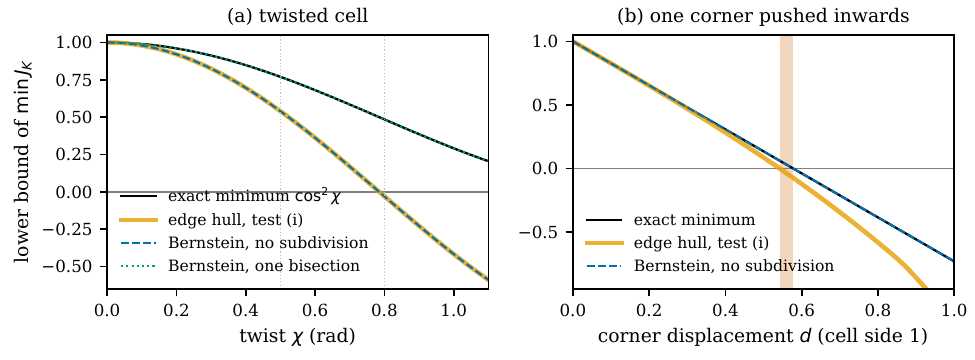}
\caption{Tightness of the lower bounds on two synthetic cells of side $1$. (a) Twisted cell: the exact minimum is
$\cos^2\chi$; the edge-hull bound and the Bernstein bound without subdivision both equal $\cos2\chi$; one bisection
along $\xi_3$ already gives the exact minimum. (b) One corner pushed inwards along the diagonal by $d$: the minimum is
attained at that corner and equals the smallest Bernstein coefficient; the edge hull becomes non-positive at
$d\approx0.542$, before the cell degenerates at $d=1/\sqrt3\approx0.577$ (shaded: valid cells that only tests (ii) and
(iii) certify). Bounds evaluated in binary64 without outward rounding; script in the ancillary files.}\label{fig:bounds}
\end{figure}

Evaluating 27 values per cell is unnecessary for most cells of a registration map, which are far from degenerate.
A much cheaper sufficient test follows from the structure of the Q1 gradient.
\begin{lemma}\label{lem:edgehull}
Let $\mathcal E_j(K)$ be the four edges of $K$ parallel to axis $j$, each with endpoints $e^-,e^+$. For every
$\xi\in[0,1]^3$ and all $r,j$,
\[
  \big(\nabla\Phi_u(\xi)\big)_{rj}=\delta_{rj}+\sum_{e\in\mathcal E_j(K)}w_e(\xi)\,\frac{u_{e^+,r}-u_{e^-,r}}{h_j},
  \qquad w_e(\xi)\ge0,\quad \sum_{e\in\mathcal E_j(K)}w_e(\xi)=1 ,
\]
where the weights $w_e$ are the bilinear shape functions of the two reference variables other than $\xi_j$.
Hence $\nabla\Phi_u(\xi)\in\Gh$ for all $\xi$, where $\Gh$ is the interval matrix with entries
$[\Gh]_{rj}=\delta_{rj}+\hull\{(u_{e^+,r}-u_{e^-,r})/h_j:e\in\mathcal E_j(K)\}$, and
$\min_\xi J_K(u;\xi)\ge\underline d$ for any interval $[\underline d,\overline d]$ that encloses
$\{\det A:A\in\Gh\}$, for instance the natural interval extension of a fixed determinant formula evaluated
with outward rounding.
\end{lemma}
The proof is in \cref{app:edgehull}. We evaluate the determinant by the cofactor expansion along the first row,
left-associated, in a fixed order. The bound is exact for affine cells (all four edge slopes coincide) and loses
tightness quickly as a cell becomes non-affine. For a unit cell whose top four corners are rotated by $+\chi$ and
bottom four corners by $-\chi$ about the cell's central vertical axis, with the trilinear map between them (not a
rotation by an interpolated angle, which would give $J=1$), $J_K=\cos^2\chi+(2\xi_3-1)^2\sin^2\chi$, so $\min J_K=\cos^2\chi$,
whereas the edge-hull bound is $\cos2\chi$, which is negative for $\chi>\pi/4$ although the cell is valid up to
$\chi=\pi/2$ (\cref{fig:bounds}(a)). For a cell with one corner pushed inwards the edge hull fails shortly before the
cell degenerates, while the Bernstein bound is exact (\cref{fig:bounds}(b)). Cells it cannot certify go to test (ii).

 Over the $26{,}681$ trial certificates of the $177$ runs of the frozen configuration on the validation and
held-out sets, test (i) decided $98.2\%$ of the cells, test (ii) $1.8\%$ and test (iii) about four cells in a
million, with subdivision depth at most $23$; no cell was uncertain in any trial, commit, handoff or final certificate.
On the final maps re-certified offline with the bound of every cell recorded (\cref{sec:results}) the shares are
$95.6\%$, $4.4\%$ and $1.3\times10^{-5}$.

\subsection{Test (ii): floating-point Bernstein coefficients with an a priori error bound}\label{sec:prescreen}
For cells left open by test (i), the 27 point values and the Bernstein coefficients are first computed in ordinary
binary64 arithmetic, $\hat\beta$, together with a cellwise bound $\hat{\mathsf B}$, itself computed in binary64, on the
rounding error, $|\hat\beta_{ijk}-\beta_{ijk}|\le\hat{\mathsf B}$ for all $i,j,k$, derived by a forward error analysis of the
fixed sequence of operations (\cref{lem:prescreen}, stated in full in \cref{app:prescreen}). The cell is certified if $\fl(\min\hat\beta-\hat{\mathsf B})>0$, which
is equivalent to $\min\hat\beta>\hat{\mathsf B}$. This test never declares a cell infeasible; cells it leaves open go to test (iii).

\subsection{Outward rounding and execution}\label{sec:rounding}
Tests (i) and (iii) use interval arithmetic in IEEE~754 binary64 with round-to-nearest-even. After every
elementary operation ($+,-,\times,\div$) the lower endpoint is moved to the next representable number towards
$-\infty$ and the upper endpoint towards $+\infty$; constants of the stencil ($0,1,\tfrac12,2$) are exact. Because
the round-to-nearest result $\hat z$ differs from the exact result $z$ by less than the distance from $\hat z$ to its
neighbour on the side of $z$, including in the subnormal range under gradual underflow, the exact result lies in the widened interval
(\cref{lem:outward}). Any non-finite intermediate result, division by an interval containing zero, or violated
precondition makes the affected cells \emph{uncertain}; the implementation never falls back to unguarded
arithmetic. The interval kernels are compiled without fast-math, reassociation or fused multiply--add contraction;
this is verified by inspecting the generated intermediate code in the regression suite, and at run time a smoke test
checks binary64 behaviour and gradual underflow. The kernel of test (ii) is compiled in the same way, but its generated
code has only been inspected by hand (\cref{app:prescreen}, remark (vi)).

A cheaper but bitwise-identical evaluation of the edge-hull bounds uses the following monotonicity.
\begin{lemma}[Hoisting]\label{lem:hoisting}
For fixed binary64 $h>0$ let $L_h(d)=\down\fl(\down d/h)$ and $U_h(d)=\up\fl(\up d/h)$, where $\down,\up$ denote
the neighbouring binary64 numbers towards $-\infty$ and $+\infty$. Under the arithmetic assumptions above, $L_h$
and $U_h$ are nondecreasing on the finite binary64 numbers. Consequently, for any finite set of binary64 edge
differences $\{d_k\}$, $\min_kL_h(d_k)=L_h(\min_kd_k)$ and $\max_kU_h(d_k)=U_h(\max_kd_k)$.
\end{lemma}
Thus the extreme raw edge differences are selected first and only two outward-rounded divisions are performed per
entry of $\Gh$, with the same endpoints as the direct evaluation.

\paragraph{Execution on the GPU} In the recommended configuration test (i) runs on the GPU with an operation sequence
that is bitwise identical to the CPU kernel: correctly rounded binary64 addition, multiplication and division
(\texttt{div.rn.f64}), no contraction, and the one-step outward moves implemented by integer increments of the bit
pattern. The first use of the GPU kernel in every run is compared bitwise with the CPU kernel; a mismatch switches
the GPU kernel off for the rest of the run, and the audit rejects any run with a recorded mismatch. Tests (ii) and (iii),
and the decision for the whole map, remain on the CPU. While this execution was introduced, every committed candidate
was in addition re-certified on the CPU; in 20 runs and 3448 such re-certifications no mismatch occurred, after which
the per-candidate re-certification was switched off; the frozen configuration keeps only the first-use comparison,
which found no mismatch in any of the $177$ runs of the validation and held-out sets.
Independently of the run, the post-hoc audit repeats the run's certificate on the CPU for every committed iterate
(requiring identical recorded bounds) and, in addition, re-certifies it with tests (i) and (iii) only, i.e.\ without
the a priori bound of test (ii), so that no returned map depends on the test-(ii) kernel, which is verified by
inspection only. For the $472$ runs reported in this article (the validation pairs, the held-out sets, the ablations,
the post-hoc time curves, the $180$\,s arm and the curvature arms of the frontier; the $38$ runs that certify
$J>\delta>0$ are reported separately in \cref{sec:results}) this covers the $13{,}401$ committed
iterates of the $470$ runs that reached a native commit (the two runs of the LM ablation arm have none): every one
was re-certified with a positive bound, and every recorded bound was reproduced bitwise.

\subsection{Soundness and global consequence}\label{sec:soundness}
\begin{theorem}[Soundness]\label{thm:soundness}
Assume IEEE~754 binary64 arithmetic with round-to-nearest-even and gradual underflow, executed without
contraction or reassociation, on the CPU and, for test (i), on the GPU. If the certificate returns \emph{certified}
for a nodal displacement $u$, then $u_a=0$ on all boundary nodes and \eqref{eq:certified-statement} holds for the real
numbers represented by the stored values of $u$ and $h$. If it returns \emph{infeasible with a Jacobian witness},
some cell contains a point at which $J_K\le0$.
\end{theorem}
\begin{proof}[Proof sketch]
By \cref{lem:outward}, every interval computed in tests (i) and (iii) contains the exact value of the corresponding
real expression of the stored inputs. A cell certified by test (i) satisfies $\min_\xi J_K>0$ by \cref{lem:edgehull},
and \cref{lem:hoisting} shows that the hoisted evaluation produces the same endpoints; the GPU kernel performs the same
correctly rounded operations in the same order and therefore produces the same endpoints. A cell certified by test
(ii) satisfies $\min\beta>0$ by \cref{lem:prescreen}, hence $\min_\xi J_K>0$ by
\eqref{eq:bernstein-bound}. A cell certified by test (iii) has a finite set of leaf boxes covering $[0,1]^3$, and on
each leaf interval Bernstein coefficients that enclose the exact coefficients of the restriction of $J_K$ and have
positive lower endpoints. A Jacobian witness is a point value computed at an actual point of a box whose enclosing
interval has upper endpoint $\le0$. Details are in \cref{app:soundness}.
\end{proof}

\begin{remark}[Domain test]\label{rem:domain-test}
A map can also be rejected because a node is found outside the image domain. This test compares floating-point
positions $\fl(X_a+u_a)$ with the domain bounds and carries no Jacobian witness. For maps that satisfy
\eqref{eq:certified-statement} with zero boundary displacement it is redundant in exact arithmetic
(\cref{thm:invertibility}); it is kept as an inexpensive fail-safe, and a rejection by it is reported separately.
\end{remark}

The following consequence turns the local statement \eqref{eq:certified-statement} into a global one.
\begin{theorem}[Global invertibility]\label{thm:invertibility}
Let $u$ be a nodal displacement with $u_a=0$ on all boundary nodes and $\min_{\xi\in[0,1]^3}J_K(u;\xi)>0$ for every
$K\in\cells$. Then $\Phi_u$ is a bi-Lipschitz homeomorphism of $\Omb$ onto $\Omb$; it maps $\Om$ onto $\Om$ and $\partial\Om$
onto $\partial\Om$.
\end{theorem}
The proof (\cref{app:invertibility}) combines a Brouwer degree argument with the cellwise inverse function theorem;
the homeomorphism part also follows from Ball's global invertibility theorem for Sobolev maps
\cite[Theorem~2]{Ball1981}, which gives an inverse in $W^{1,q}$ rather than a Lipschitz one. Together, \cref{thm:soundness,thm:invertibility} give the guarantee that is used throughout the paper.
\begin{corollary}\label{cor:certified-homeo}
Every displacement that passes the certificate defines a bi-Lipschitz homeomorphism of the image domain onto
itself. In particular every committed iterate of the solver, and every map it returns at any time, has a
positive Jacobian determinant at every point of every closed cell (one-sided on faces) and is globally one-to-one.
\end{corollary}

\begin{remark}[Grid-based folding checks measure other maps]\label{rem:other-checks}
Folding is usually reported by evaluating a finite-difference Jacobian determinant at the voxels
\cite{Hering2023,Liu2024fdj}. Such a check samples a
different discrete object and is neither necessary nor sufficient for invertibility of the trilinear map: a central
difference at a node ignores the value at the node itself, and positivity at finitely many points says nothing
about the points in between (\cref{fig:toy}). Decompositions of each voxel into tetrahedra test a piecewise-linear
map, which is also
different from $\Phi_u$; the 24-tetrahedra partition is used in the hyperelastic registration of Burger et al., who
attribute it to S.~Heldmann \cite[\S4]{Burger2013}, and Liu et al.\ derive conditions for piecewise-linear interpolation on simplices, which they
note is inconsistent with trilinear interpolation \cite{Liu2024fdj}. In the public Learn2Reg evaluation code the
number of foldings is counted after the determinant, shifted by $3$, has been clipped to at least $10^{-9}$, so the
reported count is zero for every displacement field \cite[\texttt{evaluation.py}, l.~139--147]{L2Rcode}. On our own certified maps the 24-tetrahedra test
flags cells on which the trilinear map is certified positive ($14$ to $246$ cells per map, median $148$, on every one
of the $50$ H1 maps, each certified positive on every cell and with the minimum of $J$ over the map bracketed by
subdivision at $1.1\times10^{-6}$ or above; \cref{sec:results}), and the values of $J$ at the cell centres overestimate the
minimum of $J$ over the map by one to three orders of magnitude (\cref{sec:results}). Whether maps produced by methods that report zero folding
under the voxelwise check satisfy \eqref{eq:certified-statement} is examined in \cref{sec:results}
(\cref{fig:baseline-verdicts}).
\end{remark}

%% file: figures/tikz/certificate.tex
\begin{tikzpicture}[font=\footnotesize, >=Stealth,
  test/.style={draw, rounded corners=2pt, align=center, fill=blue!8, text width=37mm, minimum height=15mm, inner sep=3pt},
  res/.style={draw, rounded corners=2pt, align=center, minimum height=7mm, inner sep=2pt}]
\node[align=center, text width=15mm] (cell) {cell $K$\\(stored\\binary64 data)};
\node[test, right=4mm of cell] (t1) {\textbf{(i) edge hull}\\interval determinant over the edge-slope hulls; constant cost; GPU, bitwise equal to CPU};
\node[test, right=7mm of t1] (t2) {\textbf{(ii) Bernstein, binary64}\\27 coefficients $\hat\beta$ and the a~priori error bound $\hat{\mathsf B}$; CPU};
\node[test, right=7mm of t2] (t3) {\textbf{(iii) Bernstein, interval}\\outward rounding, adaptive bisection (depth $\le48$, $\le255$ boxes); CPU};
\draw[->] (cell) -- (t1);
\draw[->] (t1) -- node[above, font=\scriptsize]{open} (t2);
\draw[->] (t2) -- node[above, font=\scriptsize]{open} (t3);
\node[res, fill=red!10, right=8mm of t3, yshift=7mm, text width=25mm] (bad) {\emph{infeasible}: a computed point value $\le0$};
\node[res, fill=gray!12, right=8mm of t3, yshift=-7mm, text width=25mm] (unc) {\emph{uncertain}: budget or precondition};
\draw[->] (t3.east) -- ++(3mm,0) |- (bad.west);
\draw[->] (t3.east) -- ++(3mm,0) |- (unc.west);
\path (t1.south west) ++(0,-8mm) coordinate (okw);
\path (t3.south east) ++(0,-8mm) coordinate (oke);
\node[res, fill=green!12, anchor=north, minimum width=127mm] (ok) at ($(okw)!0.5!(oke)$)
  {\emph{certified}: $J_K>0$ on the whole closed cell};
\draw[->] (t1.south) -- node[right, font=\scriptsize]{bound $>0$} (t1.south |- ok.north);
\draw[->] (t2.south) -- node[right, font=\scriptsize]{$\min\hat\beta>\hat{\mathsf B}$} (t2.south |- ok.north);
\draw[->] (t3.south) -- node[right, font=\scriptsize]{all leaves $>0$} (t3.south |- ok.north);
\end{tikzpicture}

%% file: sections/05_algorithm.tex
\section{Algorithm}\label{sec:algorithm}

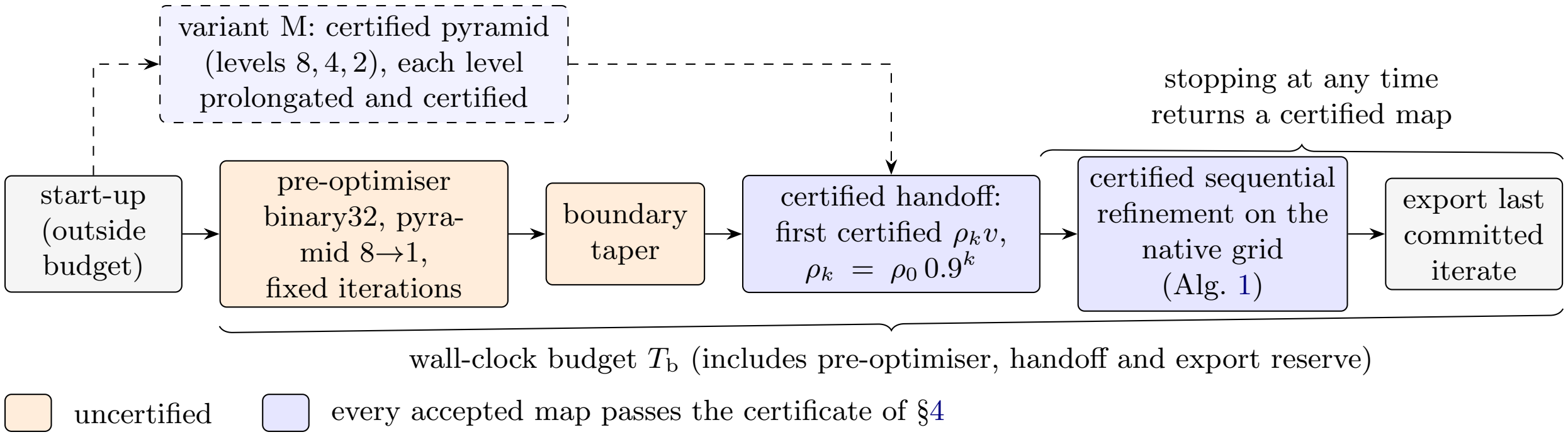
\begin{figure}[t]
\centering
\resizebox{\textwidth}{!}{\input{figures/tikz/pipeline.tex}}
\caption{One registration run. Orange: uncertified stages; blue: stages in which every accepted map is certified.
Variant M (dashed) replaces the pre-optimiser by the certified pyramid. The budget $\Tb$ includes the pre-optimiser,
the handoff and the export reserve; from the certified handoff on, stopping at any time returns a certified map.}
\label{fig:pipeline}
\end{figure}

The solver has three stages (\cref{fig:pipeline}): an initialisation stage that produces a certified initial
displacement on the
native grid, a certified sequential refinement on the native grid that runs until the time budget is exhausted,
and the export of the last committed iterate. We first describe the refinement, which carries all guarantees,
and then the two initialisation variants.

\subsection{Outer iteration}\label{sec:outer-iteration}
Let $u_0\in\adm$ be the certified native initialiser (\cref{sec:initialisation}) and
$\GF(u)=\norm{\nabla_\theta\Fout(u)}_2/\sqrt{\omega}$, where $\theta$ collects the interior nodal displacements. Given
the last committed iterate $u_i$, one outer iteration (\cref{alg:refine})
\begin{enumerate}[label=(\roman*)]
  \item freezes the references $J_K(u_i;\cdot)$ and $\curl_Ku_i(\cdot)$ at the Gauss points, which defines
  $\Ein_i$ in \eqref{eq:inner};
  \item builds a symmetric positive definite metric $P_i$ at $u_i$ (\cref{sec:metric}) and runs at most $k_{\max}$
  accepted steps of the preconditioned gradient method of \cref{sec:inner} on $\Ein_i$, starting from $u_i$;
  every trial point of the inner method is certified;
  \item if the inner method ended regularly (forcing condition met, $k_{\max}$ accepted steps, or a small-step
  stall), recomputes, independently of the inner method, the changes of $\Ein_i$ and of $\Fout$ from $u_i$ to the last
  accepted inner state $\bar u$, and commits $u_{i+1}=\bar u$ only if both decrease by more than a resolution guard
  (\cref{sec:stable-comparison}).
\end{enumerate}
The inner method ends with a failure if, in some inner step, no trial is accepted within the backtracking budget
(for instance because every trial is rejected by the certificate, falls outside the image domain or is not a
resolvable decrease), or if a non-finite value or the expiry of the budget occurs. In that case the progress of the
whole outer iteration, including inner steps accepted earlier, is discarded. If the inner method fails or the commit
test fails, the solver has no restart rule: the level ends, and on the native grid this ends the run, which then
returns the last committed iterate, or $u_0$ if nothing was committed.

\paragraph{Termination} A run is called \emph{converged} only if $\GF(u_i)\le10^{-8}+10^{-6}\GF(0)$, where $\GF(0)$
is evaluated once at the zero displacement on the native grid. Otherwise the run ends when the time budget is
exhausted and returns the last committed iterate; we call such runs \emph{budget-terminated} (every run of \cref{sec:experiments} ended this way, by the native-level budget, the global budget or the work deadline,
with the final $\GF$ at least $8\times10^{4}$ times its threshold).

The inner method also checks a forcing condition $\norm{\nabla_\theta\Ein_i}_2/\sqrt\omega\le t_i$ with
$t_i=\eta_i\GF(u_i)$, $\eta_i=\eta_0\sqrt{\min(\GF(u_i),\GF(0))/\GF(0)}\le\eta_0=10^{-2}$, and stops early if it
holds. The commit test does not require it. Its role in the analysis is discussed in \cref{sec:analysis}.

\subsection{Inner method: Poisson-preconditioned gradient with certified backtracking}\label{sec:inner}
Write $g_k=\nabla_\theta\Ein_i(\theta_k)$. The direction is $p_k=-P_i^{-1}g_k$; since $P_i$ is symmetric positive
definite, $g_k^\top p_k<0$ unless $g_k=0$. The trial step $\theta_k+t\,p_k$ is chosen as follows.

\paragraph{Initial step} The first inner step uses $t=1$. Later steps use the Barzilai--Borwein step for the
fixed metric, $t_{\mathrm{BB2}}=s^\top y/(y^\top P_i^{-1}y)$ with $s=\theta_k-\theta_{k-1}$, $y=g_k-g_{k-1}$ and
$P_i^{-1}y=p_{k-1}-p_k$, clamped to $[10^{-8},10]$; when the curvature or angle condition fails (cosine below
$10^{-12}$) it is replaced by $2\min(t_{k-1},5)$ \cite{BarzilaiBorwein1988}. The history is cleared at every new
reference. The step is then capped by $0.9$ times the largest step that keeps all nodes inside $\Om$, the domain of
the image interpolant.

\paragraph{Jacobian-aware step prediction} By \eqref{eq:detcubic}, along the search line the Jacobian
determinant at any fixed reference point is a cubic polynomial in $t$. For the points of $\Spts$ in the cells
containing the (up to $128$) most recent locations at which a trial was found infeasible, the solver computes the
first positive root of this cubic in $(0,t]$ (also at the witness locations themselves; roots with imaginary part
below $10^{-8}$ are treated as real) and caps $t$ at $0.9$ times the smallest such root. The prediction is a
heuristic: other cells may still become non-positive, and every trial is certified regardless.

\paragraph{Acceptance of a trial} A trial $\theta_k+tp_k$ is accepted if (a) the certificate of
\cref{sec:certificate} returns \emph{certified}, and (b) the stable decrease $\mathrm{ared}=\Ein_i(\theta_k)-\Ein_i(\theta_k+tp_k)$,
computed as in \cref{sec:stable-comparison}, satisfies $\mathrm{ared}-s_E\ge -c_1t\,g_k^\top p_k$ and
$\mathrm{ared}-s_E>0$ with $c_1=10^{-4}$. Otherwise $t$ is halved, at most $25$ times; backtracking also stops as soon
as $t\,|g_k^\top p_k|$ falls below the resolution of the energy, since such steps cannot be distinguished from
round-off. After a trial rejected with a Jacobian witness, the next trial uses $\min(t/2,\hat t)$, where $\hat t$ is the
step prediction above recomputed with the newly found witness. The inner method stops after $k_{\max}=5$ accepted steps, when the forcing condition holds, or with a failure when backtracking is exhausted
(\cref{sec:outer-iteration}); \cref{lem:backtracking} shows that, in exact arithmetic, backtracking cannot be
exhausted at a non-stationary point.

\subsection{Metric}\label{sec:metric}
The metric $P_i$ has three parts, each addressing one difficulty of the inner problems: a Poisson-type operator,
which is the quadratic part of the sequential penalties and makes the gradient method insensitive to the grid
spacing; a diagonal equilibration, which accounts for the data-term curvature that dominates at tissue boundaries;
and a damping near compressed cells, which keeps the step from driving cells towards the boundary of $\adm$. The
metric is built once per outer iteration at $u_i$ and kept fixed during the inner solve, so that the BB2 identity
above holds.

\paragraph{Q1 Poisson metric} For each displacement component we use
$M=\omega\,\mathsf a\,\Id+b_P\,K_Q$, where $K_Q=M_3\otimes M_2\otimes K_1+M_3\otimes K_2\otimes M_1+K_3\otimes M_2\otimes M_1$ is the
consistent Q1 stiffness matrix restricted to interior nodes ($M_s=h_s\operatorname{tridiag}(1,4,1)/6$,
$K_s=\operatorname{tridiag}(-1,2,-1)/h_s$), $b_P=(\mu+\nu)/2$ and $\mathsf a=\lambda+\bar{\mathsf a}$. The matrix is
diagonalised by the type-I discrete sine transform, with eigenvalues $\omega\big(\mathsf a+b_P\Lambda_Q\big)$,
$\Lambda_Q=d_1m_2m_3+m_1d_2m_3+m_1m_2d_3$, $d_s=4\sin^2(\vartheta_s/2)/h_s^2$,
$m_s=(2+\cos\vartheta_s)/3$, $\vartheta_s=\pi k_s/(n_s-1)$, so $M^{-1}$ costs two fast transforms. Preconditioning a gradient method by the spectral inverse of an elliptic
operator is also the basis of the Picard iteration of Mang and Biros \cite{MangBiros2015}; here the operator comes from
the quadratic part of the sequential penalties, not from a regulariser of the outer energy. The shift
$\bar{\mathsf a}=\frac{1}{|\nodesint|}\sum_{a\in\nodesint}|\nabla\widetilde I_m(X_a+u_{i,a})|^2/3$ is the isotropic mean of
the Gauss--Newton curvature of the data term; without it the lowest Poisson modes are weighted by
$\omega(\lambda+b_P\Lambda_{Q,\min})$ alone, an order of magnitude below the image curvature when $\lambda$ is small: on the
native grid $\lambda+b_P\Lambda_{Q,\min}\approx2.9\times10^{-5}$ ($\lambda=2.8\times10^{-6}$, $b_P=0.03$), while the
recorded $\bar{\mathsf a}$ of the H1 runs lies between $2.7\times10^{-4}$ and $5.6\times10^{-4}$.

\paragraph{Diagonal equilibration} Let $d$ be the exact diagonal of the Gauss--Newton matrix of $\Ein_i$ at $u_i$,
$H=K_0^\top K_0+L_\Spts^\top\diag(b''_\Spts)L_\Spts$, where $K_0$ is the Jacobian of the non-barrier residuals and
$L_\Spts$ the derivative of the Jacobian determinant at the barrier points; $d\ge\lambda\omega>0$ because of the
proximity block. Let $m$ be the constant diagonal of $M$. With $S=\diag(\sqrt{m/d})$ the equilibrated metric
$S^{-1}MS^{-1}$ has the same diagonal as $H$.

\paragraph{Affine-scaling damping} Let $m_K\in(0,\infty)$ be a computable estimate of the smallest Jacobian
determinant on cell $K$ (exact minima along the twelve cell edges, values at the points of $\Spts$, and estimates of
face and interior minima for cells where the Bernstein bound indicates a possible interior minimum), and
$r_a=\min\{1,\max(\min_{K\ni a}m_K,10^{-12})/J_{\mathrm{act}}\}$ with $J_{\mathrm{act}}=0.25$. With $R=\diag(r_a)$ the
native-level metric is
\begin{equation}\label{eq:metric}
  P_i^{-1}=S\,R\,M^{-1}R\,S .
\end{equation}
Near a strongly compressed cell, the damping scales the rate of change of its Jacobian determinant along $p$ in
proportion to its current value, in the spirit of affine-scaling interior-point methods \cite{Dikin1967}. This is
a heuristic: it does not guarantee positivity, which remains the job of the certificate. Because $R$ and $S$ are
positive diagonal matrices, $P_i$ is symmetric positive definite and the stationary points are unchanged. On the
two coarsest levels of variant M the metric is $P^{-1}=RM^{-1}R$, without the shift and the equilibration.

\subsection{Stable energy comparison}\label{sec:stable-comparison}
Differences of the energies are not computed by subtracting two large sums. For the squared-residual blocks the
difference is accumulated as $\frac12\sum(r_t-r_b)(r_t+r_b)$, and for the barrier as an algebraically rearranged
difference of $b$: with $x,y$ the values of $J$ at a barrier point before and after the step, both replaced by $J_b$
if larger, and $r_x=1-x/J_b$, $r_y=1-y/J_b$,
\[
  b(y)-b(x)=\frac{y-x}{J_b}\,(r_x+r_y)\log\frac{x}{J_b}-r_y^2\log\frac yx ,
\]
with the logarithms evaluated by $\log1p$ near $J_b$; this avoids the cancellation of two large terms near $J_b$ and
near $0$. On the
GPU the accumulation is in binary64 with a guard $4096\,\eps\max(1,E_b,E_t)$, where $\eps=2^{-52}$ is the binary64
machine epsilon and $E_b,E_t$ the energies before and after the step; whenever the decision is within eight guards of
the threshold, or for very small steps, the comparison is repeated on the CPU, accumulated in \texttt{long double}
(x87 extended precision on our platform), with the guard $s_E=16\,\eps\max(E_b,E_t)$, and likewise
$s_F=16\,\eps\max(\Fout_b,\Fout_t)$ for the outer energy. These are resolution guards, not rigorous
error bounds: the energy decrease is verified to this resolution, not proven in the sense of the certificate. In the
recommended configuration, three terms of each energy difference are first evaluated in binary32, and comparisons far
from their threshold are decided with these screened values; both the inner acceptance and the commit test are
affected (\cref{sec:implementation}).

\subsection{Initialisation}\label{sec:initialisation}
Two variants produce the native initialiser $u_0$ on the native grid (the 1\,mm image grid; variant M works on
coarser levels first). Both end with the same certified handoff: given a candidate displacement $v$ on the native
grid with $v=0$ on $\partial\Om$, the solver tries $\rho_kv$ for $\rho_k=\rho_0\,0.9^k$, $k=0,\dots,6$, and uses the first
one that passes the certificate. If none passes, the run stops, returns no map and is reported as a failed handoff.
The handoff is certified but is not a committed iterate; it is recorded and audited as the initialiser (index $0$).

\paragraph{Variant M: certified pyramid} The sequential solver above runs on a four-level pyramid with shrink
factors $8,4,2,1$; coarse images are Gaussian-smoothed and resampled from the native images, and each level $l$ has
its own outer energy $\Fout^{(l)}$, geometric weights $\mu^{(l)}=\nu^{(l)}$, curvature weight $\alpha^{(l)}$ and metric.
Each coarse level runs within a fixed time box; its last committed displacement is prolongated to the next grid
by Q1 interpolation and certified with the backoff above ($\rho_0=1$ between coarse levels, $\rho_0=0.9$ for the
native level). Monotonicity holds within each level, not across levels, since the objectives differ.

\paragraph{Variant P: uncertified binary32 pre-optimisation (recommended)} A pre-optimiser minimises, in
binary32 (single precision) on the GPU, the surrogate
\begin{multline}\label{eq:preopt}
  \mathcal Q^{(l)}(u)=\tfrac12\sum_x\big(I_m^{\mathrm{lin}}(x+u(x))-I_f(x)\big)^2
  +\tfrac{\mu^{\mathrm{pre},(l)}}{2}\sum_K\big[(J_K-1)^2+|\curl_K u|^2\big]\\
  +\tfrac{\alpha^{(l)}}{2}\sum_x|Lu|^2+\kappa_{\mathrm{pre}}\sum_K\max(0,j_0-J_K)^2
\end{multline}
on the same $8,4,2,1$ pyramid, with trilinear image sampling ($I_m^{\mathrm{lin}}$) and geometry evaluated at cell
centres ($J_K$, $\curl_K$ here denote cell-centre values). Each iteration smooths the gradient with a Gaussian of one
voxel, takes an Adam step \cite{KingmaBa2015}, scaled down if necessary so that no displacement component changes by
more than half a voxel, and smooths the displacement with a Gaussian of one voxel; as in FireANTs, the two smoothing
widths are the main hyperparameters \cite{Jena2024fireants}. The number of iterations per level is fixed ($50$, $50$, $100$, $60$; \cref{tab:params}), and a safety time cap of $60$\,s is in
place; when the cap does not bind, the output is independent of machine speed (in all $472$ reported runs the
pre-optimiser ran its configured iterations on every level, taking at most $3.3$\,s in total). There is no acceptance test and no certificate in this stage, and its geometric
penalties are absolute (deviations from $J=1$ and $\curl=0$), unlike the sequential penalties \eqref{eq:penalty}.
The resulting displacement is multiplied by a boundary taper $w(x)=\prod_k\min(1,\delta_k(x)/4)$, where $\delta_k$
is the distance in voxels to the nearer face orthogonal to axis $k$, so that it vanishes on $\partial\Om$, and handed
to the certified handoff with $\rho_0=1$ (on H1 the handoff passed at $\rho=1$ in $25$ of $49$ runs, at
$\rho=0.9$ in $23$ and at $\rho=0.81$ in one; over all $177$ runs of the frozen configuration on the validation and held-out sets it passed at
$\rho=1$, $0.9$ and $0.81$ in $79$, $90$ and $5$ runs and at $\rho=0.9^4\approx0.66$ on three H4 pairs, in at most five attempts and
$1.3$--$2.0$\,s per run on average). From then on
the run is the certified native refinement of
\cref{sec:outer-iteration}; implementation details of the pre-optimiser are in \cref{sec:implementation}.

\begin{remark}
The handoff does not depend on how the candidate displacement was produced. Any registration method whose output
can be expressed as a nodal displacement on the image grid can be used as a starting point; the guarantees of the
returned map are those of the certified refinement. (In a pre-freeze trial on the tuning pairs, a FireANTs greedy
field certified at the handoff gave a higher seg4 but a lower seg35 and a rougher map than the pre-optimiser after
$150$\,s of refinement, and was not adopted.)
\end{remark}

\subsection{Budget and anytime behaviour}\label{sec:budget}
The run has a wall-clock budget $\Tb$ (recommended configuration: $\Tb=90$\,s, of which $1$\,s is reserved for export). The budget includes the pre-optimiser and the handoff; process start-up
and a dry pass that warms the compiled kernels are outside it (on H1 the in-budget end-to-end time is $88.1$\,s on
average and the whole process takes $108$\,s of wall-clock time, $8.8$\,s of which is the compilation pre-warm). From
the certified handoff on (after about $5$\,s of the budget), the last committed iterate is certified at every moment, and its outer energy is not larger than that of
the native initialiser, up to the resolution guard of the stable comparison (\cref{prop:monotone} and the remark
after it). Stopping at any time after the handoff therefore returns a certified map; the time--accuracy curves of
\cref{sec:results} read out the committed iterates at intermediate times.

\begin{algorithm}[t]
\caption{Certified sequential refinement on the native grid}\label{alg:refine}
\begin{algorithmic}[1]
\Require certified $u_0\in\adm$, budget $\Tb$
\State $i\gets0$; compute $\GF(0)$ once at $u=0$
\While{time remains \textbf{and} $\GF(u_i)>10^{-8}+10^{-6}\GF(0)$}
  \State freeze $J_K(u_i;\Gpts)$, $\curl_Ku_i(\Gpts)$; build $P_i$ \Comment{defines $\Ein_i$}
  \State $(\bar u,\mathit{status})\gets$ \textsc{InnerPG}$(\Ein_i,P_i,u_i)$ \Comment{inner method of \cref{sec:inner}; every trial certified; $\le k_{\max}$ accepted steps}
  \If{$\mathit{status}$ regular \textbf{and} $\Ein_i(\bar u)<\Ein_i(u_i)-s_E$ \textbf{and} $\Fout(\bar u)<\Fout(u_i)-s_F$}
    \State $u_{i+1}\gets\bar u$; $i\gets i+1$ \Comment{commit}
  \Else
    \State \textbf{break}
  \EndIf
\EndWhile
\State \Return $u_i$ \Comment{certified; converged only if the GF test holds}
\end{algorithmic}
\end{algorithm}

%% file: figures/tikz/pipeline.tex
\begin{tikzpicture}[font=\footnotesize, >=Stealth,
  box/.style={draw, rounded corners=2pt, align=center, minimum height=11mm, inner sep=3pt},
  sUnc/.style={box, fill=orange!14}, sCert/.style={box, fill=blue!10}, sOut/.style={box, fill=gray!8}]
\node[sOut, text width=17mm] (start) {start-up\\(outside budget)};
\node[sUnc, right=4mm of start, text width=29mm] (pre) {pre-optimiser\\binary32, pyramid $8{\to}1$,\\fixed iterations};
\node[sUnc, right=4mm of pre, text width=15mm] (taper) {boundary\\taper};
\node[sCert, right=4mm of taper, text width=30mm] (hand) {certified handoff:\\first certified $\rho_k v$,\\$\rho_k=\rho_0\,0.9^k$};
\node[sCert, right=4mm of hand, text width=27mm] (ref) {certified sequential\\refinement on the\\native grid (Alg.~\ref{alg:refine})};
\node[sOut, right=4mm of ref, text width=17mm] (exp) {export last\\committed\\iterate};
\foreach \a/\b in {start/pre, pre/taper, taper/hand, hand/ref, ref/exp} \draw[->] (\a) -- (\b);
\node[box, dashed, above=4mm of pre, text width=42mm, fill=blue!5] (varM) {variant M: certified pyramid (levels $8,4,2$), each level prolongated and certified};
\draw[->, dashed] (start.north) |- (varM.west);
\draw[->, dashed] (varM.east) -| (hand.north);
\draw[decorate, decoration={brace, mirror, amplitude=4pt}] ([yshift=-2mm]pre.south west) -- ([yshift=-2mm]exp.south east)
  node[midway, below=5pt] {wall-clock budget $T_{\mathrm b}$ (includes pre-optimiser, handoff and export reserve)};
\draw[decorate, decoration={brace, amplitude=4pt}] ([yshift=2mm]hand.north east) -- ([yshift=2mm]exp.north east)
  node[midway, above=5pt, align=center] {stopping at any time\\returns a certified map};
\node[sUnc, minimum height=4mm, minimum width=5mm, inner sep=1pt, below=13mm of start.south west, anchor=west] (l1) {};
\node[right=1mm of l1] (t1) {uncertified};
\node[sCert, minimum height=4mm, minimum width=5mm, inner sep=1pt, right=4mm of t1] (l2) {};
\node[right=1mm of l2] {every accepted map passes the certificate of \S\ref{sec:certificate}};
\end{tikzpicture}

%% file: sections/06_analysis.tex
\section{Properties of the certified sequential scheme}\label{sec:analysis}

Throughout this section the grid, the images and the weights are fixed (one level), and $\lambda,\kappa>0$,
$\mu,\nu,\alpha\ge0$. We first record the properties that hold for every run, and then a conditional statement on
stationarity, which the budget-terminated runs of \cref{sec:experiments} do not satisfy (\cref{rem:no-convergence}).

\begin{proposition}[First-order consistency of the sequential penalties]\label{prop:consistency}
For every $u_i\in\adm$, $\Pen_i\ge0$, $\Pen_i(u_i)=0$ and $\nabla\Pen_i(u_i)=0$. Consequently
$\Ein_i(u_i)=\Fout(u_i)$ and $\nabla\Ein_i(u_i)=\nabla\Fout(u_i)$; in particular $u_i$ is a stationary point of
$\Ein_i$ if and only if it is a stationary point of $\Fout$.
\end{proposition}
\begin{proof}
Each term of \eqref{eq:penalty} has the form $\frac c2\sum_k w_k\,|\rho_k(u)-\rho_k(u_i)|^2$ with $w_k\ge0$ and
$\rho_k$ continuously differentiable on $\adm$ (the nodal displacement, $J_K$ and $\curl_K$ at a Gauss point). It is
nonnegative, vanishes at $u=u_i$, and its gradient
$c\sum_kw_k\,\nabla\rho_k(u)^\top(\rho_k(u)-\rho_k(u_i))$ vanishes at $u=u_i$.
\end{proof}
The sequential penalties therefore change the local model of $\Fout$ only at second order: they act as a
data-dependent metric on the step, not as a bias of the fixed points. This is the discrete counterpart of the
proximity principle of \cref{sec:sequential-model}.

\begin{proposition}[Monotonicity and summability]\label{prop:monotone}
Let $(u_i)_{i\ge0}\subset\adm$ satisfy $\Ein_i(u_{i+1})<\Ein_i(u_i)$ for every $i$ (in exact arithmetic the commit
test of \cref{sec:outer-iteration} implies this). Then
\begin{equation}\label{eq:descent}
  \Fout(u_{i+1})+\Pen_i(u_{i+1})<\Fout(u_i)\qquad\text{for all }i,
\end{equation}
so $\Fout(u_i)$ is strictly decreasing, and
\[
  \sum_{i\ge0}\Pen_i(u_{i+1})\le\Fout(u_0)-\inf_i\Fout(u_i)\le\Fout(u_0)<\infty .
\]
In particular, with $\norm{\cdot}_q$ and $\norm{\cdot}_{q^G}$ the weighted Euclidean norms of \eqref{eq:penalty},
$\sum_i\lambda\norm{u_{i+1}-u_i}_{q}^2<\infty$, $\sum_i\mu\norm{J(u_{i+1})-J(u_i)}_{q^G}^2<\infty$ and
$\sum_i\nu\norm{\curl u_{i+1}-\curl u_i}_{q^G}^2<\infty$; hence $u_{i+1}-u_i\to0$.
\end{proposition}
\begin{proof}
By \cref{prop:consistency}, $\Ein_i(u_i)=\Fout(u_i)$, so $\Fout(u_{i+1})+\Pen_i(u_{i+1})=\Ein_i(u_{i+1})<\Fout(u_i)$.
Summing over $i$ telescopes. Finally $\Fout\ge0$ because $\Dat,\Reg\ge0$ and $b\ge0$.
\end{proof}
The commit test of \cref{sec:outer-iteration} requires a decrease of both $\Ein_i$ and $\Fout$; only the former is
needed for \eqref{eq:descent}, and the latter is an additional safeguard against round-off. In floating point the
decrease is verified by the stable comparison of \cref{sec:stable-comparison} beyond a resolution guard, which is
not a proof; in the recommended configuration, comparisons far from their threshold use partly binary32 energies with
an empirical error band (\cref{sec:implementation}). For every reported run, the audit re-checks from the recorded
binary64 energies that each commit decreased both $\Ein_i$ and $\Fout$ by more than half the guard recorded for that
comparison, hence by at least $8\eps$ times the larger energy, and that $\Fout$ decreased strictly between consecutive
commits on the native grid; it also reports how many commits decreased by more than the full guard
(this check is part of the audit of every reported run; all $13{,}401$ commits passed it). Across levels of the pyramid variant, and across the uncertified pre-optimisation and the handoff,
there is no monotonicity statement, because the objectives differ; on the native level, \cref{prop:monotone} holds
from the native initialiser on.

\begin{lemma}[Iterates stay away from degeneracy at the barrier points]\label{lem:sublevel}
Let $u_0\in\adm$ with $0<\Fout(u_0)<\infty$. Every iterate of \cref{prop:monotone} satisfies
$J_K(u_i;\xi)\ge c_0>0$ for all $K\in\cells$, $\xi\in\Spts$, where $c_0\in(0,J_b)$ is the unique solution of
$\kappa\,\omega\,b(c_0)/27=\Fout(u_0)$. Moreover all iterates lie in
the compact set
\[
  \mathcal C_0=\Big\{u:\ u=0\text{ on }\partial\Om,\ X_a+u_a\in\Omb,\ J_K(u;\xi)\ge c_0\ \forall K,\ \forall\xi\in\Spts\Big\},
\]
on which $\Fout$ is continuously differentiable with a Lipschitz gradient.
\end{lemma}
\begin{proof}
Each barrier term $\kappa\omega\,b(J_K(u_i;\xi))/27$ is nonnegative and bounded by $\Fout(u_i)\le\Fout(u_0)$, and $b$
decreases strictly and continuously from $+\infty$ at $0^+$ to $0$ at $J_b$; points with $J_K\ge J_b>c_0$ satisfy
the bound trivially. The domain condition follows from \cref{thm:invertibility}. The set is closed and bounded; on it
$\Dat$ is $C^2$ (cubic spline), $\Reg$ is quadratic and $\Bpot$ is $C^2$ because every argument of $b$ is at least
$c_0$.
\end{proof}

\begin{proposition}[Stationarity under a forcing condition]\label{prop:stationarity}
Let $(u_i)$ be as in \cref{prop:monotone}. Suppose, in addition, that every iterate satisfies the forcing
condition
\begin{equation}\label{eq:forcing}
  \norm{\nabla\Ein_i(u_{i+1})}\le\eta\,\norm{\nabla\Fout(u_i)}\qquad\text{for some fixed }\eta<1 .
\end{equation}
Then $\norm{\nabla\Fout(u_i)}\to0$, and every accumulation point $\bar u$ of $(u_i)$ satisfies $\nabla\Fout(\bar u)=0$.
\end{proposition}
The forcing test of \cref{sec:outer-iteration} uses $\eta_i\le\eta_0<1$, so a commit at which it holds satisfies
\eqref{eq:forcing} with $\eta=\eta_0$.
\begin{proof}
On $\mathcal C_0$ the maps $u\mapsto J_K(u;\xi)$ and $u\mapsto\curl_Ku(\xi)$ are polynomial, so there is $C$ with
$\norm{\nabla\Pen_i(u)}\le C\norm{u-u_i}$ for all $u,u_i\in\mathcal C_0$. Hence, writing $\gamma_i=\norm{\nabla\Fout(u_i)}$,
\[
  \gamma_{i+1}\le\norm{\nabla\Ein_i(u_{i+1})}+\norm{\nabla\Pen_i(u_{i+1})}\le\eta\gamma_i+C\norm{u_{i+1}-u_i}.
\]
The last term tends to zero by \cref{prop:monotone}, and $(\gamma_i)$ is bounded on the compact set $\mathcal C_0$.
Taking $\limsup$ gives $\limsup\gamma_i\le\eta\limsup\gamma_i$, so $\gamma_i\to0$, and continuity of $\nabla\Fout$ on
$\mathcal C_0$ gives the last claim.
\end{proof}

\begin{remark}[Guarantees for the reported runs]\label{rem:no-convergence}
The algorithm checks the forcing condition but does not wait for it: within $k_{\max}=5$ inner steps it is
never satisfied on the native grid ($0$ of the $13{,}401$ native commits of the reported runs), and none of the runs
reported in \cref{sec:experiments} reaches the gradient threshold of \cref{sec:outer-iteration}. \Cref{prop:stationarity} therefore
explains the design (a truncated sequential method whose fixed points are those of $\Fout$) but does not certify
the returned maps as approximate stationary points. What is guaranteed for every returned map is
\cref{cor:certified-homeo} and \cref{prop:monotone}. An accumulation point may also lie on the boundary
of $\adm$ (whole-cell positivity is not preserved under limits); the barrier only keeps the sampled values away
from zero (\cref{lem:sublevel}).
\end{remark}

\begin{lemma}[Well-posedness of certified backtracking]\label{lem:backtracking}
Consider exact arithmetic, an exact feasibility test, a zero resolution guard ($s_E=0$) and no cap on the number of
halvings. If $u\in\adm$ and $\nabla\Ein_i(u)\ne0$, there is $\bar t>0$ such that every $t\in(0,\bar t]$ is accepted;
hence backtracking from any initial step accepts a step after finitely many halvings. The implemented search, with at
most $25$ halvings and the resolution exit, accepts whenever $t_02^{-25}\le\bar t$ and the exit does not fire first.
\end{lemma}
\begin{proof}
$\adm$ is open in the space of interior displacements: $(u,\xi)\mapsto J_K(u;\xi)$ is continuous and $[0,1]^3$ is
compact, so $\min_\xi J_K$ is continuous in $u$, and the domain condition is open. Hence $u+tp\in\adm$ for all
small $t>0$. Since $p$ is a descent direction and $\Ein_i$ is $C^1$ near $u$, the Armijo condition with $c_1<1$ holds for all
small $t$; take $\bar t$ as the smaller of the two thresholds. The Jacobian-determinant and domain caps only shrink the
initial step.
\end{proof}
In floating point the certificate may return \emph{uncertain} for maps that are feasible but extremely thin, and
energy differences may fall below the resolution guard; the method then stops the inner solve and reports a stall.
It never accepts such a step and never reports it as convergence.

%% file: sections/07_implementation.tex
\section{Implementation}\label{sec:implementation}

\paragraph{Precision policy} The certificate, the gradients, the metric, the directions and $\GF$ are computed in
binary64, and so are the energies recorded for the initialiser and for every committed iterate, which the audit
recomputes. Two parts use binary32. First, the pre-optimiser of variant P runs in binary32, and its output is used
only as a candidate for the certified handoff. Its smoothing convolutions are depthwise and are dispatched by PyTorch
to its own depthwise kernels rather than to cuDNN, so no TF32 arithmetic is involved; this was checked on our hardware and is recorded for every run (field \texttt{float32\_precision} of the run record). Second, in the recommended configuration the trial
energies on the native grid are screened: the Jacobian-change, curl-change and barrier terms are evaluated in
binary32 (the data, proximity and curvature terms remain binary64), together with the band
$2^{-24}\big(64\,(\hat\Pen_J+\hat\Pen_C)+4096\,\hat\Bpot\big)$ for their error, where $\hat\Pen_J$, $\hat\Pen_C$ and
$\hat\Bpot$ are the binary32 values of the two penalties and of the barrier. Both kinds of comparison of
\cref{sec:stable-comparison}, the acceptance of an inner trial and the commit test, then proceed in three stages. The
resolution is the larger of the binary64 guard and this band; a comparison whose margin exceeds eight resolutions is
decided with the screened energies; otherwise it is recomputed with binary64 energies on the GPU, and if its margin is
still within eight binary64 guards, it is repeated on the CPU in extended precision as before. The band is empirical,
not a proven bound: it exceeds the largest error measured on nine native states of a real run by factors of $32$
(penalties) and $5.6$ (barrier); it is checked against binary64 at the first three evaluations on every grid and at
every fiftieth evaluation afterwards, and a violation switches the screening off for the rest of the run. In the $472$ runs reported in this article
(validation pairs, held-out sets, ablations, the post-hoc time curves, the $180$\,s arm and the curvature arms of the
frontier) the screening decided $67{,}909$
comparisons, $1{,}951$ of which had a margin within eight resolutions and were decided after a binary64
recomputation; of the $2{,}498$ spot checks, $10$
exceeded the band, one each in $10$ runs (H1 $1/50$, LPBA40 $2/12$ and the same two pairs in their post-hoc re-run,
ablations $5/171$, none on the validation pairs, IBSR18, H4, the $180$\,s arm or the frontier); screening was then switched off for those runs, the
decisions concerned were confirmed in binary64, and the runs are listed separately where the pre-registration
requires it. In the other $462$ runs the largest spot-check error was $99.9\%$ of the band, in one H1 run; the next largest was
$69\%$. As long as the band holds, the
screening changes how quickly the comparisons are made, not what they decide, and it does not touch the certificate.

\paragraph{Division of work between GPU and CPU} Energies, residuals, gradients, Gauss--Newton diagonals and
Hessian--vector products, the spline sampling of the moving image and the spectral Poisson solves run on the GPU in
binary64. In the certificate, test (i) runs on the GPU with an operation sequence bitwise identical to the CPU kernel
and is cross-checked against the CPU at first use (\cref{sec:rounding}); tests (ii) and (iii) and the decision for the
whole map run on the CPU in compiled kernels with $8$ threads, overlapped with GPU work. Energy comparisons whose outcome is within
eight guards of the threshold are repeated on the CPU in extended precision (\cref{sec:stable-comparison}).

\paragraph{Cost of the certificate} A native map has $6{,}772{,}287$ cells. On the CPU alone (16 threads), the full
certificate of a final map takes a median of $0.59$\,s (about $1.1\times10^7$ cells per second) and the edge-hull
test alone $0.16$\,s (about $4.2\times10^7$ cells per second). In the runs, test (i) is executed on the GPU and the
CPU stages overlap with GPU work; over the $27{,}098$ full certificates of the $177$ validation and held-out runs, waiting for the CPU adds
a median of $0.2$\,ms per trial (mean $28$\,ms; an earlier instrumented run measured a median of $14$\,ms and a $99$th
percentile of $188$\,ms), except at the first use of the GPU kernel in each run, where the bitwise cross-check with the
CPU kernel took $1.6$--$3.3$\,s in that measurement; in the frozen configuration the synthetic half of this check is done in the
dry pass before the clock and the in-run part takes about $0.3$\,s (one instrumented run, \cref{sec:experiments}). Trials whose sampled Jacobian already reveals a violation are rejected on the GPU
before the full certificate is called.

\paragraph{Cost per outer iteration} In the $50$ H1 runs ($1{,}431$ native commits) a native commit takes $2.67$\,s on
average (median $2.59$\,s, $10$th to $90$th percentile $2.4$--$3.1$\,s) and contains $5.0$ accepted inner steps and
$2.8$ trials rejected by the Jacobian screen. Per commit, the recorded device-side stages take $0.84$\,s for trial
evaluation (sampling $0.32$, energies $0.12$, witness screen $0.11$, full certificate $0.07$, upload $0.05$),
$0.63$\,s for gradients, $0.35$\,s for the Poisson solves, $0.12$\,s for promoting the accepted trial, $0.09$\,s each
for the diagonal and the outer geometry and $0.10$\,s for the stable energy comparisons, $2.27$\,s in all; the
remaining $0.4$\,s is host-side work between stages. The certificate waits are hidden behind the device: over the
$27{,}098$ full certificates of these runs the wait per trial has median $0.2$\,ms and mean $28$\,ms. Test (i)
decides $98.2\%$ of the cells of a trial map (\cref{sec:certificate}). Peak device memory is $5.5$\,GiB on the OASIS
grid and $6.1$\,GiB on the IBSR18 grid, peak host memory $7.7$--$8.7$\,GiB. One-off costs in the H1 runs: the
pre-optimiser and the handoff certificate are complete $6.3$\,s after the start ($5.5$--$8.0$\,s; the handoff
certificate itself $1.3$\,s on average), the first native commit follows at $13.8$\,s after the zero-displacement
baseline and the first outer iteration, and the export of the final map takes $0.4$\,s before the end-to-end mark.
The dry pass, the loading of the compiled kernels and the process start-up before the clock, and the copy of the
result to disk after it, add $20$\,s to the wall-clock time of a run.

\paragraph{Reproducibility and audit} The initialiser and every committed iterate are stored with a hash of the
displacement array and of its certificate. After each run an independent CPU audit reloads them, recomputes every
certificate with the reference kernels (tests (i) and (iii) only), checks from the recorded binary64 energies that
every commit decreased both energies beyond its guard (\cref{sec:analysis}), and checks the recorded energies, the MSE ratio, the Dice
scores and the gradient norm ($13{,}401$ committed states over the $470$ reported runs with a native commit, all passed; \cref{sec:certificate}).
On the same machine and software the sequence of committed iterates is reproducible bit for bit: the reference arm
of the ablations, run two days later, reproduces the commit prefixes of the validation runs on all 19 pairs
(\cref{tab:abl}). The time budget only selects the index at which this sequence is read, and the number of native
commits is recorded with every run (on H1 between $25$ and $31$, mean $28.7$).
The solver, the certificate kernels, the audit and the evaluation scripts are released under the
BSD-3-Clause licence at \url{https://github.com/Zhaobh0620/seq_vp_3d}.

\paragraph{Hardware and software} All runs use one NVIDIA RTX 4090 (24\,GB) in a workstation with an Intel Core
i9-13900K and 16\,GB of RAM, of which 15\,GB are assigned to the WSL2 Linux environment the experiments run in;
Python 3.10, PyTorch 2.5.1 with CUDA 12.1, Triton 3.1.0, NumPy 2.0.2, SciPy 1.15.2 and Numba 0.65.0 (the release
carries an environment file). The run records store the source hashes, the execution profile and the options of every
run.

%% file: sections/08_experiments.tex
\section{Experiments}\label{sec:experiments}

\subsection{Data}\label{sec:data}
\paragraph{OASIS} We use the neurite-OASIS release \cite{Hoopes2021,Hoopes2022} of the OASIS-1 cross-sectional data
\cite{Marcus2007}: 414 subjects, skull-stripped and bias-corrected with FreeSurfer, affinely aligned to FreeSurfer's Talairach atlas and cropped to $160\times192\times224$
voxels at 1\,mm, with automatic segmentations into 35 structures (seg35) and 4 tissue classes (seg4).
Labels are used for evaluation only and never enter the method. We use four
disjoint sets of pairs: (i) a tuning set of 12 pairs from the Learn2Reg training range, used for all design
decisions after its introduction; (ii) the 19 validation pairs of the Learn2Reg OASIS task \cite{Hering2023} (Learn2Reg
identifiers 0395--0414, fixed $n$, moving $n+1$). We take them from the neurite-OASIS release as subjects 0438--0457,
and verified the correspondence voxel by voxel against the official Learn2Reg release (version 1.1): after a
transposition of the axes, the images and seg35 label maps of all 20 subjects coincide exactly, and the
zero-displacement seg35 Dice of the 19 pairs is $0.5718$ on both. That release contains all 414 neurite-OASIS subjects
as training data, the validation subjects included, plus 39 test subjects whose labels are withheld. These pairs were used to select among configurations and are therefore reported as
a \emph{validation} set; (iii) a held-out set of 50 pairs drawn before any run by a pre-registered random rule
from subjects that were never used in this project (pre-registration archived on 2026-10-17, before any run on these
subjects), 100 distinct subjects with none in two pairs, so that its pairs are independent; (iv) a held-out large-deformation set of 20 pairs (H4) between subjects with small and with strongly
enlarged lateral ventricles. Its pool consists of the 149 subjects that appear in no file of the project; the
lateral-ventricle volume $V$ is computed from the seg35 labels, subjects with $V$ below the pool's 25th percentile
form the small group and those above its 80th percentile the large group, and pairs are drawn at random with a fixed
seed, ten in each direction, each subject used once, keeping only pairs whose volume ratio is at least 4 (ratios
4.14--7.57, median 5.05). On H4 our method is also run with twice the budget. The ablation c$'$, with the
factor selected on the tuning pairs, is run on the same pairs to test whether absolute penalties tuned on ordinary pairs
carry over to large deformations; the twice-budget arm was part of the H4 pre-registration, and the c$'$ arm was added
to it before any H4 run. The rule was archived before the draw; a first version with the 90th percentile
yielded too few large-ventricle subjects and was amended before any pairs list existed.
\paragraph{LPBA40} 40 T1-weighted images with 56 manually delineated structures
\cite{Shattuck2008,Shattuck2009}. The labels are defined in the providers' delineation space, a rigid alignment to MNI-305 without rescaling
\cite{Shattuck2008,Shattuck2009}; we only reorder the storage axes, without resampling or padding ($181\times217\times181$ voxels at 1\,mm), and scale intensities by the 99.5th percentile. Test pairs are the 12
ordered pairs among subjects S37--S40, fixed in the same pre-registration.
\paragraph{IBSR18} 18 T1-weighted images of the Internet Brain Segmentation Repository, distributed rotated to Talairach
orientation; we use the skull-stripped release with its structure labels (34--36 labels per subject)
\cite{Rohlfing2012}, which differs from the 84-label version evaluated by
Klein et al.\ \cite{Klein2009}. Following their protocol, every image is aligned to FSL's MNI152 1\,mm brain template by
FLIRT with 9 degrees of freedom and the correlation-ratio cost, and for each pair the moving image is then aligned
rigidly (6 degrees of freedom) to the fixed one. The moving image and its labels are resampled once from native
space with the composed transformation (trilinear for the image, nearest neighbour for the labels), the template grid
($182\times218\times182$) is zero-padded to $184\times224\times184$, and intensities are divided by the 99.5th
percentile of the brain voxels. The same aligned images are given to all methods. We use FLIRT 6.0 (FSL conda package fsl-flirt 2111.4)
\cite{JenkinsonSmith2001,Jenkinson2002}. The 36 ordered pairs (each subject fixed twice and
moving twice) were fixed in an amendment to the pre-registration before any processing. No parameter is tuned on
IBSR18.

\subsection{Evaluation measures}\label{sec:metrics}
Accuracy: mean Dice over the 35 OASIS structures (seg35, computed as in the Learn2Reg evaluation code
\cite{L2Rcode}), over the 4 tissue classes (seg4), and, for LPBA40, over the 54 delineated structures other than the cerebellum and the brainstem, which the skull-stripped LPBA40 images do not contain (the mean over all 56 structures is also reported); 95th-percentile Hausdorff
distance (HD95); the Learn2Reg DSC30 score \cite{Hering2023}. Regularity: the Learn2Reg SDlogJ, which by the public
evaluation code is the standard deviation of $\log\max(10^{-9},D_{\mathrm{cd}}+3)$, where $D_{\mathrm{cd}}$ is the
determinant of the central-difference Jacobian matrix of the map (identity plus the central-difference gradient of
the displacement; the shift by $3$ is part of the code) \cite{L2Rcode}, and, as an unshifted measure, the standard
deviation of $\log J$ over the brain cells, with $J$ the Jacobian determinant of the trilinear map at the cell centres,
clipped at $10^{-12}$. A single non-positive cell contributes $\log10^{-12}\approx-27.6$ to the latter, so for maps
with folds it is dominated by the folds; we therefore also report robust measures over the brain cells: the fractions of
cells with $J\le0$, $J<0.1$ and $J>10$, the 1st, 5th, 50th, 95th and 99th percentiles of $J$, the standard deviation of
$\log J$ after trimming 1\% at each end, and the interquartile range and the median absolute deviation of $\log J$.

Folding, three ways: (a) the fraction of voxels with
$D_{\mathrm{cd}}\le0$, counted before any clipping (the count reported by the public evaluation code is zero by
construction, \cref{rem:other-checks}); (b) our certificate applied to the trilinear map of the method's displacement
field, the same object for every method (the displacement exported on the fixed grid, interpreted as nodal displacements),
reported as the number and fraction of cells with a certified non-positive Jacobian, split into brain and background,
together with the smallest certified bound and the bracketed minimum of $J$; (c) a 24-tetrahedra piecewise-linear proxy \cite{Burger2013}, reported only as a proxy. For certified maps we
also report the minimum of $J$ over the map. The lower bound recorded with a certificate is the bound on its decision
path; it is valid but can be far from the minimum, so we bisect the cells with the smallest bounds until the minimum
is bracketed to a width below $10^{-12}$. Cost: wall-clock time end to end and of the
solve, on the same machine. Image fit: MSE relative to the identity.

All differences are paired over image pairs; we report means, bootstrap 95\% intervals ($10^4$ resamples) and two-sided Wilcoxon signed-rank tests, and the
number of pairs in which one method is better. The primary endpoint is seg35 on the held-out set H1; seg4, HD95 and
SDlogJ are secondary endpoints. For each measure, $p$-values of the comparisons against the baselines are adjusted
with Holm's procedure \cite{Holm1979}, separately for each endpoint. Pairs of LPBA40 and IBSR18 share subjects; for
these sets the bootstrap resamples subjects with replacement and weights each pair by the product of the resampling
counts of its two subjects, and the Wilcoxon tests, which treat pairs as independent, are marked as not adjusted for
this dependence. Each method is run once per pair. A run that fails for a reason of the method (for our method a failed certified handoff;
for any method an error raised by the method or its time-out) is a result and is reported as such; a run interrupted by the
infrastructure (host GPU allocation error, out-of-memory kill from outside the run, machine restart, disk guard) is
repeated once in a new directory, and both are reported.

\subsection{Baselines}\label{sec:baselines}
Each baseline uses published or default settings, without tuning on our evaluation pairs, and is run on the
same machine: FireANTs (greedy and SyN; for OASIS the authors' released OASIS
script rather than the package defaults \cite{Jena2024fireants}), ANTs SyN through the official \texttt{antsRegistrationSyN.sh} script with its default settings (rigid, affine and
SyN stages, cross-correlation metric with radius 4, iterations $100\times70\times50\times20$ over four resolutions,
double precision, no histogram matching; labels transformed with \texttt{GenericLabel}; 8 threads, as in our own
runs), with the SyN-only variant (\texttt{-t so}) as a sensitivity check (version 2.6.5, official binary; $188$\,s per
pair on H1),
NiftyReg \cite{Modat2010} (\texttt{reg\_f3d} 1.5.69, defaults), ConvexAdam without label features
\cite{Siebert2025} (package 0.2.0, defaults), deedsBCV \cite{Heinrich2013} (program defaults, no affine stage), and
learning-based methods with released weights: uniGradICON \cite{Tian2024} (1.0.6), with and without its
instance optimisation (unsupervised; the Learn2Reg OASIS validation pairs are among its test sets, not its training
data) and, as a reference point, VoxelMorph \cite{Balakrishnan2019}. SynthMorph~2 \cite{Hoffmann2022} (trained on
synthetic images and label maps only) was planned but could not be run on our machine with its default settings
(input extent 256): on the GPU (24\,GB) the forward pass failed to allocate memory, and on the CPU the process
exceeded a 10\,GiB memory limit within about 20\,s, in both its deformable-only and its joint mode. Its weights
(3.27\,GiB) and a single 4\,GiB concatenated feature map with its two 2\,GiB inputs already need about 11.3\,GiB,
while the machine has 16\,GB of RAM, 15\,GB of which are available to the Linux (WSL2) environment the experiments run in. As pre-registered, we did not change its settings (for example a smaller input
extent), so it is absent from the comparison. VPreg \cite{VPreg}, earlier work of three of the present authors, is run from the authors' MATLAB implementation
(\texttt{VPreg\_gpu}, commit 3402d8a, used unchanged under MATLAB R2023b on the same GPU) with the constants fixed in the released code, which are the setting its authors used for OASIS; its time is the code's own
timer of the forward registration (the inverse map, which it also computes, is excluded), and its map is evaluated
after conversion to our voxel convention, checked on every pair against the authors' own warped image
(correlation $0.9992$ or higher on all $69$ pairs). TransMorph \cite{Chen2022transmorph} is reported separately as a
label-supervised method: its released OASIS models were trained with a Dice loss on the 35 OASIS labels, on the
Learn2Reg training data, which contain all 414 subjects and hence every subject of our validation and held-out sets;
this favours it strongly. We evaluate its raw output flow, as in the authors' Learn2Reg submission; its network warps
images with a slightly different coordinate normalisation, equivalent to a field that differs from the raw flow by at most about half a voxel (measured maximum $0.49$
voxels on the OASIS validation, H1, LPBA40 and IBSR18 runs and $0.51$ on one H4 pair, where a foreground voxel is
mapped outside the image range). To separate the effect of optimisation time, FireANTs
(greedy and SyN) and uniGradICON with instance optimisation are also run with their iteration counts scaled so that
one registration takes about $90$\,s, our budget; the scaling is calibrated by time alone on one tuning pair, every
other setting is unchanged, and these equal-time arms are reported next to the default ones on the OASIS sets.
For FireANTs the iteration caps are multiplied by 22 (greedy) and 27 (SyN) with the package's own convergence test
left in place, so its runs still stop early on most pairs (on average after $3.6$ and $10.9$\,s on H1); uniGradICON
runs 123 instead of 50 instance-optimisation steps ($85$\,s). On LPBA40 and IBSR18 the learning-based methods receive the images in the anatomical orientation of their OASIS training data (for LPBA40, whose headers carry no orientation, left and right are fixed by the label names; for IBSR18 the label positions confirmed the header orientation in all 18 subjects, assuming the FreeSurfer numbering of its left and right structures), and TransMorph and VoxelMorph receive them cropped or zero-padded to their input size of $160\times192\times224$ voxels around the fixed image's brain; the displacement is mapped back to the original grid. The classical methods receive the images unchanged. Versions, commits, checkpoint hashes and configurations are fixed in a pre-registration
before any baseline is run on evaluation pairs (FireANTs 1.5.1 with its fused operations 1.2.0, ANTs 2.6.5,
NiftyReg 1.5.69, deedsBCV and ConvexAdam at the archived commits, uniGradICON 1.0.6, TransMorph's released Base and
Large OASIS checkpoints, VoxelMorph's released T1 brain model).

\subsection{Results}\label{sec:results}
\begin{figure}[t]
\centering
\includegraphics[width=\textwidth]{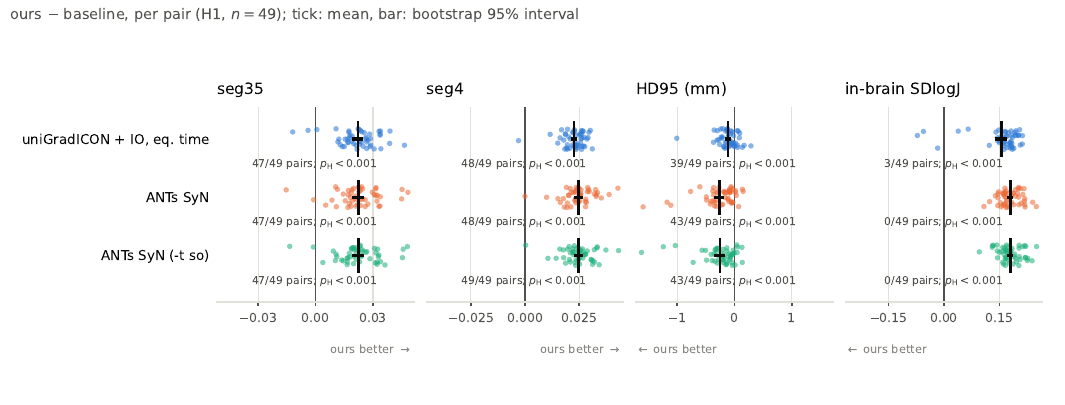}
\caption{Per-pair differences ours $-$ baseline on H1 ($n=49$) against the three label-free baselines with the
highest mean seg35: seg35, seg4, HD95 and in-brain SDlogJ. Dots: pairs; tick: mean; bar: bootstrap $95\%$ interval
of the mean; under each strip: pairs on which ours is better and the Holm-adjusted Wilcoxon $p$.}\label{fig:paired}
\end{figure}

\begin{figure}[t]
\centering
\includegraphics[width=\textwidth]{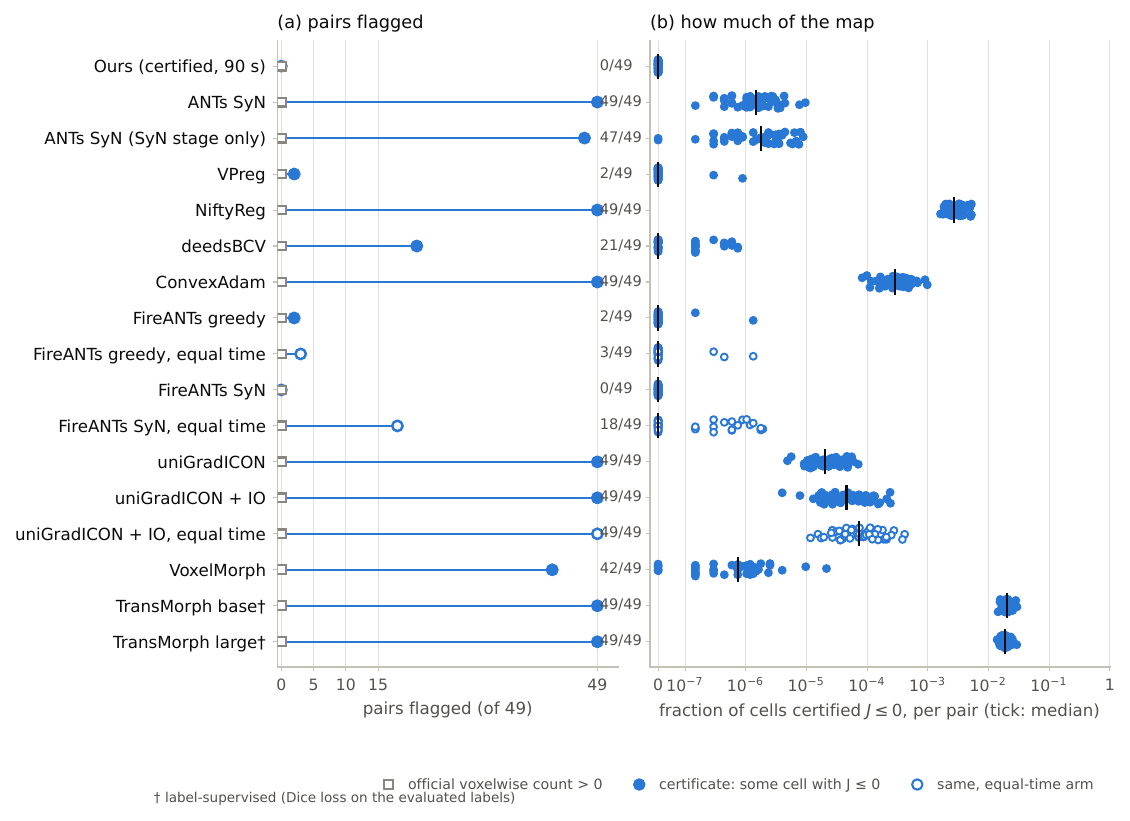}
\caption{One check that cannot see folding and one that proves it (49 held-out pairs, H1; frozen baseline runs).
(a) For each method, the number of pairs on which the official Learn2Reg count reports a folding (grey squares; zero
for every method, because the count is taken after the determinant has been clipped) and the number of pairs on which
the certificate proves a non-positive Jacobian in at least one cell of the trilinear map (blue). Open markers:
equal-time arms. (b) Per pair, the fraction of the $6{,}772{,}287$ cells certified non-positive; the tick is the
median over pairs. Baseline fields are tested with the Jacobian test only; our maps are fully certified. \dag:
trained with a Dice loss on the evaluated labels.}\label{fig:baseline-verdicts}
\end{figure}

\paragraph{Accuracy and folding on OASIS} \Cref{tab:main} reports every method on the held-out set H1 and on the
19 validation pairs, \cref{tab:main-diff} the paired differences on H1, and \cref{fig:paired} the per-pair
differences against the three label-free baselines with the highest mean seg35. The H1 table has 49 pairs: on one of
the 50 runs the binary32 energy screening recorded a comparison outside its empirical error band
(\cref{sec:implementation}); the comparison was repeated in binary64 and the run completed, but the pre-registered
rule keeps such runs out of the main table (with it, seg35 is $0.8202$ and seg4 $0.8886$; the 50-pair tables are in
the ancillary files). On H1 our method reaches seg35 $0.8205$ and seg4 $0.8888$, with every returned map certified.
The strongest label-free baselines are ANTs SyN ($0.7982$ / $0.8641$, $188$\,s on 8 CPU threads) and uniGradICON
with instance optimisation at equal time ($0.7984$ / $0.8661$, $85$\,s); the differences to ours are $+0.022$ seg35
and $+0.025$ seg4 against ANTs ($95\%$ intervals $[+0.019,+0.025]$ and $[+0.023,+0.027]$; ours higher on $47$ and
$48$ of the $49$ pairs) and $+0.022$ / $+0.023$ against uniGradICON at equal time ($47/49$ and $48/49$), all with
Holm-adjusted $p<0.001$; HD95 is lower by $0.26$ and $0.11$\,mm. The seg35 differences to FireANTs greedy, NiftyReg,
VoxelMorph and uniGradICON without instance optimisation are $0.027$--$0.032$, those to FireANTs SyN (default and
equal time), ConvexAdam, VPreg and deedsBCV $0.047$--$0.084$. VPreg, the closest predecessor of our model,
reaches $0.7589$ / $0.7952$ in $47$\,s on the same GPU (its forward registration only) and returns the smoothest
maps of all methods (in-brain SDlogJ $0.24$, Learn2Reg SDlogJ $0.035$); ours is higher by $0.062$ seg35 and
$0.094$ seg4 on every one of the 49 pairs, with HD95 lower by $0.47$\,mm. The validation pairs give the same picture ($0.8194$ /
$0.8823$; $+0.023$ / $+0.026$ against ANTs, $18/19$ and $19/19$ pairs). Decomposed over structures (equal weights of
$1/35$ per label), the $+0.0224$ against ANTs comes from the ventricles ($+0.0064$), the vessels and choroid plexus
($+0.0063$), the subcortical grey matter ($+0.0042$) and the cortex ($+0.0033$), while cerebellum and brainstem
contribute nothing measurable; the decomposition against uniGradICON at equal time is the same to within $0.0015$
per group. The price is regularity: our maps are rougher than those of every classical baseline (in-brain SDlogJ
$0.62$ against $0.44$ for ANTs and $0.47$ for uniGradICON at equal time; Learn2Reg SDlogJ $0.079$ against $0.065$),
a point taken up in \cref{sec:discussion}. The label-supervised TransMorph models, trained with a Dice loss on these
labels and on data that contain these subjects, reach seg35 $0.88$; their seg4 is level with ours ($-0.0004$ and
$-0.0012$, $p=0.99$), and their maps have about $1.3\times10^{5}$ certified folded cells each.
\begin{table}[t]
\centering\scriptsize
\caption{OASIS: all methods on the held-out set H1 (49 pairs, top) and on the 19 validation pairs (bottom); mean
$\pm$ standard deviation over pairs. seg35, seg4: mean Dice over 35 structures and 4 tissue classes; HD95 over the
35 structures (TransMorph: 48 pairs, one structure missing from one warped label map); SDlogJ (L2R): the Learn2Reg
measure; SDlogJ (brain): over the brain cells of the trilinear map; folds (c.d.): fraction of voxels with a
non-positive central-difference determinant before clipping; certified $J\le0$ cells: cells of the trilinear map
proven non-positive by the certificate (Jacobian test for the baselines, full certificate for ours; of
$6{,}772{,}287$ cells); time: for ours end to end from the start of the task clock, for the baselines the
registration call only (the pre-registered time measure), on the same machine (RTX 4090; ANTs on 8 CPU threads). Equal-time
arms (eq.\ time): iteration counts calibrated to about $90$\,s on one tuning pair; uniGradICON takes $85$\,s, while
FireANTs, whose iteration caps are multiplied by 22 and 27 with its own convergence test left in place, still stops
after a few seconds (time column). $\dagger$: label-supervised (Dice
loss on these labels; training data contain these subjects).}\label{tab:main}
\resizebox{\textwidth}{!}{\input{analysis/h1/table1_H1_means.tex}}\\[4pt]
\resizebox{\textwidth}{!}{\input{analysis/val/table1_val_means.tex}}
\end{table}
\begin{table}[t]
\centering\scriptsize
\caption{H1 (49 pairs): paired differences ours $-$ method (positive favours ours for seg35 and seg4, negative for
HD95, SDlogJ and certified cells), with the bootstrap $95\%$ interval ($10^4$ resamples), the Holm-adjusted
two-sided Wilcoxon $p$ over the 16 baselines within each measure, and the number of pairs on which ours is
better.}\label{tab:main-diff}
\resizebox{\textwidth}{!}{\input{analysis/h1/table1_H1_diffs.tex}}
\end{table}
\paragraph{Certificate applied to other methods} \Cref{fig:baseline-verdicts} applies the certificate to the
displacement field of every method on the 49 H1 pairs (Jacobian test only for the baselines, whose exported fields
need not vanish on the boundary; the full certificate for ours). The official folding count is zero for every field,
as it must be (\cref{rem:other-checks}); the unclipped central-difference fraction (a) is zero on every pair for
ANTs, deedsBCV and FireANTs and positive on most pairs for NiftyReg, ConvexAdam and the learned methods. The
certificate separates the methods. FireANTs SyN is certified positive on all 49 pairs, FireANTs greedy on 47 (10
cells in total on the other two) and deedsBCV on 28 (on the other 21 it proves a non-positive cell, one per map at the
median). VPreg, whose energy penalises the Jacobian determinant, is
certified positive on 47 pairs; on the other two the certificate proves $J<0$ in six and in two cells (lower bounds
$-0.038$ and $-0.081$) while the cell-centre determinant is positive everywhere and the voxelwise check finds
nothing. ANTs SyN has a certified non-positive cell on every pair, a
median of $10$ cells per map and at most $65$, on maps where the voxelwise check finds no non-positive voxel at all;
its SyN-only variant on 47 pairs. The learned methods fold on every pair (uniGradICON $136$ cells per map at the
median, $310$ with instance optimisation and $505$ at equal time; VoxelMorph on 42 pairs, $5$ cells), NiftyReg and
ConvexAdam on every pair over a median of $0.27\%$ and $0.03\%$ of the cells, and the label-supervised TransMorph over $2\%$. A
handful of folded cells and a map folded over a sizeable fraction of its cells are different situations, and the
figure shows both. In the other direction, the 24-tetrahedra proxy flags cells on all 49 of our certified maps
(\cref{rem:other-checks}).

\begin{figure}[t]
\centering
\includegraphics[width=\textwidth]{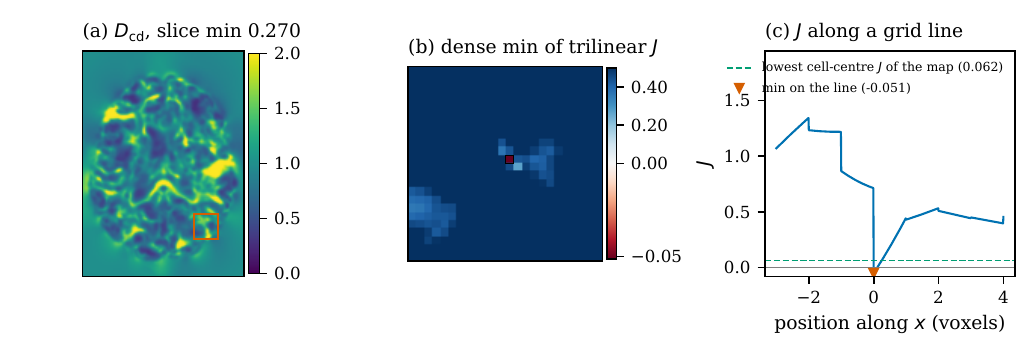}
\caption{One displacement field, different folding verdicts (ANTs SyN on H1 pair 0197\,$\leftarrow$\,0371, from the
frozen baseline runs). (a) Central-difference determinant $D_{\mathrm{cd}}$ on the axial node plane through the
certified cell; it is positive at every voxel of the map, and the official folding count is $0$. (b) Minimum of the
trilinear $J$, sampled at $9^3$ points per cell, in the window marked in (a); outlined: the cell with a negative
sample. (c) Trilinear $J$ along the grid line through the sampled minimum; $J$ is defined cellwise and jumps across
cell faces. The dashed line is the smallest cell-centre value of $J$ over the whole map ($0.062$). Dense sampling only
confirms what the certificate proves (lower bound $-0.051$ on this cell).}\label{fig:verdicts}
\end{figure}

\begin{figure}[t]
\centering
\includegraphics[width=\textwidth]{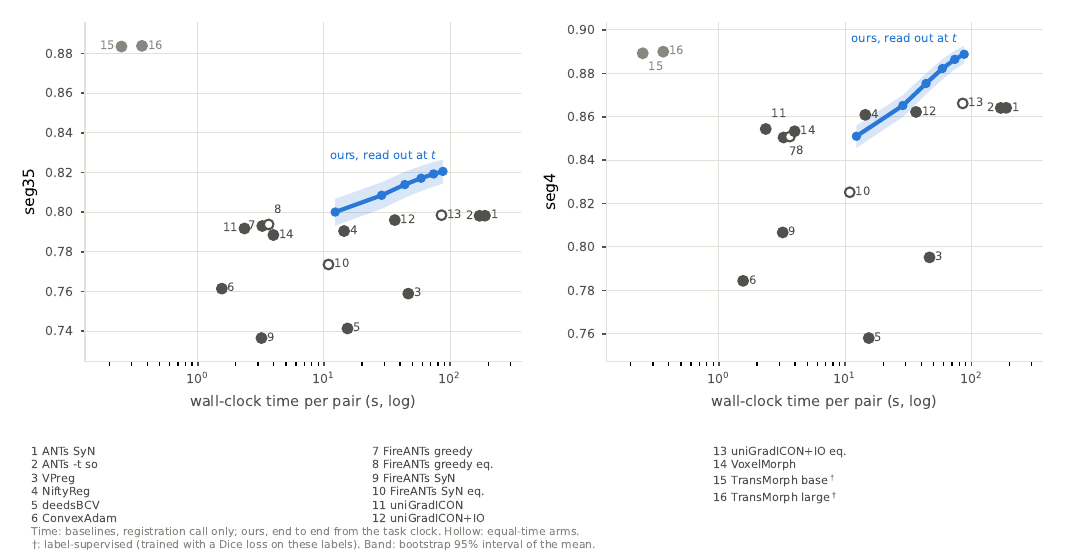}
\caption{Time--accuracy on H1 (49 pairs): mean seg35 (left) and seg4 (right) against wall-clock time per pair
(logarithmic). Ours: the last committed map at $t=15,30,45,60,75,89$\,s of the main $90$\,s runs, plotted at the mean
time at which that map was committed; band: bootstrap $95\%$ interval of the mean (at $15$\,s, ten pairs are at the
certified handoff map). Baselines: numbered points at their mean registration time (hollow: equal-time arms; grey:
label-supervised). Time: for ours end to end from the task clock, for the baselines the registration call
only.}\label{fig:time}
\end{figure}

\begin{figure}[t]
\centering
\includegraphics[width=\textwidth]{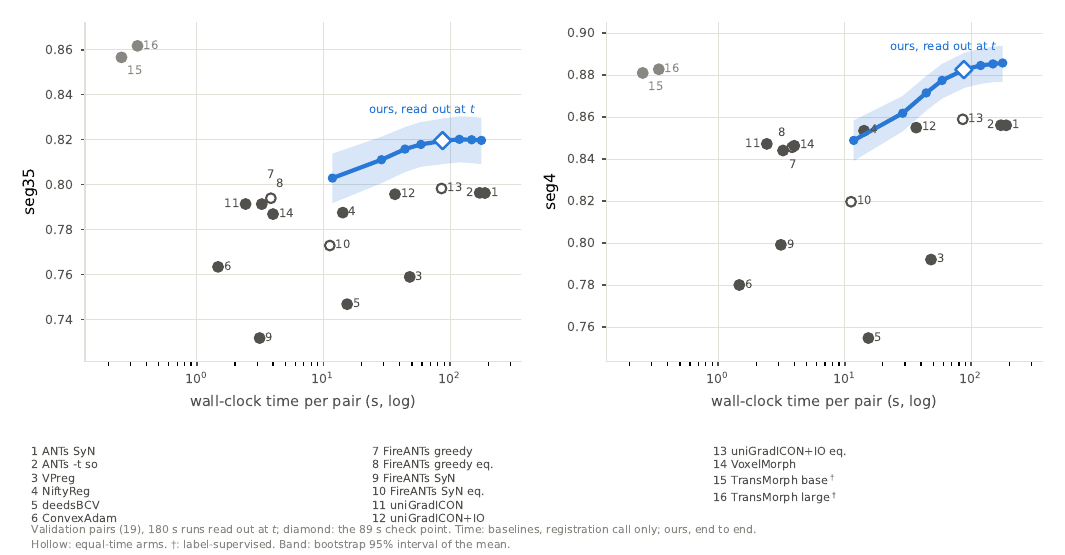}
\caption{Time--accuracy on the 19 validation pairs with a $180$\,s budget: mean seg35 (left) and seg4 (right) of the
last committed map at $t=15,30,45,60,90,120,150,179$\,s (diamond: the $89$\,s check point), plotted at the mean time
at which that map was committed; band: bootstrap $95\%$ interval of the mean. Baselines: numbered points at their
mean registration time on the same pairs (hollow: equal-time arms; grey: label-supervised).}\label{fig:valtime}
\end{figure}

\begin{figure}[t]
\centering
\includegraphics[width=\textwidth]{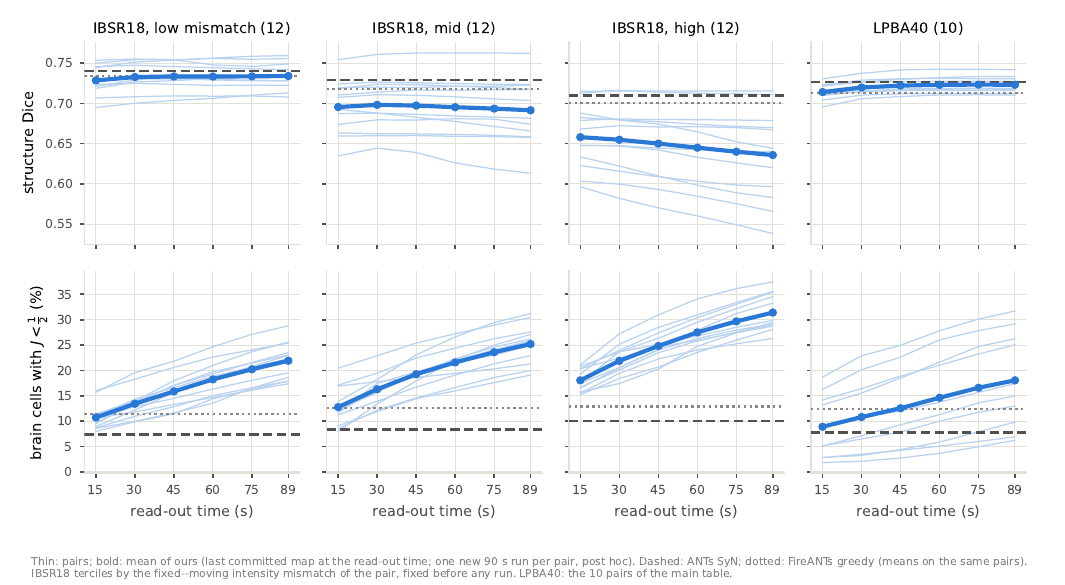}
\caption{Post-hoc time curves of the frozen configuration on IBSR18 (H3; 36 pairs, split by the intensity mismatch of
the pair into the terciles of \cref{sec:discussion}, fixed before any run) and LPBA40 (H2; the 10 pairs of
\cref{tab:gen}). Top: structure Dice of the last committed map at $t=15,\dots,89$\,s; bottom: fraction of brain cells
with $J<\tfrac12$. Thin: pairs; bold: mean; dashed: ANTs SyN and dotted: FireANTs greedy, means on the same pairs. One
new $90$\,s run per pair, read out before its deletion; the three possible outcomes and the prediction per tercile
were written down before the runs (archived plan).}\label{fig:h3tc}
\end{figure}

\begin{figure}[t]
\centering
\includegraphics[width=\textwidth]{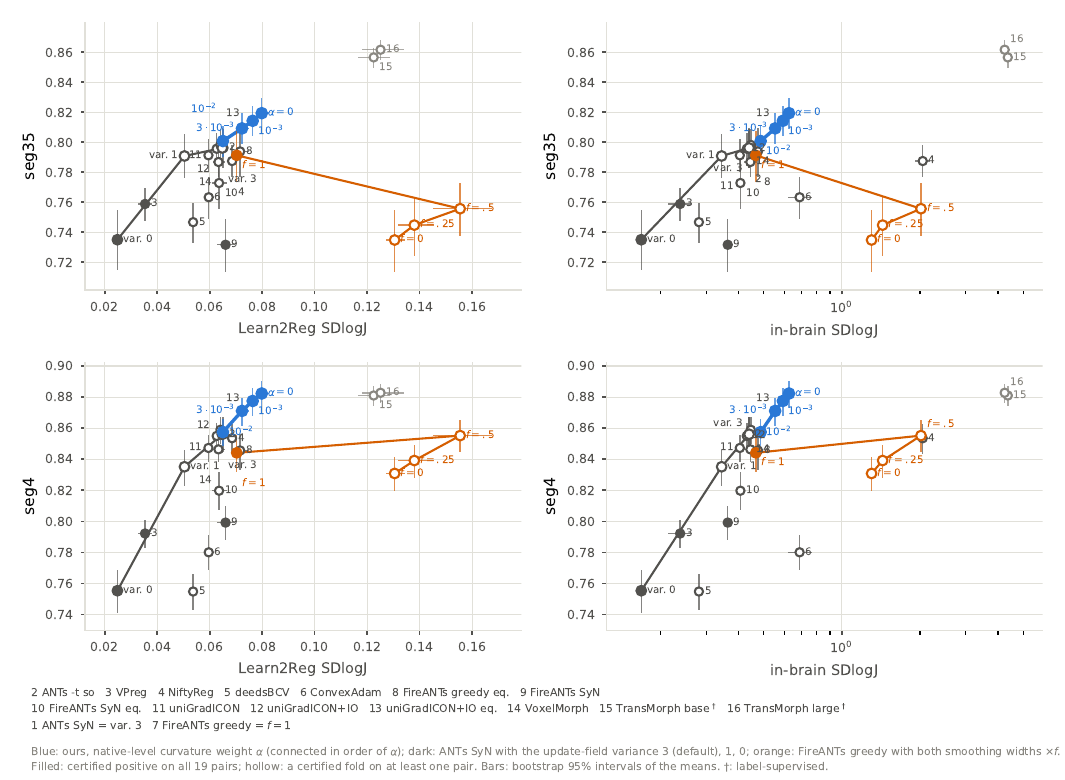}
\caption{Accuracy--regularity frontier on the 19 validation pairs: mean seg35 (top) and seg4 (bottom) against the
Learn2Reg SDlogJ (left) and the in-brain SDlogJ (right, logarithmic), with bootstrap $95\%$ intervals in both
directions. Blue: ours with the native-level curvature weight $\alpha=0$ (frozen), $10^{-3}$, $3\cdot10^{-3}$,
$10^{-2}$; dark: ANTs SyN with the update-field variance $3$ (default), $1$, $0$; orange: FireANTs greedy with both
smoothing widths multiplied by $f=1$ (default), $0.5$, $0.25$, $0$; numbered grey points: the other default and
equal-time arms of \cref{tab:main}. Filled: certified positive on all 19 pairs; hollow: a certified fold on at least
one pair.}\label{fig:frontier}
\end{figure}

\begin{figure}[t]
\centering
\includegraphics[width=\textwidth]{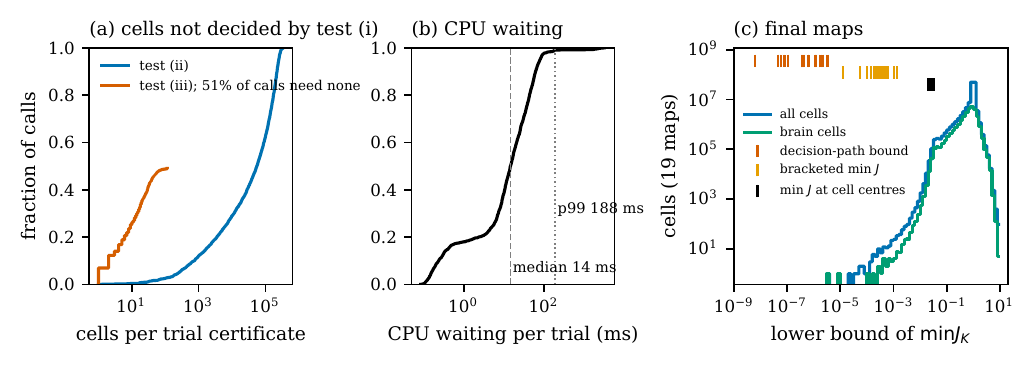}
\caption{Certificate cost and tightness on native maps (validation runs of a pre-freeze execution version, 19
pairs; the tightness of the frozen runs' final maps is given in the text). (a) For each of the 2329 trial certificates, the number of cells (out of
$6{,}772{,}287$) left open by test (i) and passed to test (ii), and to test (iii); 51\% of the calls need test (iii)
for no cell. (b) CPU waiting per trial evaluation after the GPU work, over 2374 evaluations. (c) Tightest valid lower
bound of $\min J_K$ per cell over the 19 final maps (larger of the edge-hull and test-(ii) bounds, and the test-(iii)
bound where used), with, per map, the bound recorded on the certificate's decision path, the minimum of $J$ bracketed
by subdivision, and the smallest value of $J$ at cell centres.}\label{fig:cost}
\end{figure}

\begin{figure}[t]
\centering
\includegraphics[width=\textwidth]{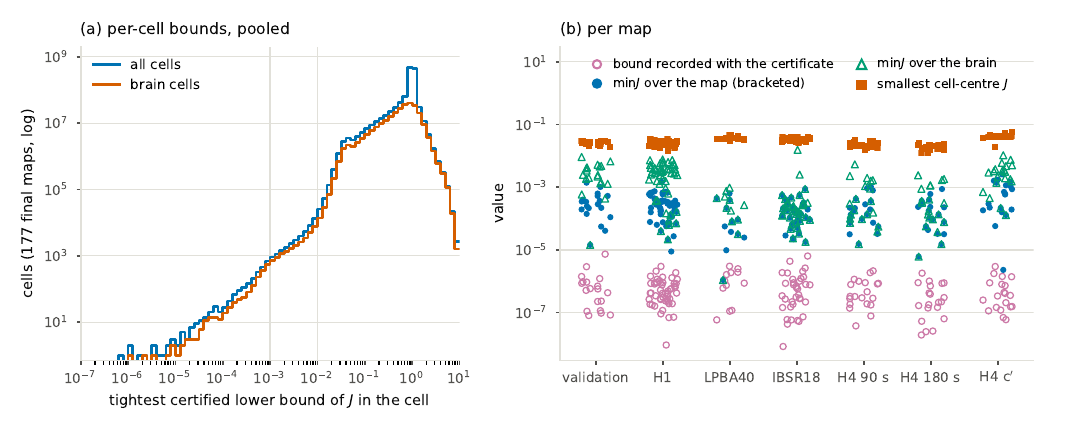}
\caption{Certificate tightness on the $177$ final maps of the reported runs (frozen configuration; validation pairs,
H1, LPBA40, IBSR18 and the three H4 arms), re-certified offline with the lower bound of every cell recorded. (a)
Histogram over all $1.23\times10^{9}$ cells of the tightest valid lower bound of $J$ over the cell (the larger of the
edge-hull and prescreen bounds, and the subdivision bound where used), all cells and brain cells. (b) Per map: the
bound recorded with the certificate (its decision path; in $176$ maps a loosely decided edge-hull cell), the minimum
of $J$ over the map and over the brain (bracketed by subdivision to within $2.2\times10^{-13}$), and the smallest
cell-centre value.}\label{fig:tight}
\end{figure}

The folds are neither marginal nor an artefact of the export (\cref{tab:fold-depth}). Each baseline field is stored
in binary32 and certified exactly as stored, and the stored field is the map the method applies: resampling the
moving image with it reproduces the method's own warped image, with a correlation of at least $0.9993$ on every pair
(for ANTs, whose field composes the affine and the SyN stage, $1.0000$ with a largest difference of
$4\times10^{-7}$), except for TransMorph ($0.994$, the coordinate convention of \cref{sec:baselines}; the field its
network uses folds as well). On every folded map the minimum of $J$ is at most $-0.0012$ (deedsBCV), more than four
orders of magnitude beyond the unit roundoff $6\times10^{-8}$ of binary32, and its median over the folded maps of a
method lies between $-0.012$ (FireANTs greedy at equal time) and $-1.7$ (NiftyReg; TransMorph about $-10$).
\begin{table}[t]
\centering\scriptsize
\caption{Depth of the certified folds of the baselines (49 H1 pairs, Jacobian test). Maps with a fold: maps with at
least one cell certified non-positive; cells: median over those maps; min $J$: upper end of the bracket of the
minimum of $J$ over the map (the minimum is at most this value), median over the folded maps and the value closest
to zero; export corr.: smallest correlation over the pairs between the moving image resampled with the exported
field and the method's own warped image. Field used by the network: TransMorph's raw flow corrected for its
coordinate normalisation (\cref{sec:baselines}). Our 49 maps are certified positive.}\label{tab:fold-depth}
\resizebox{\textwidth}{!}{\input{analysis/h1/fold_magnitudes.tex}}
\end{table}
\Cref{fig:verdicts} shows the mechanism on one of the frozen baseline fields. In the ANTs SyN field of H1 pair
0197\,$\leftarrow$\,0371 the official evaluation counts no folding and the central-difference determinant is positive at
every voxel (its minimum on the displayed slice is $0.27$), yet the certificate proves $J<0$ in one cell, with lower
bound $-0.051$, and dense sampling finds $J=-0.051$ at a corner of that cell. At that grid node
$D_{\mathrm{cd}}=1.06$, while the eight cells meeting there have $J$ between $-0.051$ and $2.18$: the central
difference averages over the cells around the node and does not see the one-sided value, as in
\cref{fig:toy}(a)--(b). The smallest cell-centre value of $J$ over the whole map is $0.062$.
\paragraph{Time--accuracy} \Cref{fig:time} shows seg35 and seg4 against wall-clock time on H1. Our curves are read
out from the main $90$\,s runs: at $t=15,30,45,60,75,89$\,s we evaluate the last committed iterate (the certified
handoff map if no commit has occurred yet, and no map before the handoff), so that every pair is still registered
only once; at $15$\,s ten of the 49 pairs are at the handoff map, from $30$\,s on all have committed. Baselines
contribute points at their registration time: the default configuration and, where present, the equal-time arm. At
the $15$\,s read-out seg35 is $0.800$ ($[0.793,0.807]$), the level of ANTs SyN at $188$\,s and of uniGradICON with
instance optimisation at equal time ($0.798$ both); from $30$\,s ($0.808$, $[0.802,0.815]$) the interval excludes
them. seg4 reaches the strongest baseline later: $0.851$ at $15$\,s, $0.865$ at $30$\,s and $0.875$ at $45$\,s,
against $0.866$ for uniGradICON at equal time; VPreg ($47$\,s) lies below the $15$\,s read-out on both measures.
Neither curve has flattened at the end of the budget ($+0.0013$ seg35
and $+0.0024$ seg4 over the last $14$\,s); the $90$\,s budget was fixed before the held-out runs
(\cref{sec:discussion}). The corresponding curves on IBSR18 and LPBA40 are in \cref{fig:h3tc}. On the validation pairs with a $180$\,s
budget (\cref{fig:valtime}; the same runs read out at $89$\,s agree with the $90$\,s runs to within $0.0003$), seg35
reaches its plateau at about $90$\,s ($0.8195$ at $90$\,s, $0.8201$ at $120$\,s, $0.8197$ at $179$\,s) while seg4
still rises slowly ($0.8827$ at $90$\,s to $0.8858$ at $179$\,s) and the map keeps roughening (in-brain SDlogJ $0.63$
to $0.68$, brain cells with $J<\tfrac12$ from $17\%$ to $19\%$); HD95 improves by $0.02$\,mm. Doubling the budget
therefore buys little accuracy at a rougher map.
\paragraph{Generalisation} \Cref{tab:gen,tab:gen-diff} report the OASIS-frozen configuration, with no parameter
changed, on the LPBA40 test pairs (H2; 54 structures; 10 pairs in the main table, two runs listed separately for
their binary32 screening record, \cref{sec:implementation}) and on IBSR18 (H3; 36 pairs). On these two sets the
intervals are subject-block bootstraps and the Wilcoxon tests are not adjusted for the shared subjects
(\cref{sec:data}). On LPBA40 our method is level with the strongest baselines: structure Dice $0.7231$ against
$0.7270$ for ANTs SyN (difference $-0.004$ $[-0.015,+0.011]$, Holm $p=0.64$, ours higher on $3/10$ pairs), $0.7221$
for NiftyReg and $0.7181$ for uniGradICON with instance optimisation; only the SyN-only ANTs variant is ahead
($0.7325$; $-0.009$ $[-0.016,-0.002]$, $p=0.023$, $0/10$), while FireANTs greedy ($+0.010$, $p=0.029$), deedsBCV,
ConvexAdam, VoxelMorph and TransMorph are behind. With the two separately listed runs included the means are
$0.7247$ (ours) and $0.7292$ (ANTs SyN). On IBSR18 the result is negative: $0.6867$ against $0.7263$ for ANTs SyN
($-0.040$ $[-0.068,-0.015]$, $p<0.001$, ours higher on $7/36$ pairs), $0.7174$ for FireANTs greedy ($-0.031$,
$p=0.005$) and $0.7171$ for the label-supervised TransMorph ($p<0.001$); against NiftyReg, deedsBCV, ConvexAdam,
FireANTs SyN and uniGradICON the differences lie within $\pm0.02$ and are not significant. On both sets our maps are
rougher than those of ANTs (in-brain SDlogJ $0.59$ against $0.38$ on LPBA40 and $0.77$ against $0.44$ on IBSR18, on
every pair) and HD95 is larger by about $0.2$\,mm. Our maps are certified on every pair, at $88$\,s against
$193$--$208$\,s for ANTs. Of the baselines, FireANTs SyN returns no certified fold on any pair of either set and
FireANTs greedy and deedsBCV on most pairs; ANTs folds a few cells on $5/10$ and $30/36$ pairs; uniGradICON folds
tens of cells on nearly every pair, and NiftyReg, ConvexAdam, VoxelMorph and TransMorph hundreds to tens of thousands
(\cref{tab:gen}). The IBSR18 deficit is analysed post hoc in \cref{sec:discussion}: it is graded by the intensity
mismatch between the fixed and the moving image, to which the correlation-type data terms of the baselines are
insensitive. \Cref{fig:h3tc} reads the same configuration out against time on both sets (post hoc: one new run per
pair with the frozen configuration after the results above were known; the $89$\,s read-out
reproduces the held-out Dice of each pair to $+0.0004$ $[-0.0001,+0.0010]$ on IBSR18 and $+0.0001$ on LPBA40). On
LPBA40 Dice rises to a plateau from about $60$\,s ($0.714$ at $15$\,s, $0.723$ at $89$\,s; ANTs SyN $0.727$). On IBSR18
the curves separate by the intensity mismatch of the pair: on the twelve pairs with the smallest mismatch Dice rises to
a plateau ($0.729$ at $15$\,s, $0.734$ from $45$\,s on; ANTs $0.740$), on the twelve with the largest it is highest at
the first read-out ($0.658$ at $15$\,s) and falls to $0.636$ at $89$\,s ($-0.022$ $[-0.034,-0.011]$, $11/12$ pairs)
while the fraction of brain cells with $J<\tfrac12$ rises from $18\%$ to $31\%$; even at its best time it stays
$0.052$ $[0.031,0.071]$ below the ANTs value of the same pairs ($0.710$). The mean over all $36$ pairs peaks at
$30$\,s, $0.008$ $[0.004,0.013]$ above its $89$\,s value, so an earlier stop would have narrowed the deficit by about a
fifth and removed none of it.
\begin{table}[tbp]
\centering\footnotesize
\caption{Generalisation: the OASIS-frozen configuration on LPBA40 (H2; 54 structures; 10 pairs) and IBSR18 (H3;
36 pairs), mean $\pm$ SD over pairs. Dice: macro-averaged structure Dice; SDlogJ (L2R): the official Learn2Reg measure;
SDlogJ (brain): over brain cells, cell-centre $J$; folds (c.d.): central-difference fraction before clipping;
certified $J\le0$ cells: cells of the trilinear map certified non-positive (Jacobian test for the baselines, full
certificate for ours); time: for the baselines the registration call (the pre-registered measure), for ours end
to end. $\dagger$: trained with a Dice loss on OASIS labels.}\label{tab:gen}
\resizebox{\textwidth}{!}{\input{tables/table2_H2_means.tex}}\\[4pt]
\resizebox{\textwidth}{!}{\input{tables/table2_H3_means.tex}}
\end{table}
\begin{table}[tbp]
\centering\footnotesize
\caption{Paired differences ours $-$ method on LPBA40 (top) and IBSR18 (bottom): mean, subject-block bootstrap 95\%
interval, Holm-adjusted two-sided Wilcoxon $p$ (not adjusted for the shared subjects) and the number of pairs in which
ours is better. Positive favours ours for Dice; negative favours ours for HD95, SDlogJ, folds, certified cells and
time.}\label{tab:gen-diff}
\resizebox{\textwidth}{!}{\input{tables/table2_H2_diffs.tex}}\\[4pt]
\resizebox{\textwidth}{!}{\input{tables/table2_H3_diffs.tex}}
\end{table}
\paragraph{Large deformations} On the 20 H4 pairs (\cref{tab:h4}) the frozen configuration reaches seg35 $0.7632$ and
seg4 $0.8496$ at $90$\,s, above every label-free baseline: $+0.017$ $[+0.010,+0.024]$ and $+0.019$ $[+0.013,+0.025]$
against the strongest, uniGradICON with instance optimisation at equal time (Holm $p\le0.002$; ours higher on $18/20$
and $19/20$ pairs), and $+0.029$ and $+0.034$ against ANTs SyN ($20/20$ and $19/20$); HD95 is $2.93$\,mm against
$3.02$\,mm (uniGradICON, $p=0.25$) and $3.38$\,mm (ANTs, $p=0.007$). Twice the budget adds $+0.005$ seg35 and
$+0.008$ seg4 ($15/20$ and $20/20$ pairs) and makes the maps rougher (in-brain SDlogJ $0.72\to0.80$); the arm with
absolute penalties (c$'$, factor $10^{-2}$) is below the frozen configuration by $0.008$ and $0.010$ ($p\le0.017$) and
smoother ($0.66$). As on the other sets our maps are rougher than the baselines' (ANTs $0.54$, uniGradICON $0.59$) and
certified on every pair, while every baseline except FireANTs and deedsBCV returns certified folds on most pairs.
TransMorph, trained with the labels of these subjects, reaches $0.874$ and $0.877$ with about $1.6\times10^5$
certified folded cells per map. On the CSF class of seg4, where the ventricular compression is measured, the $90$\,s
configuration reaches $0.805$ against $0.795$ for ANTs ($+0.010$ $[-0.006,+0.026]$, $13/20$ pairs) and $0.821$ for
uniGradICON at equal time ($-0.015$ $[-0.032,-0.001]$, $8/20$; neither difference significant after Holm); on the
lateral ventricles themselves uniGradICON is ahead by $0.023$--$0.026$, on the six ventricular labels together our
configuration is ahead of every label-free baseline ($0.677$ against $0.650$ for uniGradICON at equal time and $0.639$
for ANTs, $18/20$ and $19/20$ pairs), so its advantage on H4 comes from the smaller ventricular spaces and from the
other structures rather than from the lateral ventricles (\cref{sec:discussion}). Summaries of $\log J$ over the brain
cells that are insensitive to the clipped folded cells of the baselines (defined in an amendment to the
pre-registration after the model freeze) rank the methods as the plain in-brain SDlogJ does: the median-absolute-deviation estimate of the standard
deviation of $\log J$ is $0.52$ for ours against $0.44$ (ANTs), $0.42$ (uniGradICON at equal time), $0.45$ (NiftyReg)
and $0.58$ (TransMorph), the standard deviation over the cells between the 1st and 99th percentiles $0.66$ against
$0.48$, $0.50$, $0.64$ and $0.84$, and the fraction of brain cells with $J<0.1$ is $1.8\%$ against $0.5\%$, $1.1\%$,
$2.5\%$ and $6.0\%$ (paired differences with Holm $p\le0.013$ except NiftyReg's trimmed standard deviation,
$p=0.18$); twice the budget raises all three ($0.62$, $0.74$, $2.4\%$) and the arm with absolute penalties lowers the
fraction below $0.1$ to $0.6\%$ at an unchanged robust standard deviation.
\begin{table}[tbp]
\centering\footnotesize
\caption{Large deformations (H4; 20 pairs with a ventricle volume ratio of at least 4): mean $\pm$ SD, columns as in
\cref{tab:gen} (top), and paired differences of the 90\,s configuration minus each arm with bootstrap 95\% interval,
Holm-adjusted Wilcoxon $p$ and pairs in which the 90\,s configuration is better (bottom).}\label{tab:h4}
\resizebox{\textwidth}{!}{\input{tables/table2_H4_means.tex}}\\[4pt]
\resizebox{\textwidth}{!}{\input{tables/table2_H4_diffs.tex}}
\end{table}
\paragraph{Ablations} Each ablation changes one component of the certified native refinement and leaves the
pre-optimiser and the handoff unchanged, so all variants start from the same certified initialiser; the
pre-optimiser's own penalties and smoothing are not ablated. The ablations are run on the 19 validation pairs and
compared with the frozen configuration on the same pairs. Components: (a) Jacobian-change penalty off ($\mu=0$) and (b) curl-change penalty off ($\nu=0$), with
the metric following the reduced weights, $b_P=(\mu+\nu)/2$; (c) fixed references, i.e.\ the penalties of
\eqref{eq:penalty} measured from the native initialiser instead of the last committed iterate, and (c$'$)
references at the identity ($J=1$, $\curl=0$, zero displacement), i.e.\ the usual absolute penalties at the same
weights; since the inner energy is then itself the objective, commits are accepted on its decrease alone in both
cases. At equal weights the absolute penalties of (c$'$) are large at the initialiser, which is far from the identity,
so the method pulls the map back towards the identity and the outer energy rises. For (c$'$) we therefore also scale
$\lambda,\mu,\nu$ by a common factor in $\{10^{-1},10^{-2},10^{-3}\}$, select the factor on the 12 tuning pairs as the one with the highest mean seg4, taking the larger factor
when means differ by at most $0.002$ (seg35 is reported), and report the selected factor together with the
equal-weight version (the rule selected $10^{-2}$); (d) the metric replaced by its diagonal
version $r^2/d$ (no Poisson coupling) and, as a second level, by the identity, whose first trial step is capped by
a fixed largest node displacement because the unit step of the Poisson metric has no meaning there; (e) the
Levenberg--Marquardt inner solver of the library in place of the preconditioned gradient method, with the same budget;
an inner Levenberg--Marquardt step cannot be interrupted, so its runs overrun the budget (126.5\,s on the usability
check), and their actual wall time is reported; this arm is therefore not wall-time-equal.
\Cref{tab:abl} gives the result. Removing the Jacobian-change penalty (a) costs $0.0023$ seg35 and $0.0079$ seg4
(Holm $p\le0.006$; the frozen configuration is higher on $16/19$ and $19/19$ pairs) and yields smoother maps
(in-brain SDlogJ $-0.077$); removing the curl-change penalty (b) changes seg35 by $+0.0009$ ($p=0.10$) and seg4 by
$+0.0006$ ($p=0.023$), below $0.001$ either way, and moves the two regularity measures by small amounts in opposite
directions (in-brain SDlogJ $+0.009$, $p=0.005$, rougher on $14/19$ pairs; Learn2Reg SDlogJ $-0.001$, $p=0.003$);
this arm also halves the Poisson coupling $b_P$ of the metric. Fixed references (c) cost $0.007$ and $0.021$ and are smoother
($-0.15$); references at the identity at equal weights (c$'\!\times1$) cost $0.086$ and $0.117$, because the method
pulls the map back towards the identity, whereas with the weights scaled by $10^{-2}$ the arm is $+0.003$ seg35 and
$-0.003$ seg4 from the frozen configuration and smoother ($-0.04$). The diagonal metric (d) is likewise within
$0.0035$ of it ($+0.0035$ seg35, $-0.0028$ seg4) and smoother ($-0.05$); the identity metric costs $0.016$ and
$0.033$ ($14$ pairs; five runs are listed separately for their binary32 screening record); and the
Levenberg--Marquardt inner solver (e) costs $0.017$ and $0.035$ at $111$\,s instead of $88$\,s, with two pairs without
any native commit within the budget. At the $90$\,s budget, then, the accuracy rests on the Jacobian-change penalty at
its native weight, on the reference being the last commit rather than the initialiser, on the metric being scaled by
the image curvature (diagonal or Poisson) and on the inner solver; the curl-change penalty contributes nothing
measurable to accuracy and the Poisson coupling of the metric less than $0.004$ in either measure, and the incremental form of the penalties matters through the
reference it moves (c) and the weights it permits (c$'$). Every map of every arm is certified, and the same-batch
frozen configuration reproduces the validation runs of \cref{tab:main} bitwise on their commit prefixes.
\begin{table}[tbp]
\centering\footnotesize
\caption{Ablations on the 19 validation pairs (90\,s; C: the frozen configuration run in the same batch). $n$: pairs in
the main table; means of seg35, seg4 and in-brain SDlogJ; differences arm $-$ C with bootstrap 95\% interval,
Holm-adjusted two-sided Wilcoxon $p$ over the eight arms and the number of pairs in which the arm is better; time: end
to end; the last column counts the runs without a native commit within the budget (method failures, not in the means)
and, in parentheses, the runs listed separately for their binary32 screening record.}\label{tab:abl}
\resizebox{\textwidth}{!}{\input{tables/table3_ablation.tex}}
\end{table}
\paragraph{Accuracy--regularity frontier} The rougher maps of \cref{tab:main} raise the question whether the
accuracy is bought with the roughness. \Cref{fig:frontier} moves our regularisation up and the baselines' down on the
validation pairs (pre-registered): three new arms of ours with the native-level curvature weight
$\alpha\in\{10^{-3},3\cdot10^{-3},10^{-2}\}$, ANTs SyN with its update-field variance lowered from $3$ to $1$ and $0$,
and FireANTs greedy with both smoothing widths multiplied by $f\in\{0.5,0.25,0\}$, everything else as frozen. Raising
$\alpha$ moves our maps monotonically along a frontier: in-brain SDlogJ $0.63$, $0.59$, $0.55$, $0.49$ with seg35
$0.819$, $0.814$, $0.809$, $0.801$ and seg4 $0.882$, $0.878$, $0.871$, $0.858$ (every difference to $\alpha=0$ on
$19/19$ pairs), all maps certified. No $\alpha$ reaches the in-brain SDlogJ of ANTs SyN ($0.44$), FireANTs greedy
($0.47$) or uniGradICON with instance optimisation at equal time ($0.45$); at the closest arm ($\alpha=10^{-2}$,
$0.49$) the paired seg35 differences are $+0.005$ to ANTs, $+0.003$ to uniGradICON and $+0.010$ to FireANTs greedy
(seg4 $+0.001$, $-0.002$ and $+0.013$), of which only the seg4 difference to FireANTs greedy remains significant after
Holm adjustment ($p<0.001$, $18/19$ pairs). In the other direction, loosening the baselines does not catch up: with a
lower update-field variance ANTs SyN becomes less accurate and, at the same time, smoother (seg35 $0.791$ and
$0.735$, in-brain SDlogJ $0.34$ and $0.17$ at variance $1$ and $0$), and FireANTs greedy becomes rougher and less
accurate together (seg35 $0.756$, $0.745$, $0.735$; in-brain SDlogJ $2.0$, $1.4$, $1.3$) with a certified fold on
every pair; every loosened arm lies $0.03$ to $0.08$ seg35 below the frozen configuration on $19/19$ pairs. At matched
regularity, then, the accuracy advantage over the classical baselines shrinks to a few thousandths and is no longer
significant, while the certificate is retained; the advantage of the frozen configuration comes together with its
roughness. (The trimmed in-brain SDlogJ named in the pre-registration is a post-hoc measure that was not computed for
the new arms; the plain in-brain SDlogJ is used throughout.)
\paragraph{Certificate cost and tightness} The $177$ final maps of the reported runs (validation pairs, H1, H2, H3
and the three H4 arms) were re-certified offline on the CPU with the lower bound of every cell recorded; the
certificates reproduce the recorded ones field by field. The bound recorded with a certificate, $8.3\times10^{-9}$
to $7.4\times10^{-6}$ (median $4\times10^{-7}$), is the bound on its decision path: in $176$ of the $177$ maps it comes
from a cell decided loosely by the edge hull, whose true minimum is $0.10$ to $0.28$ at the median. The minimum of
$J$ over a map, bracketed by subdivision to within $2.2\times10^{-13}$, is $1.1\times10^{-6}$ to $2.6\times10^{-3}$
(median $2.5\times10^{-4}$ on H1, $6.9\times10^{-5}$ on LPBA40 and $1.4\times10^{-4}$ on IBSR18), while the smallest cell-centre value is $0.012$ to $0.058$ (median $0.028$), one to four orders of magnitude
larger. On H1 about five cells per map (of
$6.8$ million; pooled fraction $7.4\times10^{-7}$) have a certified bound below $10^{-3}$ and about $140$
($2.0\times10^{-5}$) below $10^{-2}$; $49$ of the $50$ maps have such a cell, $12$ within the brain. The fractions are
within a factor of three on the other sets and six to eight times larger for the $180$\,s arm of H4. The maps are
therefore invertible everywhere, but a handful of cells per map are compressed by more than a factor of $10^3$
(\cref{sec:discussion}). Of the $27{,}098$ full certificates run in these $177$ runs, $41$ (in $22$ runs) rejected a
trial map that the device witness screen had passed; none concerned a committed map. In one additional instrumented
run on a tuning pair, whose commits agree bitwise with an uninstrumented run of the same pair over their common
prefix of $25$, the $149$ trial certificates took $65$\,ms at the median and $162$\,ms at the 90th percentile on the
certificate thread ($51$\,ms when the prescreen decided every remaining cell, $74$ calls; $111$\,ms when the interval
fallback was needed, $75$ calls), and the handoff certificate $1.05$\,s; most of this overlaps with the GPU, and the
time the run waits for a certificate is far smaller ($14$\,ms at the median in the pre-freeze measurement of
\cref{fig:cost}). \Cref{fig:tight} shows the bounds of the $177$ final maps; \cref{fig:cost} summarises cost and tightness on the
pre-freeze validation runs.
\paragraph{Certifying a positive lower bound} The same tests decide $J>\delta$ on every closed cell for any
$\delta\ge0$. A post-freeze arm (pre-registered, report only) ran the frozen configuration on the 19 validation pairs
with $\delta\in\{10^{-3},10^{-2}\}$ in every certificate that decides acceptance (trial, commit and handoff), in the
device screen and the step prediction, and in the barrier, where $J/J_b$ in \cref{eq:barrier} is replaced by
$(J-\delta)/(J_b-\delta)$ so that $b$ blows up at $J=\delta$, with the same $J_b$ and $\kappa$; the pre-optimiser was
unchanged. These $38$ runs use an option outside the frozen configuration and
are not counted in \cref{sec:implementation}. All ended on the budget with native commits (mean $26.3$ and $25.1$,
against $27.7$ for $\delta=0$), every handoff was certified ($\rho=0.9$ on $8$ and $12$ pairs, against $8$), the
$976$ committed states were re-certified offline at their $\delta$, and no spot check of the binary32 screen left
its band. With $\delta=10^{-2}$, six runs ended when the deadline interrupted a CPU operation started late in the run
close to the constraint: in five the repetition on the CPU of an energy comparison after a very small step or near
its threshold (\cref{sec:stable-comparison}; once also with $\delta=10^{-3}$), in one the CPU part of a trial evaluation. This cost
them $6$ to $32$\,s and contributes to the lower number of commits; their result is the last certified commit, as in
every budget-terminated run. Against the frozen runs on the same pairs, the paired differences for $\delta=10^{-3}$ and $10^{-2}$ are
seg35 $-0.0003$ and $-0.0011$ (Holm $p=0.03$ for both, $14/19$ pairs lower), seg4 $-0.0007$ ($p=0.11$) and
$-0.0029$ ($p<10^{-4}$, $18/19$), and in-brain SDlogJ $-0.007$ ($p=0.10$) and $-0.026$ ($p<10^{-4}$, $18/19$
smoother); the mean HD95 rises by $0.008$ and $0.027$\,mm. Both costs in accuracy are smaller than that of the smallest
curvature weight of the frontier ($\alpha=10^{-3}$: seg35 $-0.005$, seg4 $-0.005$, in-brain SDlogJ $-0.033$). The
minimum of $J$ over a map, bracketed as above to within $10^{-13}$ (for $\delta=10^{-2}$ with the subdivided cells
picked below $2\cdot10^{-2}$, since no cell has a bound below $10^{-2}$), moves to just above $\delta$: median
$1.4\times10^{-3}$ (range $1.0\times10^{-3}$ to $2.1\times10^{-3}$) and $1.001\times10^{-2}$ ($1.000\times10^{-2}$ to
$1.22\times10^{-2}$), against $2.9\times10^{-4}$ ($1.5\times10^{-5}$ to $1.4\times10^{-3}$, $17$ of $19$ maps with a
cell below $10^{-3}$) for $\delta=0$. The few critical cells therefore stay at the certified threshold, wherever it
is placed, and a threshold of $10^{-2}$ removes every cell compressed by more than a factor of $100$ at a cost of
about $0.001$ seg35 and $0.003$ seg4 on these pairs (\cref{tab:delta}).
\begin{table}[t]
\centering
\caption{Certified lower bound $\delta$ on $J$ (post-freeze arm, report only): the frozen configuration on the 19
validation pairs with $J>\delta$ certified and the barrier shifted to $\delta$, against the frozen runs ($\delta=0$).
Means over the pairs; differences arm $-$ frozen with bootstrap $95\%$ interval and Holm-adjusted two-sided Wilcoxon
$p$ over the two arms, in units of $10^{-3}$ (for HD95, $10^{-3}$\,mm); native commits within the $90$\,s budget;
handoffs certified at $\rho=0.9$ instead of $1$; min $J$: minimum of $J$ over the final map, bracketed by subdivision,
median and smallest over the 19 maps.}\label{tab:delta}
\resizebox{\textwidth}{!}{\input{analysis/delta/delta_table.tex}}
\end{table}
\paragraph{Local conditioning} The certificate proves that every returned map is a homeomorphism; it does not bound
how strongly a cell is distorted. A post-freeze offline measure (pre-registered, report only, computed on the CPU from
the stored maps of the 19 validation pairs) evaluates $F=\nabla\Phi_u$ with the cell polynomial at the centre and the
eight corners of every cell and computes its singular values $\sigma_1\ge\sigma_2\ge\sigma_3$ in binary64
(\cref{tab:cond}): per brain cell the condition number $\kappa=\sigma_1/\sigma_3$, the largest over its samples with
$\det F>0$, and per map the sampled Lipschitz constants $L(\Phi)=\max\sigma_1$ and $L(\Phi^{-1})=1/\min\sigma_3$ and
the upper tail of $J$. These are values at samples, not bounds. No sample of our maps has $\det F\le0$, as the
certificate requires, while every map of ANTs SyN, uniGradICON, NiftyReg and TransMorph has such samples. The typical
cell of our maps is about as well conditioned as with the baselines (median $\kappa$ $2.56$ against $2.1$ to $2.2$
for ANTs SyN, uniGradICON at equal time, FireANTs greedy and NiftyReg and $1.37$ for VPreg, although higher on $19/19$
pairs against each of them), but the tail is much heavier: the 99th percentile of $\kappa$ is $44$ against $6$ to $9$
for ANTs SyN, uniGradICON and FireANTs greedy, $33$ for NiftyReg and $2.3$ for VPreg, and $6.7\%$ of the brain cells
have $\kappa>10$ against $0.1\%$ to $0.7\%$ for the first three (every difference to these five baselines on $19/19$
pairs, Holm $p<10^{-4}$, except the 99th percentile against NiftyReg, $15/19$ pairs, $p=0.003$). The local expansion
is larger as well (99th percentile of $J$ $4.6$ against $2.8$ to $3.2$ for ANTs SyN, uniGradICON, FireANTs greedy
and NiftyReg). $L(\Phi^{-1})$ is comparable only between
maps without a fold, since next to a fold $\sigma_3\to0$: it is $855$ for ours against $34$ for FireANTs greedy, $3.1$
for VPreg and $2.4$ for ANTs SyN at variance $0$ (higher on $19/19$ pairs against each). Only TransMorph (99th
percentile of $\kappa$ $192$) and FireANTs greedy with reduced smoothing ($163$ to $454$) have a heavier tail than ours,
on $19/19$ pairs. The two options of this section act on different parts of the distribution. Certifying $J>10^{-2}$
lowers $L(\Phi^{-1})$ from $855$ to $237$ and the 99th percentile of $\kappa$ from $44$ to $31$ (both on $19/19$
pairs), consistent with $\sigma_3\ge J/\sigma_1^2$; the native curvature weight shortens the tail of $\kappa$
monotonically ($44$, $31$, $20$, $11$ for $\alpha=0$, $10^{-3}$, $3\cdot10^{-3}$, $10^{-2}$, each on $19/19$ pairs) but
changes $L(\Phi^{-1})$ little ($658$, $615$, $531$; only the last difference is significant, $p=0.017$, $15/19$ pairs),
because it does not raise the smallest values of $J$. The certified maps are therefore invertible everywhere, but in a
few percent of the brain cells they are more anisotropic than the maps of the smooth baselines, the same finding as
their higher SDlogJ seen in a quantity closer to the inverse map; a certified lower bound on $J$ reduces this at a
small cost in accuracy (\cref{tab:delta}), a curvature weight at a larger one (\cref{fig:frontier}).
\begin{table}[t]
\centering
\caption{Local conditioning of the returned maps on the 19 validation pairs (post-freeze offline measure, report
only). $F=\nabla\Phi_u$ is evaluated at the centre and the eight corners of every cell; per brain cell
$\kappa=\sigma_1/\sigma_3$ is the largest over its samples with $\det F>0$ and $J$ the largest over its samples;
percentiles, maxima and the percentage of brain cells with $\kappa>10$ are taken per map, $L(\Phi)=\max\sigma_1$ and
$L(\Phi^{-1})=1/\min\sigma_3$ over the samples of the brain cells, and every entry is the median over the 19 maps.
$L(\Phi^{-1})$ is given only where all 19 maps are certified positive, since next to a fold $\sigma_3\to0$; $\det
F\le0$: samples with $\det F\le0$ (left out of $\kappa$ and $L$), summed over the 19 maps; certified: maps certified
positive. Values at samples, not bounds. $\delta$, $\alpha$: our arms of \cref{tab:delta} and \cref{fig:frontier};
variance, $f$: the loosened baselines of \cref{fig:frontier}. $\dagger$: trained with a Dice loss on OASIS labels; raw
flow (its grid-sample-equivalent field differs by less than $2\%$ in every column except the maximum of $\kappa$).}
\label{tab:cond}
\resizebox{\textwidth}{!}{\input{analysis/cond/cond_table.tex}}
\end{table}
\paragraph{Failure cases} No reported run stalled: every H1 run made between $25$ and $31$ native commits, and the
only runs without a native commit are the two Levenberg--Marquardt ablation runs. The pair on which the frozen
configuration loses most against ANTs is H1 pair 0218\,$\leftarrow$\,0222 (seg35 $0.804$ against $0.820$ for ANTs and
$0.816$ for uniGradICON at equal time; seg4 $0.889$ against $0.869$): the run is ordinary ($29$ commits, handoff at
$\rho=1$, first native commit at $13.0$\,s) and gains on the cortex ($+0.05$ on both hemispheres) and the white
matter, but loses on small deep structures, the left inferior lateral ventricle ($0.44$ against $0.57$), the right
vessel ($0.56$ against $0.67$), the ventral diencephalon ($-0.05$ to $-0.07$), the right accumbens and the left choroid
plexus ($-0.06$), with a rougher map (in-brain SDlogJ $0.61$ against $0.40$). The hardest pair for every method is
0299\,$\leftarrow$\,0140 (our seg35 $0.759$, ANTs $0.751$, uniGradICON at equal time $0.759$; HD95 $3.7$\,mm): there
the frozen configuration is ahead on the vessels and the inferior lateral ventricles ($+0.09$ to $+0.20$) and behind
on the caudate ($-0.12$ and $-0.14$) and the thalamus ($-0.07$). Both cases show the same pattern as the label
decomposition of \cref{sec:discussion}: the cortex is matched to within a voxel, and the residual error sits in small
deep structures, where the map over-displaces.

%% file: analysis/h1/table1_H1_means.tex
\begin{tabular}{lccccccccc}
\toprule
Method & $n$ & seg35 $\uparrow$ & seg4 $\uparrow$ & HD95 (mm) $\downarrow$ & SDlogJ (L2R) $\downarrow$ & SDlogJ (brain) $\downarrow$ & folds (c.d.) $\downarrow$ & certified $J\le0$ cells $\downarrow$ & time (s) $\downarrow$ \\
\midrule
Ours (certified, 90 s) & 49 & 0.8205 $\pm$ 0.0218 & 0.8888 $\pm$ 0.0143 & 1.88 $\pm$ 0.49 & 0.079 $\pm$ 0.004 & 0.623 $\pm$ 0.051 & 0.0e+00 $\pm$ 0.0e+00 & 0 $\pm$ 0 & 88.1 $\pm$ 0.7 \\
ANTs SyN & 49 & 0.7982 $\pm$ 0.0220 & 0.8641 $\pm$ 0.0154 & 2.14 $\pm$ 0.58 & 0.065 $\pm$ 0.004 & 0.443 $\pm$ 0.046 & 0.0e+00 $\pm$ 0.0e+00 & 13 $\pm$ 12 & 188.0 $\pm$ 4.4 \\
ANTs SyN (-t so) & 49 & 0.7981 $\pm$ 0.0223 & 0.8641 $\pm$ 0.0153 & 2.13 $\pm$ 0.59 & 0.066 $\pm$ 0.005 & 0.443 $\pm$ 0.046 & 0.0e+00 $\pm$ 0.0e+00 & 16 $\pm$ 15 & 170.5 $\pm$ 4.2 \\
VPreg & 49 & 0.7589 $\pm$ 0.0257 & 0.7952 $\pm$ 0.0185 & 2.35 $\pm$ 0.54 & 0.035 $\pm$ 0.006 & 0.242 $\pm$ 0.064 & 0.0e+00 $\pm$ 0.0e+00 & 0 $\pm$ 1 & 46.5 $\pm$ 7.0 \\
NiftyReg & 49 & 0.7904 $\pm$ 0.0209 & 0.8609 $\pm$ 0.0133 & 2.10 $\pm$ 0.51 & 0.069 $\pm$ 0.005 & 2.146 $\pm$ 0.446 & 1.9e-03 $\pm$ 6.8e-04 & 20460 $\pm$ 6627 & 14.4 $\pm$ 0.3 \\
deedsBCV & 49 & 0.7413 $\pm$ 0.0326 & 0.7580 $\pm$ 0.0274 & 2.47 $\pm$ 0.66 & 0.055 $\pm$ 0.004 & 0.286 $\pm$ 0.027 & 0.0e+00 $\pm$ 0.0e+00 & 1 $\pm$ 1 & 15.4 $\pm$ 0.2 \\
ConvexAdam & 49 & 0.7614 $\pm$ 0.0303 & 0.7844 $\pm$ 0.0262 & 2.43 $\pm$ 0.70 & 0.061 $\pm$ 0.004 & 0.758 $\pm$ 0.198 & 1.9e-04 $\pm$ 1.4e-04 & 2229 $\pm$ 1300 & 1.6 $\pm$ 0.1 \\
FireANTs greedy & 49 & 0.7930 $\pm$ 0.0352 & 0.8504 $\pm$ 0.0252 & 2.17 $\pm$ 0.64 & 0.073 $\pm$ 0.008 & 0.470 $\pm$ 0.031 & 0.0e+00 $\pm$ 0.0e+00 & 0 $\pm$ 1 & 3.2 $\pm$ 0.3 \\
FireANTs greedy, eq. time & 49 & 0.7938 $\pm$ 0.0337 & 0.8509 $\pm$ 0.0243 & 2.17 $\pm$ 0.63 & 0.073 $\pm$ 0.007 & 0.475 $\pm$ 0.032 & 0.0e+00 $\pm$ 0.0e+00 & 0 $\pm$ 1 & 3.6 $\pm$ 0.8 \\
FireANTs SyN & 49 & 0.7365 $\pm$ 0.0320 & 0.8067 $\pm$ 0.0223 & 2.58 $\pm$ 0.69 & 0.066 $\pm$ 0.008 & 0.372 $\pm$ 0.030 & 0.0e+00 $\pm$ 0.0e+00 & 0 $\pm$ 0 & 3.2 $\pm$ 0.2 \\
FireANTs SyN, eq. time & 49 & 0.7736 $\pm$ 0.0364 & 0.8252 $\pm$ 0.0270 & 2.40 $\pm$ 0.73 & 0.064 $\pm$ 0.006 & 0.408 $\pm$ 0.035 & 0.0e+00 $\pm$ 0.0e+00 & 2 $\pm$ 4 & 10.9 $\pm$ 4.2 \\
uniGradICON & 49 & 0.7917 $\pm$ 0.0210 & 0.8544 $\pm$ 0.0131 & 2.10 $\pm$ 0.51 & 0.060 $\pm$ 0.005 & 0.412 $\pm$ 0.047 & 2.9e-06 $\pm$ 4.5e-06 & 177 $\pm$ 113 & 2.3 $\pm$ 0.1 \\
uniGradICON + IO & 49 & 0.7960 $\pm$ 0.0199 & 0.8622 $\pm$ 0.0118 & 2.03 $\pm$ 0.47 & 0.063 $\pm$ 0.005 & 0.443 $\pm$ 0.058 & 1.5e-05 $\pm$ 2.4e-05 & 450 $\pm$ 393 & 36.4 $\pm$ 0.2 \\
uniGradICON + IO, eq. time & 49 & 0.7984 $\pm$ 0.0189 & 0.8661 $\pm$ 0.0113 & 1.99 $\pm$ 0.45 & 0.065 $\pm$ 0.005 & 0.467 $\pm$ 0.081 & 2.9e-05 $\pm$ 4.0e-05 & 690 $\pm$ 606 & 85.2 $\pm$ 0.1 \\
VoxelMorph & 49 & 0.7884 $\pm$ 0.0213 & 0.8533 $\pm$ 0.0118 & 2.08 $\pm$ 0.47 & 0.064 $\pm$ 0.006 & 0.460 $\pm$ 0.092 & 9.5e-08 $\pm$ 6.6e-07 & 10 $\pm$ 22 & 4.0 $\pm$ 0.2 \\
TransMorph base$^\dagger$ & 49 & 0.8835 $\pm$ 0.0098 & 0.8892 $\pm$ 0.0071 & 1.39 $\pm$ 0.33 & 0.124 $\pm$ 0.021 & 4.323 $\pm$ 0.403 & 8.3e-03 $\pm$ 1.7e-03 & 135001 $\pm$ 21468 & 0.2 $\pm$ 0.0 \\
TransMorph large$^\dagger$ & 49 & 0.8838 $\pm$ 0.0097 & 0.8900 $\pm$ 0.0071 & 1.39 $\pm$ 0.32 & 0.124 $\pm$ 0.024 & 4.215 $\pm$ 0.402 & 7.9e-03 $\pm$ 1.7e-03 & 130548 $\pm$ 21014 & 0.4 $\pm$ 0.1 \\
\bottomrule
\end{tabular}

%% file: analysis/val/table1_val_means.tex
\begin{tabular}{lccccccccc}
\toprule
Method & $n$ & seg35 $\uparrow$ & seg4 $\uparrow$ & HD95 (mm) $\downarrow$ & SDlogJ (L2R) $\downarrow$ & SDlogJ (brain) $\downarrow$ & folds (c.d.) $\downarrow$ & certified $J\le0$ cells $\downarrow$ & time (s) $\downarrow$ \\
\midrule
Ours (certified, 90 s) & 19 & 0.8194 $\pm$ 0.0232 & 0.8823 $\pm$ 0.0191 & 1.82 $\pm$ 0.46 & 0.080 $\pm$ 0.005 & 0.625 $\pm$ 0.047 & 0.0e+00 $\pm$ 0.0e+00 & 0 $\pm$ 0 & 87.8 $\pm$ 0.8 \\
ANTs SyN & 19 & 0.7962 $\pm$ 0.0295 & 0.8561 $\pm$ 0.0248 & 2.08 $\pm$ 0.55 & 0.065 $\pm$ 0.003 & 0.439 $\pm$ 0.027 & 0.0e+00 $\pm$ 0.0e+00 & 20 $\pm$ 24 & 188.4 $\pm$ 4.7 \\
ANTs SyN (-t so) & 19 & 0.7963 $\pm$ 0.0296 & 0.8561 $\pm$ 0.0248 & 2.08 $\pm$ 0.56 & 0.065 $\pm$ 0.003 & 0.439 $\pm$ 0.027 & 0.0e+00 $\pm$ 0.0e+00 & 24 $\pm$ 26 & 171.3 $\pm$ 4.7 \\
VPreg & 19 & 0.7589 $\pm$ 0.0241 & 0.7922 $\pm$ 0.0196 & 2.31 $\pm$ 0.46 & 0.035 $\pm$ 0.006 & 0.238 $\pm$ 0.056 & 0.0e+00 $\pm$ 0.0e+00 & 0 $\pm$ 0 & 48.0 $\pm$ 9.9 \\
NiftyReg & 19 & 0.7875 $\pm$ 0.0238 & 0.8535 $\pm$ 0.0217 & 2.05 $\pm$ 0.47 & 0.069 $\pm$ 0.003 & 2.049 $\pm$ 0.297 & 1.7e-03 $\pm$ 5.2e-04 & 18515 $\pm$ 4994 & 14.2 $\pm$ 0.4 \\
deedsBCV & 19 & 0.7468 $\pm$ 0.0295 & 0.7548 $\pm$ 0.0255 & 2.39 $\pm$ 0.56 & 0.054 $\pm$ 0.002 & 0.281 $\pm$ 0.018 & 0.0e+00 $\pm$ 0.0e+00 & 1 $\pm$ 2 & 15.4 $\pm$ 0.2 \\
ConvexAdam & 19 & 0.7634 $\pm$ 0.0312 & 0.7801 $\pm$ 0.0246 & 2.32 $\pm$ 0.55 & 0.060 $\pm$ 0.003 & 0.687 $\pm$ 0.158 & 1.5e-04 $\pm$ 1.2e-04 & 1779 $\pm$ 1219 & 1.5 $\pm$ 0.1 \\
FireANTs greedy & 19 & 0.7913 $\pm$ 0.0386 & 0.8441 $\pm$ 0.0261 & 2.12 $\pm$ 0.61 & 0.070 $\pm$ 0.005 & 0.466 $\pm$ 0.024 & 0.0e+00 $\pm$ 0.0e+00 & 0 $\pm$ 0 & 3.3 $\pm$ 0.3 \\
FireANTs greedy, eq. time & 19 & 0.7939 $\pm$ 0.0363 & 0.8456 $\pm$ 0.0255 & 2.10 $\pm$ 0.58 & 0.072 $\pm$ 0.006 & 0.475 $\pm$ 0.031 & 0.0e+00 $\pm$ 0.0e+00 & 0 $\pm$ 0 & 3.8 $\pm$ 0.8 \\
FireANTs SyN & 19 & 0.7317 $\pm$ 0.0394 & 0.7992 $\pm$ 0.0249 & 2.50 $\pm$ 0.59 & 0.066 $\pm$ 0.007 & 0.363 $\pm$ 0.019 & 0.0e+00 $\pm$ 0.0e+00 & 0 $\pm$ 0 & 3.1 $\pm$ 0.2 \\
FireANTs SyN, eq. time & 19 & 0.7729 $\pm$ 0.0380 & 0.8197 $\pm$ 0.0272 & 2.30 $\pm$ 0.60 & 0.064 $\pm$ 0.006 & 0.406 $\pm$ 0.034 & 0.0e+00 $\pm$ 0.0e+00 & 4 $\pm$ 10 & 11.2 $\pm$ 4.8 \\
uniGradICON & 19 & 0.7913 $\pm$ 0.0242 & 0.8472 $\pm$ 0.0191 & 2.03 $\pm$ 0.47 & 0.060 $\pm$ 0.003 & 0.404 $\pm$ 0.029 & 2.0e-06 $\pm$ 1.9e-06 & 177 $\pm$ 110 & 2.4 $\pm$ 0.1 \\
uniGradICON + IO & 19 & 0.7957 $\pm$ 0.0238 & 0.8550 $\pm$ 0.0191 & 1.97 $\pm$ 0.47 & 0.063 $\pm$ 0.003 & 0.430 $\pm$ 0.037 & 1.0e-05 $\pm$ 2.0e-05 & 422 $\pm$ 390 & 36.7 $\pm$ 0.2 \\
uniGradICON + IO, eq. time & 19 & 0.7982 $\pm$ 0.0241 & 0.8589 $\pm$ 0.0191 & 1.94 $\pm$ 0.45 & 0.064 $\pm$ 0.003 & 0.446 $\pm$ 0.042 & 1.5e-05 $\pm$ 2.4e-05 & 578 $\pm$ 474 & 85.6 $\pm$ 0.2 \\
VoxelMorph & 19 & 0.7868 $\pm$ 0.0222 & 0.8464 $\pm$ 0.0169 & 2.04 $\pm$ 0.44 & 0.063 $\pm$ 0.005 & 0.444 $\pm$ 0.051 & 0.0e+00 $\pm$ 0.0e+00 & 7 $\pm$ 9 & 4.0 $\pm$ 0.1 \\
TransMorph base$^\dagger$ & 19 & 0.8566 $\pm$ 0.0143 & 0.8810 $\pm$ 0.0138 & 1.46 $\pm$ 0.33 & 0.122 $\pm$ 0.013 & 4.359 $\pm$ 0.397 & 8.5e-03 $\pm$ 1.5e-03 & 137121 $\pm$ 17802 & 0.3 $\pm$ 0.0 \\
TransMorph large$^\dagger$ & 19 & 0.8617 $\pm$ 0.0146 & 0.8828 $\pm$ 0.0138 & 1.42 $\pm$ 0.29 & 0.125 $\pm$ 0.018 & 4.241 $\pm$ 0.397 & 8.0e-03 $\pm$ 1.5e-03 & 132149 $\pm$ 17520 & 0.3 $\pm$ 0.1 \\
\bottomrule
\end{tabular}

%% file: analysis/h1/table1_H1_diffs.tex
\begin{tabular}{lcccccc}
\toprule
Ours (certified, 90 s) $-$ method & seg35 $\uparrow$ & seg4 $\uparrow$ & HD95 (mm) $\downarrow$ & SDlogJ (L2R) $\downarrow$ & SDlogJ (brain) $\downarrow$ & certified $J\le0$ cells $\downarrow$ \\
\midrule
ANTs SyN & \makecell{+0.0224 [+0.0192, +0.0253]\\ $p_{\mathrm{H}}$ $<$0.001; 47/49} & \makecell{+0.0247 [+0.0227, +0.0266]\\ $p_{\mathrm{H}}$ $<$0.001; 48/49} & \makecell{-0.26 [-0.36, -0.18]\\ $p_{\mathrm{H}}$ $<$0.001; 43/49} & \makecell{+0.014 [+0.013, +0.015]\\ $p_{\mathrm{H}}$ $<$0.001; 0/49} & \makecell{+0.179 [+0.171, +0.188]\\ $p_{\mathrm{H}}$ $<$0.001; 0/49} & \makecell{-13 [-17, -10]\\ $p_{\mathrm{H}}$ $<$0.001; 49/49} \\
ANTs SyN (-t so) & \makecell{+0.0224 [+0.0192, +0.0255]\\ $p_{\mathrm{H}}$ $<$0.001; 47/49} & \makecell{+0.0247 [+0.0227, +0.0266]\\ $p_{\mathrm{H}}$ $<$0.001; 49/49} & \makecell{-0.25 [-0.35, -0.17]\\ $p_{\mathrm{H}}$ $<$0.001; 43/49} & \makecell{+0.013 [+0.013, +0.014]\\ $p_{\mathrm{H}}$ $<$0.001; 0/49} & \makecell{+0.180 [+0.171, +0.188]\\ $p_{\mathrm{H}}$ $<$0.001; 0/49} & \makecell{-16 [-20, -12]\\ $p_{\mathrm{H}}$ $<$0.001; 47/49} \\
VPreg & \makecell{+0.0616 [+0.0566, +0.0669]\\ $p_{\mathrm{H}}$ $<$0.001; 49/49} & \makecell{+0.0936 [+0.0901, +0.0973]\\ $p_{\mathrm{H}}$ $<$0.001; 49/49} & \makecell{-0.47 [-0.53, -0.42]\\ $p_{\mathrm{H}}$ $<$0.001; 49/49} & \makecell{+0.044 [+0.043, +0.045]\\ $p_{\mathrm{H}}$ $<$0.001; 0/49} & \makecell{+0.381 [+0.370, +0.391]\\ $p_{\mathrm{H}}$ $<$0.001; 0/49} & \makecell{-0 [-0, +0]\\ $p_{\mathrm{H}}$ 0.539; 2/49} \\
NiftyReg & \makecell{+0.0301 [+0.0274, +0.0327]\\ $p_{\mathrm{H}}$ $<$0.001; 48/49} & \makecell{+0.0279 [+0.0263, +0.0294]\\ $p_{\mathrm{H}}$ $<$0.001; 49/49} & \makecell{-0.22 [-0.30, -0.16]\\ $p_{\mathrm{H}}$ $<$0.001; 45/49} & \makecell{+0.010 [+0.009, +0.011]\\ $p_{\mathrm{H}}$ $<$0.001; 1/49} & \makecell{-1.523 [-1.642, -1.414]\\ $p_{\mathrm{H}}$ $<$0.001; 49/49} & \makecell{-20460 [-22358, -18688]\\ $p_{\mathrm{H}}$ $<$0.001; 49/49} \\
deedsBCV & \makecell{+0.0792 [+0.0736, +0.0849]\\ $p_{\mathrm{H}}$ $<$0.001; 49/49} & \makecell{+0.1308 [+0.1262, +0.1356]\\ $p_{\mathrm{H}}$ $<$0.001; 49/49} & \makecell{-0.59 [-0.68, -0.51]\\ $p_{\mathrm{H}}$ $<$0.001; 49/49} & \makecell{+0.024 [+0.023, +0.025]\\ $p_{\mathrm{H}}$ $<$0.001; 0/49} & \makecell{+0.337 [+0.327, +0.347]\\ $p_{\mathrm{H}}$ $<$0.001; 0/49} & \makecell{-1 [-1, -0]\\ $p_{\mathrm{H}}$ $<$0.001; 21/49} \\
ConvexAdam & \makecell{+0.0591 [+0.0545, +0.0637]\\ $p_{\mathrm{H}}$ $<$0.001; 49/49} & \makecell{+0.1044 [+0.1005, +0.1090]\\ $p_{\mathrm{H}}$ $<$0.001; 49/49} & \makecell{-0.55 [-0.65, -0.46]\\ $p_{\mathrm{H}}$ $<$0.001; 49/49} & \makecell{+0.018 [+0.017, +0.019]\\ $p_{\mathrm{H}}$ $<$0.001; 0/49} & \makecell{-0.135 [-0.187, -0.087]\\ $p_{\mathrm{H}}$ $<$0.001; 35/49} & \makecell{-2229 [-2603, -1887]\\ $p_{\mathrm{H}}$ $<$0.001; 49/49} \\
FireANTs greedy & \makecell{+0.0275 [+0.0224, +0.0331]\\ $p_{\mathrm{H}}$ $<$0.001; 47/49} & \makecell{+0.0384 [+0.0347, +0.0425]\\ $p_{\mathrm{H}}$ $<$0.001; 49/49} & \makecell{-0.30 [-0.38, -0.22]\\ $p_{\mathrm{H}}$ $<$0.001; 46/49} & \makecell{+0.006 [+0.004, +0.008]\\ $p_{\mathrm{H}}$ $<$0.001; 4/49} & \makecell{+0.152 [+0.143, +0.162]\\ $p_{\mathrm{H}}$ $<$0.001; 0/49} & \makecell{-0 [-1, +0]\\ $p_{\mathrm{H}}$ 0.539; 2/49} \\
FireANTs greedy, eq. time & \makecell{+0.0268 [+0.0220, +0.0318]\\ $p_{\mathrm{H}}$ $<$0.001; 47/49} & \makecell{+0.0379 [+0.0346, +0.0417]\\ $p_{\mathrm{H}}$ $<$0.001; 49/49} & \makecell{-0.29 [-0.38, -0.22]\\ $p_{\mathrm{H}}$ $<$0.001; 47/49} & \makecell{+0.006 [+0.005, +0.008]\\ $p_{\mathrm{H}}$ $<$0.001; 4/49} & \makecell{+0.148 [+0.138, +0.158]\\ $p_{\mathrm{H}}$ $<$0.001; 0/49} & \makecell{-0 [-1, +0]\\ $p_{\mathrm{H}}$ 0.435; 3/49} \\
FireANTs SyN & \makecell{+0.0840 [+0.0791, +0.0894]\\ $p_{\mathrm{H}}$ $<$0.001; 49/49} & \makecell{+0.0821 [+0.0789, +0.0855]\\ $p_{\mathrm{H}}$ $<$0.001; 49/49} & \makecell{-0.71 [-0.81, -0.62]\\ $p_{\mathrm{H}}$ $<$0.001; 49/49} & \makecell{+0.013 [+0.011, +0.015]\\ $p_{\mathrm{H}}$ $<$0.001; 2/49} & \makecell{+0.251 [+0.241, +0.260]\\ $p_{\mathrm{H}}$ $<$0.001; 0/49} & \makecell{+0 [+0, +0]\\ $p_{\mathrm{H}}$ 1.000; 0/49} \\
FireANTs SyN, eq. time & \makecell{+0.0469 [+0.0412, +0.0529]\\ $p_{\mathrm{H}}$ $<$0.001; 49/49} & \makecell{+0.0636 [+0.0593, +0.0681]\\ $p_{\mathrm{H}}$ $<$0.001; 49/49} & \makecell{-0.52 [-0.63, -0.43]\\ $p_{\mathrm{H}}$ $<$0.001; 48/49} & \makecell{+0.015 [+0.014, +0.016]\\ $p_{\mathrm{H}}$ $<$0.001; 1/49} & \makecell{+0.215 [+0.203, +0.227]\\ $p_{\mathrm{H}}$ $<$0.001; 0/49} & \makecell{-2 [-3, -1]\\ $p_{\mathrm{H}}$ $<$0.001; 18/49} \\
uniGradICON & \makecell{+0.0288 [+0.0257, +0.0318]\\ $p_{\mathrm{H}}$ $<$0.001; 47/49} & \makecell{+0.0344 [+0.0330, +0.0357]\\ $p_{\mathrm{H}}$ $<$0.001; 49/49} & \makecell{-0.22 [-0.27, -0.17]\\ $p_{\mathrm{H}}$ $<$0.001; 44/49} & \makecell{+0.019 [+0.018, +0.020]\\ $p_{\mathrm{H}}$ $<$0.001; 0/49} & \makecell{+0.210 [+0.203, +0.217]\\ $p_{\mathrm{H}}$ $<$0.001; 0/49} & \makecell{-177 [-209, -147]\\ $p_{\mathrm{H}}$ $<$0.001; 49/49} \\
uniGradICON + IO & \makecell{+0.0246 [+0.0215, +0.0274]\\ $p_{\mathrm{H}}$ $<$0.001; 47/49} & \makecell{+0.0266 [+0.0250, +0.0280]\\ $p_{\mathrm{H}}$ $<$0.001; 49/49} & \makecell{-0.15 [-0.21, -0.11]\\ $p_{\mathrm{H}}$ $<$0.001; 42/49} & \makecell{+0.016 [+0.015, +0.017]\\ $p_{\mathrm{H}}$ $<$0.001; 0/49} & \makecell{+0.180 [+0.170, +0.189]\\ $p_{\mathrm{H}}$ $<$0.001; 0/49} & \makecell{-450 [-566, -348]\\ $p_{\mathrm{H}}$ $<$0.001; 49/49} \\
uniGradICON + IO, eq. time & \makecell{+0.0221 [+0.0191, +0.0249]\\ $p_{\mathrm{H}}$ $<$0.001; 47/49} & \makecell{+0.0227 [+0.0210, +0.0241]\\ $p_{\mathrm{H}}$ $<$0.001; 48/49} & \makecell{-0.11 [-0.17, -0.07]\\ $p_{\mathrm{H}}$ $<$0.001; 39/49} & \makecell{+0.014 [+0.014, +0.015]\\ $p_{\mathrm{H}}$ $<$0.001; 0/49} & \makecell{+0.156 [+0.137, +0.171]\\ $p_{\mathrm{H}}$ $<$0.001; 3/49} & \makecell{-690 [-865, -532]\\ $p_{\mathrm{H}}$ $<$0.001; 49/49} \\
VoxelMorph & \makecell{+0.0321 [+0.0298, +0.0344]\\ $p_{\mathrm{H}}$ $<$0.001; 49/49} & \makecell{+0.0355 [+0.0342, +0.0367]\\ $p_{\mathrm{H}}$ $<$0.001; 49/49} & \makecell{-0.20 [-0.26, -0.16]\\ $p_{\mathrm{H}}$ $<$0.001; 45/49} & \makecell{+0.015 [+0.015, +0.016]\\ $p_{\mathrm{H}}$ $<$0.001; 0/49} & \makecell{+0.162 [+0.145, +0.176]\\ $p_{\mathrm{H}}$ $<$0.001; 1/49} & \makecell{-10 [-16, -5]\\ $p_{\mathrm{H}}$ $<$0.001; 42/49} \\
TransMorph base$^\dagger$ & \makecell{-0.0630 [-0.0683, -0.0581]\\ $p_{\mathrm{H}}$ $<$0.001; 0/49} & \makecell{-0.0004 [-0.0031, +0.0020]\\ $p_{\mathrm{H}}$ 0.992; 30/49} & \makecell{+0.49 [+0.40, +0.58]\\ $p_{\mathrm{H}}$ $<$0.001; 1/48} & \makecell{-0.045 [-0.050, -0.040]\\ $p_{\mathrm{H}}$ $<$0.001; 49/49} & \makecell{-3.700 [-3.803, -3.598]\\ $p_{\mathrm{H}}$ $<$0.001; 49/49} & \makecell{-135001 [-141017, -129004]\\ $p_{\mathrm{H}}$ $<$0.001; 49/49} \\
TransMorph large$^\dagger$ & \makecell{-0.0633 [-0.0685, -0.0583]\\ $p_{\mathrm{H}}$ $<$0.001; 0/49} & \makecell{-0.0012 [-0.0040, +0.0013]\\ $p_{\mathrm{H}}$ 0.992; 29/49} & \makecell{+0.49 [+0.41, +0.58]\\ $p_{\mathrm{H}}$ $<$0.001; 1/48} & \makecell{-0.045 [-0.052, -0.040]\\ $p_{\mathrm{H}}$ $<$0.001; 49/49} & \makecell{-3.592 [-3.694, -3.489]\\ $p_{\mathrm{H}}$ $<$0.001; 49/49} & \makecell{-130548 [-136713, -124820]\\ $p_{\mathrm{H}}$ $<$0.001; 49/49} \\
\bottomrule
\end{tabular}

%% file: analysis/h1/fold_magnitudes.tex
\begin{tabular}{lrrrrr}
\toprule
Method & maps with a fold & cells per folded map & min $J$, median & min $J$, closest to $0$ & export corr.\ (min) \\
\midrule
ANTs SyN & 49/49 & 10 & -0.322 & -0.0086 & 1.0000 \\
ANTs SyN (-t so) & 47/49 & 12 & -0.338 & -0.015 & 1.0000 \\
NiftyReg & 49/49 & 18218 & -1.72 & -0.981 & 1.0000 \\
deedsBCV & 21/49 & 1 & -0.077 & -0.0012 & 1.0000 \\
ConvexAdam & 49/49 & 1948 & -0.534 & -0.246 & 1.0000 \\
FireANTs greedy & 2/49 & 5 & -0.017 & -0.0035 & 1.0000 \\
FireANTs greedy, eq.\ time & 3/49 & 3 & -0.012 & -0.0047 & 1.0000 \\
FireANTs SyN & 0/49 & -- & -- & -- & 1.0000 \\
FireANTs SyN, eq.\ time & 18/49 & 4 & -0.307 & -0.0041 & 1.0000 \\
VPreg & 2/49 & 4 & -0.060 & -0.038 & 0.9993 \\
uniGradICON & 49/49 & 136 & -0.243 & -0.066 & 1.0000 \\
uniGradICON + IO & 49/49 & 310 & -0.268 & -0.100 & 1.0000 \\
uniGradICON + IO, eq.\ time & 49/49 & 505 & -0.359 & -0.120 & 1.0000 \\
VoxelMorph & 42/49 & 5 & -0.095 & -0.0016 & 1.0000 \\
TransMorph base$^\dagger$ & 49/49 & 136174 & -9.17 & -6.26 & 0.9940 \\
\quad field used by the network & 49/49 & 136174 & -9.32 & -6.37 & -- \\
TransMorph large$^\dagger$ & 49/49 & 129133 & -10.29 & -6.06 & 0.9940 \\
\quad field used by the network & 49/49 & 129133 & -10.46 & -6.16 & -- \\
\bottomrule
\end{tabular}

%% file: tables/table2_H2_means.tex
\begin{tabular}{lcccccccc}
\toprule
Method & $n$ & Dice $\uparrow$ & HD95 (mm) $\downarrow$ & SDlogJ (L2R) $\downarrow$ & SDlogJ (brain) $\downarrow$ & folds (c.d.) $\downarrow$ & certified $J\le0$ cells $\downarrow$ & time (s) $\downarrow$ \\
\midrule
Ours (certified, 90 s) & 10 & 0.7231 $\pm$ 0.0100 & 6.18 $\pm$ 0.40 & 0.061 $\pm$ 0.004 & 0.593 $\pm$ 0.061 & 0.0e+00 $\pm$ 0.0e+00 & 0 $\pm$ 0 & 88.4 $\pm$ 0.6 \\
ANTs SyN & 10 & 0.7270 $\pm$ 0.0103 & 5.99 $\pm$ 0.40 & 0.042 $\pm$ 0.004 & 0.384 $\pm$ 0.012 & 0.0e+00 $\pm$ 0.0e+00 & 2 $\pm$ 3 & 193.2 $\pm$ 6.7 \\
ANTs SyN (-t so) & 10 & 0.7325 $\pm$ 0.0088 & 5.91 $\pm$ 0.39 & 0.050 $\pm$ 0.006 & 0.387 $\pm$ 0.009 & 0.0e+00 $\pm$ 0.0e+00 & 2 $\pm$ 2 & 176.6 $\pm$ 3.8 \\
NiftyReg & 10 & 0.7221 $\pm$ 0.0120 & 5.94 $\pm$ 0.42 & 0.052 $\pm$ 0.004 & 1.634 $\pm$ 0.491 & 6.2e-04 $\pm$ 3.9e-04 & 7471 $\pm$ 4422 & 13.6 $\pm$ 0.7 \\
deedsBCV & 10 & 0.7015 $\pm$ 0.0101 & 6.16 $\pm$ 0.43 & 0.042 $\pm$ 0.004 & 0.239 $\pm$ 0.008 & 0.0e+00 $\pm$ 0.0e+00 & 0 $\pm$ 0 & 15.3 $\pm$ 0.2 \\
ConvexAdam & 10 & 0.7063 $\pm$ 0.0067 & 6.21 $\pm$ 0.45 & 0.046 $\pm$ 0.006 & 0.539 $\pm$ 0.136 & 4.0e-05 $\pm$ 4.1e-05 & 541 $\pm$ 473 & 1.8 $\pm$ 0.1 \\
FireANTs greedy & 10 & 0.7132 $\pm$ 0.0158 & 6.27 $\pm$ 0.50 & 0.067 $\pm$ 0.019 & 0.479 $\pm$ 0.022 & 0.0e+00 $\pm$ 0.0e+00 & 0 $\pm$ 1 & 3.8 $\pm$ 0.3 \\
FireANTs SyN & 10 & 0.6831 $\pm$ 0.0172 & 6.62 $\pm$ 0.50 & 0.064 $\pm$ 0.016 & 0.385 $\pm$ 0.017 & 0.0e+00 $\pm$ 0.0e+00 & 0 $\pm$ 0 & 3.6 $\pm$ 0.1 \\
uniGradICON & 10 & 0.7111 $\pm$ 0.0064 & 6.05 $\pm$ 0.32 & 0.041 $\pm$ 0.005 & 0.317 $\pm$ 0.015 & 1.5e-08 $\pm$ 4.7e-08 & 17 $\pm$ 16 & 2.4 $\pm$ 0.1 \\
uniGradICON + IO & 10 & 0.7181 $\pm$ 0.0059 & 5.94 $\pm$ 0.28 & 0.043 $\pm$ 0.005 & 0.346 $\pm$ 0.019 & 9.0e-08 $\pm$ 1.6e-07 & 36 $\pm$ 28 & 36.9 $\pm$ 0.2 \\
VoxelMorph & 10 & 0.6587 $\pm$ 0.0342 & 6.79 $\pm$ 0.65 & 0.381 $\pm$ 0.468 & 0.501 $\pm$ 0.064 & 4.5e-03 $\pm$ 5.4e-03 & 22064 $\pm$ 24102 & 3.9 $\pm$ 0.3 \\
TransMorph base$^\dagger$ & 10 & 0.6537 $\pm$ 0.0344 & 7.02 $\pm$ 0.74 & 0.075 $\pm$ 0.013 & 3.456 $\pm$ 0.962 & 5.6e-03 $\pm$ 3.8e-03 & 72949 $\pm$ 34027 & 0.3 $\pm$ 0.0 \\
TransMorph large$^\dagger$ & 10 & 0.6570 $\pm$ 0.0345 & 6.95 $\pm$ 0.71 & 0.072 $\pm$ 0.013 & 3.277 $\pm$ 1.375 & 5.0e-03 $\pm$ 3.4e-03 & 64312 $\pm$ 32339 & 0.4 $\pm$ 0.1 \\
\bottomrule
\end{tabular}

%% file: tables/table2_H3_means.tex
\begin{tabular}{lcccccccc}
\toprule
Method & $n$ & Dice $\uparrow$ & HD95 (mm) $\downarrow$ & SDlogJ (L2R) $\downarrow$ & SDlogJ (brain) $\downarrow$ & folds (c.d.) $\downarrow$ & certified $J\le0$ cells $\downarrow$ & time (s) $\downarrow$ \\
\midrule
Ours (certified, 90 s) & 36 & 0.6867 $\pm$ 0.0573 & 4.20 $\pm$ 1.72 & 0.086 $\pm$ 0.010 & 0.774 $\pm$ 0.091 & 0.0e+00 $\pm$ 0.0e+00 & 0 $\pm$ 0 & 88.3 $\pm$ 0.9 \\
ANTs SyN & 36 & 0.7263 $\pm$ 0.0248 & 3.95 $\pm$ 1.66 & 0.058 $\pm$ 0.004 & 0.438 $\pm$ 0.030 & 0.0e+00 $\pm$ 0.0e+00 & 7 $\pm$ 8 & 207.9 $\pm$ 6.5 \\
ANTs SyN (-t so) & 36 & 0.7262 $\pm$ 0.0249 & 3.95 $\pm$ 1.65 & 0.058 $\pm$ 0.004 & 0.438 $\pm$ 0.032 & 0.0e+00 $\pm$ 0.0e+00 & 7 $\pm$ 9 & 190.4 $\pm$ 5.8 \\
NiftyReg & 36 & 0.7019 $\pm$ 0.0361 & 4.16 $\pm$ 1.58 & 0.066 $\pm$ 0.004 & 2.189 $\pm$ 0.388 & 1.6e-03 $\pm$ 6.2e-04 & 19742 $\pm$ 6929 & 14.9 $\pm$ 0.5 \\
deedsBCV & 36 & 0.6923 $\pm$ 0.0257 & 4.08 $\pm$ 1.63 & 0.049 $\pm$ 0.002 & 0.277 $\pm$ 0.015 & 0.0e+00 $\pm$ 0.0e+00 & 1 $\pm$ 1 & 16.8 $\pm$ 0.8 \\
ConvexAdam & 36 & 0.7023 $\pm$ 0.0267 & 4.09 $\pm$ 1.61 & 0.053 $\pm$ 0.003 & 0.668 $\pm$ 0.219 & 1.2e-04 $\pm$ 1.2e-04 & 1510 $\pm$ 1244 & 1.8 $\pm$ 0.1 \\
FireANTs greedy & 36 & 0.7174 $\pm$ 0.0291 & 4.02 $\pm$ 1.62 & 0.068 $\pm$ 0.003 & 0.509 $\pm$ 0.026 & 0.0e+00 $\pm$ 0.0e+00 & 1 $\pm$ 3 & 3.8 $\pm$ 0.2 \\
FireANTs SyN & 36 & 0.6860 $\pm$ 0.0290 & 4.27 $\pm$ 1.60 & 0.061 $\pm$ 0.003 & 0.395 $\pm$ 0.016 & 0.0e+00 $\pm$ 0.0e+00 & 0 $\pm$ 0 & 3.5 $\pm$ 0.2 \\
uniGradICON & 36 & 0.6993 $\pm$ 0.0290 & 4.09 $\pm$ 1.59 & 0.054 $\pm$ 0.006 & 0.414 $\pm$ 0.050 & 2.3e-07 $\pm$ 1.0e-06 & 21 $\pm$ 31 & 2.5 $\pm$ 0.1 \\
uniGradICON + IO & 36 & 0.7060 $\pm$ 0.0277 & 4.01 $\pm$ 1.61 & 0.056 $\pm$ 0.006 & 0.439 $\pm$ 0.051 & 1.5e-07 $\pm$ 3.7e-07 & 64 $\pm$ 60 & 37.3 $\pm$ 0.5 \\
VoxelMorph & 36 & 0.6541 $\pm$ 0.0604 & 4.54 $\pm$ 1.78 & 0.091 $\pm$ 0.017 & 0.702 $\pm$ 0.191 & 1.7e-03 $\pm$ 9.4e-04 & 20919 $\pm$ 7599 & 4.1 $\pm$ 0.2 \\
TransMorph base$^\dagger$ & 36 & 0.7129 $\pm$ 0.0229 & 4.03 $\pm$ 1.60 & 0.096 $\pm$ 0.013 & 4.419 $\pm$ 0.497 & 6.5e-03 $\pm$ 1.7e-03 & 117066 $\pm$ 23769 & 0.3 $\pm$ 0.0 \\
TransMorph large$^\dagger$ & 36 & 0.7171 $\pm$ 0.0249 & 4.01 $\pm$ 1.60 & 0.084 $\pm$ 0.009 & 3.632 $\pm$ 0.563 & 4.3e-03 $\pm$ 1.4e-03 & 84065 $\pm$ 22716 & 0.4 $\pm$ 0.1 \\
\bottomrule
\end{tabular}

%% file: tables/table2_H2_diffs.tex
\begin{tabular}{lccccc}
\toprule
Ours (certified, 90 s) $-$ method & Dice $\uparrow$ & HD95 (mm) $\downarrow$ & SDlogJ (L2R) $\downarrow$ & SDlogJ (brain) $\downarrow$ & certified $J\le0$ cells $\downarrow$ \\
\midrule
ANTs SyN & \makecell{-0.0039 [-0.0153, +0.0112]\\ $p_{\mathrm{H}}$ 0.645; 3/10} & \makecell{+0.20 [-0.06, +0.44]\\ $p_{\mathrm{H}}$ 0.186; 2/10} & \makecell{+0.019 [+0.012, +0.024]\\ $p_{\mathrm{H}}$ 0.023; 0/10} & \makecell{+0.209 [+0.195, +0.231]\\ $p_{\mathrm{H}}$ 0.023; 0/10} & \makecell{-2 [-6, +0]\\ $p_{\mathrm{H}}$ 0.250; 5/10} \\
ANTs SyN (-t so) & \makecell{-0.0094 [-0.0156, -0.0021]\\ $p_{\mathrm{H}}$ 0.023; 0/10} & \makecell{+0.27 [+0.14, +0.41]\\ $p_{\mathrm{H}}$ 0.023; 0/10} & \makecell{+0.011 [+0.009, +0.015]\\ $p_{\mathrm{H}}$ 0.023; 0/10} & \makecell{+0.206 [+0.188, +0.223]\\ $p_{\mathrm{H}}$ 0.023; 0/10} & \makecell{-2 [-4, +0]\\ $p_{\mathrm{H}}$ 0.156; 6/10} \\
NiftyReg & \makecell{+0.0010 [-0.0043, +0.0052]\\ $p_{\mathrm{H}}$ 1.000; 5/10} & \makecell{+0.24 [+0.13, +0.42]\\ $p_{\mathrm{H}}$ 0.023; 0/10} & \makecell{+0.008 [+0.007, +0.012]\\ $p_{\mathrm{H}}$ 0.023; 0/10} & \makecell{-1.040 [-1.234, -0.451]\\ $p_{\mathrm{H}}$ 0.023; 10/10} & \makecell{-7471 [-9060, -2246]\\ $p_{\mathrm{H}}$ 0.023; 10/10} \\
deedsBCV & \makecell{+0.0216 [+0.0146, +0.0291]\\ $p_{\mathrm{H}}$ 0.023; 10/10} & \makecell{+0.02 [-0.24, +0.21]\\ $p_{\mathrm{H}}$ 1.000; 4/10} & \makecell{+0.018 [+0.013, +0.022]\\ $p_{\mathrm{H}}$ 0.023; 0/10} & \makecell{+0.355 [+0.326, +0.386]\\ $p_{\mathrm{H}}$ 0.023; 0/10} & \makecell{+0 [+0, +0]\\ $p_{\mathrm{H}}$ 1.000; 0/10} \\
ConvexAdam & \makecell{+0.0169 [+0.0093, +0.0254]\\ $p_{\mathrm{H}}$ 0.023; 10/10} & \makecell{-0.03 [-0.58, +0.18]\\ $p_{\mathrm{H}}$ 1.000; 3/10} & \makecell{+0.015 [+0.011, +0.019]\\ $p_{\mathrm{H}}$ 0.023; 0/10} & \makecell{+0.055 [-0.087, +0.139]\\ $p_{\mathrm{H}}$ 0.160; 2/10} & \makecell{-541 [-1037, -265]\\ $p_{\mathrm{H}}$ 0.023; 10/10} \\
FireANTs greedy & \makecell{+0.0100 [+0.0007, +0.0233]\\ $p_{\mathrm{H}}$ 0.029; 9/10} & \makecell{-0.08 [-0.27, +0.23]\\ $p_{\mathrm{H}}$ 0.697; 7/10} & \makecell{-0.006 [-0.021, +0.004]\\ $p_{\mathrm{H}}$ 1.000; 4/10} & \makecell{+0.114 [+0.094, +0.137]\\ $p_{\mathrm{H}}$ 0.023; 0/10} & \makecell{-0 [-2, +0]\\ $p_{\mathrm{H}}$ 1.000; 1/10} \\
FireANTs SyN & \makecell{+0.0400 [+0.0287, +0.0579]\\ $p_{\mathrm{H}}$ 0.023; 10/10} & \makecell{-0.43 [-0.91, -0.09]\\ $p_{\mathrm{H}}$ 0.023; 10/10} & \makecell{-0.003 [-0.014, +0.008]\\ $p_{\mathrm{H}}$ 1.000; 4/10} & \makecell{+0.209 [+0.180, +0.248]\\ $p_{\mathrm{H}}$ 0.023; 0/10} & \makecell{+0 [+0, +0]\\ $p_{\mathrm{H}}$ 1.000; 0/10} \\
uniGradICON & \makecell{+0.0120 [+0.0007, +0.0193]\\ $p_{\mathrm{H}}$ 0.029; 9/10} & \makecell{+0.13 [-0.05, +0.36]\\ $p_{\mathrm{H}}$ 0.258; 2/10} & \makecell{+0.020 [+0.018, +0.022]\\ $p_{\mathrm{H}}$ 0.023; 0/10} & \makecell{+0.276 [+0.262, +0.294]\\ $p_{\mathrm{H}}$ 0.023; 0/10} & \makecell{-17 [-30, +0]\\ $p_{\mathrm{H}}$ 0.023; 9/10} \\
uniGradICON + IO & \makecell{+0.0050 [-0.0047, +0.0121]\\ $p_{\mathrm{H}}$ 0.193; 7/10} & \makecell{+0.24 [+0.05, +0.50]\\ $p_{\mathrm{H}}$ 0.035; 1/10} & \makecell{+0.018 [+0.016, +0.020]\\ $p_{\mathrm{H}}$ 0.023; 0/10} & \makecell{+0.248 [+0.235, +0.257]\\ $p_{\mathrm{H}}$ 0.023; 0/10} & \makecell{-36 [-68, -2]\\ $p_{\mathrm{H}}$ 0.023; 10/10} \\
VoxelMorph & \makecell{+0.0645 [+0.0310, +0.1127]\\ $p_{\mathrm{H}}$ 0.023; 10/10} & \makecell{-0.61 [-1.18, -0.28]\\ $p_{\mathrm{H}}$ 0.023; 10/10} & \makecell{-0.320 [-0.567, -0.014]\\ $p_{\mathrm{H}}$ 0.023; 10/10} & \makecell{+0.093 [+0.007, +0.169]\\ $p_{\mathrm{H}}$ 0.023; 1/10} & \makecell{-22064 [-29202, -715]\\ $p_{\mathrm{H}}$ 0.023; 10/10} \\
TransMorph base$^\dagger$ & \makecell{+0.0694 [+0.0382, +0.1219]\\ $p_{\mathrm{H}}$ 0.023; 10/10} & \makecell{-0.83 [-1.63, -0.39]\\ $p_{\mathrm{H}}$ 0.023; 10/10} & \makecell{-0.014 [-0.026, -0.006]\\ $p_{\mathrm{H}}$ 0.023; 10/10} & \makecell{-2.862 [-3.902, -2.139]\\ $p_{\mathrm{H}}$ 0.023; 10/10} & \makecell{-72949 [-121635, -28331]\\ $p_{\mathrm{H}}$ 0.023; 10/10} \\
TransMorph large$^\dagger$ & \makecell{+0.0661 [+0.0350, +0.1196]\\ $p_{\mathrm{H}}$ 0.023; 10/10} & \makecell{-0.77 [-1.48, -0.41]\\ $p_{\mathrm{H}}$ 0.023; 10/10} & \makecell{-0.012 [-0.023, -0.002]\\ $p_{\mathrm{H}}$ 0.023; 10/10} & \makecell{-2.684 [-4.323, -1.409]\\ $p_{\mathrm{H}}$ 0.023; 10/10} & \makecell{-64312 [-110422, -19944]\\ $p_{\mathrm{H}}$ 0.023; 10/10} \\
\bottomrule
\end{tabular}

%% file: tables/table2_H3_diffs.tex
\begin{tabular}{lccccc}
\toprule
Ours (certified, 90 s) $-$ method & Dice $\uparrow$ & HD95 (mm) $\downarrow$ & SDlogJ (L2R) $\downarrow$ & SDlogJ (brain) $\downarrow$ & certified $J\le0$ cells $\downarrow$ \\
\midrule
ANTs SyN & \makecell{-0.0396 [-0.0679, -0.0152]\\ $p_{\mathrm{H}}$ $<$0.001; 7/36} & \makecell{+0.23 [+0.05, +0.44]\\ $p_{\mathrm{H}}$ 0.042; 14/35} & \makecell{+0.028 [+0.024, +0.033]\\ $p_{\mathrm{H}}$ $<$0.001; 0/36} & \makecell{+0.336 [+0.293, +0.381]\\ $p_{\mathrm{H}}$ $<$0.001; 0/36} & \makecell{-7 [-12, -3]\\ $p_{\mathrm{H}}$ $<$0.001; 30/36} \\
ANTs SyN (-t so) & \makecell{-0.0395 [-0.0671, -0.0157]\\ $p_{\mathrm{H}}$ $<$0.001; 7/36} & \makecell{+0.23 [+0.05, +0.44]\\ $p_{\mathrm{H}}$ 0.042; 13/35} & \makecell{+0.028 [+0.024, +0.033]\\ $p_{\mathrm{H}}$ $<$0.001; 0/36} & \makecell{+0.336 [+0.292, +0.381]\\ $p_{\mathrm{H}}$ $<$0.001; 0/36} & \makecell{-7 [-13, -3]\\ $p_{\mathrm{H}}$ $<$0.001; 31/36} \\
NiftyReg & \makecell{-0.0152 [-0.0485, +0.0121]\\ $p_{\mathrm{H}}$ 1.000; 19/36} & \makecell{+0.02 [-0.28, +0.31]\\ $p_{\mathrm{H}}$ 1.000; 19/35} & \makecell{+0.020 [+0.016, +0.026]\\ $p_{\mathrm{H}}$ $<$0.001; 0/36} & \makecell{-1.415 [-1.663, -1.183]\\ $p_{\mathrm{H}}$ $<$0.001; 36/36} & \makecell{-19742 [-24385, -15368]\\ $p_{\mathrm{H}}$ $<$0.001; 36/36} \\
deedsBCV & \makecell{-0.0056 [-0.0373, +0.0224]\\ $p_{\mathrm{H}}$ 1.000; 21/36} & \makecell{+0.10 [-0.09, +0.32]\\ $p_{\mathrm{H}}$ 1.000; 17/35} & \makecell{+0.038 [+0.033, +0.042]\\ $p_{\mathrm{H}}$ $<$0.001; 0/36} & \makecell{+0.497 [+0.451, +0.539]\\ $p_{\mathrm{H}}$ $<$0.001; 0/36} & \makecell{-1 [-2, -0]\\ $p_{\mathrm{H}}$ 0.005; 12/36} \\
ConvexAdam & \makecell{-0.0156 [-0.0471, +0.0126]\\ $p_{\mathrm{H}}$ 1.000; 18/36} & \makecell{+0.09 [-0.11, +0.31]\\ $p_{\mathrm{H}}$ 1.000; 19/35} & \makecell{+0.034 [+0.028, +0.038]\\ $p_{\mathrm{H}}$ $<$0.001; 0/36} & \makecell{+0.106 [-0.070, +0.251]\\ $p_{\mathrm{H}}$ 0.010; 9/36} & \makecell{-1510 [-2434, -796]\\ $p_{\mathrm{H}}$ $<$0.001; 36/36} \\
FireANTs greedy & \makecell{-0.0307 [-0.0611, -0.0058]\\ $p_{\mathrm{H}}$ 0.005; 10/36} & \makecell{+0.17 [-0.02, +0.37]\\ $p_{\mathrm{H}}$ 0.147; 14/35} & \makecell{+0.019 [+0.015, +0.024]\\ $p_{\mathrm{H}}$ $<$0.001; 0/36} & \makecell{+0.265 [+0.227, +0.304]\\ $p_{\mathrm{H}}$ $<$0.001; 0/36} & \makecell{-1 [-3, -0]\\ $p_{\mathrm{H}}$ 0.022; 8/36} \\
FireANTs SyN & \makecell{+0.0007 [-0.0315, +0.0277]\\ $p_{\mathrm{H}}$ 1.000; 23/36} & \makecell{-0.10 [-0.30, +0.12]\\ $p_{\mathrm{H}}$ 1.000; 24/35} & \makecell{+0.026 [+0.021, +0.031]\\ $p_{\mathrm{H}}$ $<$0.001; 0/36} & \makecell{+0.379 [+0.339, +0.420]\\ $p_{\mathrm{H}}$ $<$0.001; 0/36} & \makecell{+0 [+0, +0]\\ $p_{\mathrm{H}}$ 1.000; 0/36} \\
uniGradICON & \makecell{-0.0125 [-0.0381, +0.0102]\\ $p_{\mathrm{H}}$ 1.000; 17/36} & \makecell{+0.10 [-0.07, +0.28]\\ $p_{\mathrm{H}}$ 1.000; 18/35} & \makecell{+0.032 [+0.028, +0.036]\\ $p_{\mathrm{H}}$ $<$0.001; 0/36} & \makecell{+0.360 [+0.316, +0.402]\\ $p_{\mathrm{H}}$ $<$0.001; 0/36} & \makecell{-21 [-41, -8]\\ $p_{\mathrm{H}}$ $<$0.001; 30/36} \\
uniGradICON + IO & \makecell{-0.0193 [-0.0461, +0.0038]\\ $p_{\mathrm{H}}$ 0.513; 15/36} & \makecell{+0.17 [+0.01, +0.36]\\ $p_{\mathrm{H}}$ 0.147; 13/35} & \makecell{+0.030 [+0.026, +0.035]\\ $p_{\mathrm{H}}$ $<$0.001; 0/36} & \makecell{+0.335 [+0.292, +0.378]\\ $p_{\mathrm{H}}$ $<$0.001; 0/36} & \makecell{-64 [-104, -31]\\ $p_{\mathrm{H}}$ $<$0.001; 36/36} \\
VoxelMorph & \makecell{+0.0327 [+0.0262, +0.0384]\\ $p_{\mathrm{H}}$ $<$0.001; 36/36} & \makecell{-0.33 [-0.43, -0.24]\\ $p_{\mathrm{H}}$ $<$0.001; 33/35} & \makecell{-0.004 [-0.014, +0.002]\\ $p_{\mathrm{H}}$ 0.189; 22/36} & \makecell{+0.071 [-0.025, +0.145]\\ $p_{\mathrm{H}}$ 0.002; 7/36} & \makecell{-20919 [-25093, -16307]\\ $p_{\mathrm{H}}$ $<$0.001; 36/36} \\
TransMorph base$^\dagger$ & \makecell{-0.0262 [-0.0524, -0.0053]\\ $p_{\mathrm{H}}$ 0.003; 8/36} & \makecell{+0.16 [-0.04, +0.35]\\ $p_{\mathrm{H}}$ 0.094; 11/35} & \makecell{-0.009 [-0.021, -0.000]\\ $p_{\mathrm{H}}$ 0.004; 26/36} & \makecell{-3.645 [-3.904, -3.350]\\ $p_{\mathrm{H}}$ $<$0.001; 36/36} & \makecell{-117066 [-130263, -103627]\\ $p_{\mathrm{H}}$ $<$0.001; 36/36} \\
TransMorph large$^\dagger$ & \makecell{-0.0304 [-0.0570, -0.0097]\\ $p_{\mathrm{H}}$ $<$0.001; 7/36} & \makecell{+0.18 [-0.01, +0.38]\\ $p_{\mathrm{H}}$ 0.060; 14/35} & \makecell{+0.002 [-0.006, +0.009]\\ $p_{\mathrm{H}}$ 0.189; 13/36} & \makecell{-2.858 [-3.201, -2.477]\\ $p_{\mathrm{H}}$ $<$0.001; 36/36} & \makecell{-84065 [-97325, -68965]\\ $p_{\mathrm{H}}$ $<$0.001; 36/36} \\
\bottomrule
\end{tabular}

%% file: tables/table2_H4_means.tex
\begin{tabular}{lcccccccc}
\toprule
Method & $n$ & seg35 $\uparrow$ & seg4 $\uparrow$ & HD95 (mm) $\downarrow$ & SDlogJ (L2R) $\downarrow$ & SDlogJ (brain) $\downarrow$ & certified $J\le0$ cells $\downarrow$ & time (s) $\downarrow$ \\
\midrule
Ours, 90 s & 20 & 0.7632 $\pm$ 0.0291 & 0.8496 $\pm$ 0.0194 & 2.93 $\pm$ 0.95 & 0.086 $\pm$ 0.004 & 0.723 $\pm$ 0.072 & 0 $\pm$ 0 & 88.0 $\pm$ 0.7 \\
Ours, 180 s & 20 & 0.7680 $\pm$ 0.0264 & 0.8574 $\pm$ 0.0159 & 2.85 $\pm$ 0.95 & 0.095 $\pm$ 0.003 & 0.799 $\pm$ 0.076 & 0 $\pm$ 0 & 177.5 $\pm$ 1.3 \\
Ours c$^{\prime}$ ($\times10^{-2}$), 90 s & 20 & 0.7553 $\pm$ 0.0441 & 0.8400 $\pm$ 0.0285 & 3.07 $\pm$ 1.06 & 0.085 $\pm$ 0.004 & 0.662 $\pm$ 0.065 & 0 $\pm$ 0 & 88.3 $\pm$ 0.9 \\
ANTs SyN & 20 & 0.7347 $\pm$ 0.0240 & 0.8160 $\pm$ 0.0175 & 3.38 $\pm$ 0.72 & 0.073 $\pm$ 0.003 & 0.536 $\pm$ 0.072 & 11 $\pm$ 12 & 189.2 $\pm$ 2.5 \\
ANTs SyN (-t so) & 20 & 0.7345 $\pm$ 0.0246 & 0.8160 $\pm$ 0.0176 & 3.37 $\pm$ 0.71 & 0.074 $\pm$ 0.003 & 0.534 $\pm$ 0.074 & 7 $\pm$ 9 & 173.8 $\pm$ 3.9 \\
NiftyReg & 20 & 0.7340 $\pm$ 0.0205 & 0.8231 $\pm$ 0.0180 & 3.19 $\pm$ 0.67 & 0.076 $\pm$ 0.003 & 2.822 $\pm$ 0.672 & 29573 $\pm$ 7789 & 14.3 $\pm$ 0.6 \\
deedsBCV & 20 & 0.6618 $\pm$ 0.0312 & 0.6854 $\pm$ 0.0200 & 3.82 $\pm$ 0.71 & 0.059 $\pm$ 0.004 & 0.334 $\pm$ 0.027 & 0 $\pm$ 1 & 15.4 $\pm$ 0.2 \\
ConvexAdam & 20 & 0.6771 $\pm$ 0.0317 & 0.7092 $\pm$ 0.0232 & 3.89 $\pm$ 0.82 & 0.066 $\pm$ 0.003 & 0.871 $\pm$ 0.233 & 2762 $\pm$ 1603 & 1.6 $\pm$ 0.1 \\
FireANTs greedy & 20 & 0.6989 $\pm$ 0.0344 & 0.7795 $\pm$ 0.0196 & 3.57 $\pm$ 0.78 & 0.080 $\pm$ 0.009 & 0.533 $\pm$ 0.048 & 0 $\pm$ 0 & 3.3 $\pm$ 0.2 \\
FireANTs greedy, eq. time & 20 & 0.7020 $\pm$ 0.0316 & 0.7804 $\pm$ 0.0189 & 3.55 $\pm$ 0.77 & 0.080 $\pm$ 0.007 & 0.542 $\pm$ 0.049 & 0 $\pm$ 0 & 4.1 $\pm$ 0.8 \\
FireANTs SyN & 20 & 0.6448 $\pm$ 0.0267 & 0.7389 $\pm$ 0.0179 & 4.04 $\pm$ 0.78 & 0.073 $\pm$ 0.009 & 0.427 $\pm$ 0.046 & 0 $\pm$ 0 & 3.3 $\pm$ 0.1 \\
FireANTs SyN, eq. time & 20 & 0.6747 $\pm$ 0.0291 & 0.7485 $\pm$ 0.0197 & 3.88 $\pm$ 0.79 & 0.070 $\pm$ 0.005 & 0.459 $\pm$ 0.034 & 2 $\pm$ 5 & 12.3 $\pm$ 3.8 \\
uniGradICON & 20 & 0.7293 $\pm$ 0.0255 & 0.8097 $\pm$ 0.0123 & 3.28 $\pm$ 0.87 & 0.069 $\pm$ 0.004 & 0.509 $\pm$ 0.081 & 256 $\pm$ 169 & 2.4 $\pm$ 0.1 \\
uniGradICON + IO & 20 & 0.7388 $\pm$ 0.0193 & 0.8234 $\pm$ 0.0114 & 3.13 $\pm$ 0.78 & 0.072 $\pm$ 0.004 & 0.557 $\pm$ 0.104 & 755 $\pm$ 696 & 36.4 $\pm$ 0.2 \\
uniGradICON + IO, eq. time & 20 & 0.7458 $\pm$ 0.0176 & 0.8306 $\pm$ 0.0115 & 3.02 $\pm$ 0.74 & 0.074 $\pm$ 0.004 & 0.591 $\pm$ 0.116 & 1001 $\pm$ 878 & 85.1 $\pm$ 0.2 \\
VoxelMorph & 20 & 0.7288 $\pm$ 0.0312 & 0.8206 $\pm$ 0.0130 & 3.07 $\pm$ 0.86 & 0.074 $\pm$ 0.004 & 0.643 $\pm$ 0.202 & 68 $\pm$ 164 & 4.0 $\pm$ 0.2 \\
TransMorph base$^\dagger$ & 20 & 0.8744 $\pm$ 0.0089 & 0.8765 $\pm$ 0.0060 & 1.78 $\pm$ 0.63 & 0.136 $\pm$ 0.032 & 5.092 $\pm$ 0.286 & 166463 $\pm$ 17898 & 0.2 $\pm$ 0.0 \\
TransMorph large$^\dagger$ & 20 & 0.8736 $\pm$ 0.0095 & 0.8771 $\pm$ 0.0064 & 1.80 $\pm$ 0.62 & 0.140 $\pm$ 0.037 & 4.996 $\pm$ 0.276 & 161922 $\pm$ 18375 & 0.3 $\pm$ 0.0 \\
\bottomrule
\end{tabular}

%% file: tables/table2_H4_diffs.tex
\begin{tabular}{lccccc}
\toprule
Ours, 90 s $-$ method & seg35 $\uparrow$ & seg4 $\uparrow$ & HD95 (mm) $\downarrow$ & SDlogJ (L2R) $\downarrow$ & SDlogJ (brain) $\downarrow$ \\
\midrule
Ours, 180 s & \makecell{-0.0048 [-0.0079, -0.0024]\\ $p_{\mathrm{H}}$ 0.002; 5/20} & \makecell{-0.0078 [-0.0107, -0.0054]\\ $p_{\mathrm{H}}$ $<$0.001; 0/20} & \makecell{+0.08 [+0.04, +0.11]\\ $p_{\mathrm{H}}$ 0.001; 1/20} & \makecell{-0.009 [-0.011, -0.008]\\ $p_{\mathrm{H}}$ $<$0.001; 20/20} & \makecell{-0.077 [-0.091, -0.064]\\ $p_{\mathrm{H}}$ $<$0.001; 20/20} \\
Ours c$^{\prime}$ ($\times10^{-2}$), 90 s & \makecell{+0.0080 [+0.0016, +0.0184]\\ $p_{\mathrm{H}}$ 0.017; 14/20} & \makecell{+0.0097 [+0.0054, +0.0161]\\ $p_{\mathrm{H}}$ $<$0.001; 19/20} & \makecell{-0.14 [-0.27, -0.05]\\ $p_{\mathrm{H}}$ 0.012; 15/20} & \makecell{+0.001 [-0.000, +0.003]\\ $p_{\mathrm{H}}$ 0.105; 6/20} & \makecell{+0.061 [+0.051, +0.071]\\ $p_{\mathrm{H}}$ $<$0.001; 0/20} \\
ANTs SyN & \makecell{+0.0286 [+0.0224, +0.0349]\\ $p_{\mathrm{H}}$ $<$0.001; 20/20} & \makecell{+0.0337 [+0.0268, +0.0405]\\ $p_{\mathrm{H}}$ $<$0.001; 19/20} & \makecell{-0.45 [-0.71, -0.22]\\ $p_{\mathrm{H}}$ 0.007; 16/20} & \makecell{+0.014 [+0.011, +0.016]\\ $p_{\mathrm{H}}$ $<$0.001; 0/20} & \makecell{+0.187 [+0.169, +0.202]\\ $p_{\mathrm{H}}$ $<$0.001; 0/20} \\
ANTs SyN (-t so) & \makecell{+0.0287 [+0.0230, +0.0347]\\ $p_{\mathrm{H}}$ $<$0.001; 20/20} & \makecell{+0.0336 [+0.0268, +0.0402]\\ $p_{\mathrm{H}}$ $<$0.001; 19/20} & \makecell{-0.44 [-0.70, -0.21]\\ $p_{\mathrm{H}}$ 0.012; 16/20} & \makecell{+0.013 [+0.011, +0.015]\\ $p_{\mathrm{H}}$ $<$0.001; 0/20} & \makecell{+0.189 [+0.172, +0.205]\\ $p_{\mathrm{H}}$ $<$0.001; 0/20} \\
NiftyReg & \makecell{+0.0293 [+0.0218, +0.0368]\\ $p_{\mathrm{H}}$ $<$0.001; 19/20} & \makecell{+0.0265 [+0.0196, +0.0336]\\ $p_{\mathrm{H}}$ $<$0.001; 19/20} & \makecell{-0.26 [-0.47, -0.07]\\ $p_{\mathrm{H}}$ 0.145; 13/20} & \makecell{+0.010 [+0.008, +0.012]\\ $p_{\mathrm{H}}$ $<$0.001; 1/20} & \makecell{-2.099 [-2.364, -1.841]\\ $p_{\mathrm{H}}$ $<$0.001; 20/20} \\
deedsBCV & \makecell{+0.1014 [+0.0934, +0.1090]\\ $p_{\mathrm{H}}$ $<$0.001; 20/20} & \makecell{+0.1643 [+0.1587, +0.1701]\\ $p_{\mathrm{H}}$ $<$0.001; 20/20} & \makecell{-0.89 [-1.12, -0.69]\\ $p_{\mathrm{H}}$ $<$0.001; 20/20} & \makecell{+0.027 [+0.024, +0.030]\\ $p_{\mathrm{H}}$ $<$0.001; 0/20} & \makecell{+0.389 [+0.360, +0.415]\\ $p_{\mathrm{H}}$ $<$0.001; 0/20} \\
ConvexAdam & \makecell{+0.0861 [+0.0798, +0.0924]\\ $p_{\mathrm{H}}$ $<$0.001; 20/20} & \makecell{+0.1404 [+0.1346, +0.1465]\\ $p_{\mathrm{H}}$ $<$0.001; 20/20} & \makecell{-0.96 [-1.11, -0.82]\\ $p_{\mathrm{H}}$ $<$0.001; 20/20} & \makecell{+0.020 [+0.018, +0.023]\\ $p_{\mathrm{H}}$ $<$0.001; 0/20} & \makecell{-0.148 [-0.242, -0.059]\\ $p_{\mathrm{H}}$ 0.024; 15/20} \\
FireANTs greedy & \makecell{+0.0643 [+0.0561, +0.0728]\\ $p_{\mathrm{H}}$ $<$0.001; 20/20} & \makecell{+0.0701 [+0.0641, +0.0760]\\ $p_{\mathrm{H}}$ $<$0.001; 20/20} & \makecell{-0.64 [-0.79, -0.50]\\ $p_{\mathrm{H}}$ $<$0.001; 19/20} & \makecell{+0.006 [+0.001, +0.011]\\ $p_{\mathrm{H}}$ 0.059; 4/20} & \makecell{+0.190 [+0.171, +0.207]\\ $p_{\mathrm{H}}$ $<$0.001; 0/20} \\
FireANTs greedy, eq. time & \makecell{+0.0613 [+0.0531, +0.0695]\\ $p_{\mathrm{H}}$ $<$0.001; 20/20} & \makecell{+0.0692 [+0.0634, +0.0750]\\ $p_{\mathrm{H}}$ $<$0.001; 20/20} & \makecell{-0.62 [-0.77, -0.48]\\ $p_{\mathrm{H}}$ $<$0.001; 19/20} & \makecell{+0.006 [+0.002, +0.010]\\ $p_{\mathrm{H}}$ 0.052; 5/20} & \makecell{+0.181 [+0.163, +0.197]\\ $p_{\mathrm{H}}$ $<$0.001; 0/20} \\
FireANTs SyN & \makecell{+0.1184 [+0.1107, +0.1264]\\ $p_{\mathrm{H}}$ $<$0.001; 20/20} & \makecell{+0.1107 [+0.1056, +0.1157]\\ $p_{\mathrm{H}}$ $<$0.001; 20/20} & \makecell{-1.11 [-1.30, -0.94]\\ $p_{\mathrm{H}}$ $<$0.001; 20/20} & \makecell{+0.014 [+0.011, +0.016]\\ $p_{\mathrm{H}}$ $<$0.001; 1/20} & \makecell{+0.295 [+0.275, +0.313]\\ $p_{\mathrm{H}}$ $<$0.001; 0/20} \\
FireANTs SyN, eq. time & \makecell{+0.0886 [+0.0830, +0.0943]\\ $p_{\mathrm{H}}$ $<$0.001; 20/20} & \makecell{+0.1012 [+0.0955, +0.1066]\\ $p_{\mathrm{H}}$ $<$0.001; 20/20} & \makecell{-0.96 [-1.11, -0.82]\\ $p_{\mathrm{H}}$ $<$0.001; 20/20} & \makecell{+0.016 [+0.015, +0.018]\\ $p_{\mathrm{H}}$ $<$0.001; 0/20} & \makecell{+0.264 [+0.242, +0.286]\\ $p_{\mathrm{H}}$ $<$0.001; 0/20} \\
uniGradICON & \makecell{+0.0339 [+0.0281, +0.0391]\\ $p_{\mathrm{H}}$ $<$0.001; 19/20} & \makecell{+0.0399 [+0.0353, +0.0440]\\ $p_{\mathrm{H}}$ $<$0.001; 20/20} & \makecell{-0.36 [-0.47, -0.26]\\ $p_{\mathrm{H}}$ $<$0.001; 20/20} & \makecell{+0.018 [+0.015, +0.020]\\ $p_{\mathrm{H}}$ $<$0.001; 0/20} & \makecell{+0.213 [+0.192, +0.231]\\ $p_{\mathrm{H}}$ $<$0.001; 0/20} \\
uniGradICON + IO & \makecell{+0.0244 [+0.0179, +0.0303]\\ $p_{\mathrm{H}}$ $<$0.001; 18/20} & \makecell{+0.0262 [+0.0199, +0.0313]\\ $p_{\mathrm{H}}$ $<$0.001; 19/20} & \makecell{-0.20 [-0.33, -0.06]\\ $p_{\mathrm{H}}$ 0.033; 15/20} & \makecell{+0.014 [+0.012, +0.016]\\ $p_{\mathrm{H}}$ $<$0.001; 0/20} & \makecell{+0.165 [+0.137, +0.190]\\ $p_{\mathrm{H}}$ $<$0.001; 1/20} \\
uniGradICON + IO, eq. time & \makecell{+0.0174 [+0.0103, +0.0239]\\ $p_{\mathrm{H}}$ 0.002; 18/20} & \makecell{+0.0191 [+0.0125, +0.0247]\\ $p_{\mathrm{H}}$ $<$0.001; 19/20} & \makecell{-0.09 [-0.25, +0.08]\\ $p_{\mathrm{H}}$ 0.246; 13/20} & \makecell{+0.012 [+0.010, +0.014]\\ $p_{\mathrm{H}}$ $<$0.001; 0/20} & \makecell{+0.132 [+0.098, +0.162]\\ $p_{\mathrm{H}}$ $<$0.001; 1/20} \\
VoxelMorph & \makecell{+0.0344 [+0.0281, +0.0414]\\ $p_{\mathrm{H}}$ $<$0.001; 20/20} & \makecell{+0.0290 [+0.0233, +0.0340]\\ $p_{\mathrm{H}}$ $<$0.001; 19/20} & \makecell{-0.14 [-0.33, +0.04]\\ $p_{\mathrm{H}}$ 0.246; 15/20} & \makecell{+0.012 [+0.010, +0.015]\\ $p_{\mathrm{H}}$ $<$0.001; 1/20} & \makecell{+0.079 [+0.011, +0.141]\\ $p_{\mathrm{H}}$ 0.036; 8/20} \\
TransMorph base$^\dagger$ & \makecell{-0.1111 [-0.1216, -0.1016]\\ $p_{\mathrm{H}}$ $<$0.001; 0/20} & \makecell{-0.0269 [-0.0339, -0.0207]\\ $p_{\mathrm{H}}$ $<$0.001; 0/20} & \makecell{+1.15 [+0.96, +1.36]\\ $p_{\mathrm{H}}$ $<$0.001; 0/20} & \makecell{-0.049 [-0.065, -0.037]\\ $p_{\mathrm{H}}$ $<$0.001; 20/20} & \makecell{-4.369 [-4.487, -4.248]\\ $p_{\mathrm{H}}$ $<$0.001; 20/20} \\
TransMorph large$^\dagger$ & \makecell{-0.1103 [-0.1202, -0.1013]\\ $p_{\mathrm{H}}$ $<$0.001; 0/20} & \makecell{-0.0275 [-0.0342, -0.0214]\\ $p_{\mathrm{H}}$ $<$0.001; 0/20} & \makecell{+1.12 [+0.93, +1.33]\\ $p_{\mathrm{H}}$ $<$0.001; 0/20} & \makecell{-0.053 [-0.071, -0.040]\\ $p_{\mathrm{H}}$ $<$0.001; 20/20} & \makecell{-4.273 [-4.389, -4.154]\\ $p_{\mathrm{H}}$ $<$0.001; 20/20} \\
\bottomrule
\end{tabular}

%% file: tables/table3_ablation.tex
\begin{tabular}{lccccccccc}
\toprule
Arm & $n$ & seg35 & seg4 & SDlogJ (brain) & seg35 $-$ C & seg4 $-$ C & SDlogJ (brain) $-$ C & time (s) & no native commit \\
\midrule
frozen configuration (same batch) & 19 & 0.8194 & 0.8824 & 0.628 & -- & -- & -- & 88.0 & 0 \\
(a) $\mu=0$ & 19 & 0.8171 & 0.8746 & 0.550 & \makecell{-0.0023 [-0.0033, -0.0012]\\ $p_{\mathrm{H}}$ 0.006; 3/19} & \makecell{-0.0079 [-0.0090, -0.0068]\\ $p_{\mathrm{H}}$ $<$0.001; 0/19} & \makecell{-0.077 [-0.086, -0.070]\\ $p_{\mathrm{H}}$ $<$0.001; 19/19} & 88.2 & 0 \\
(b) $\nu=0$ & 19 & 0.8203 & 0.8831 & 0.637 & \makecell{+0.0009 [+0.0001, +0.0018]\\ $p_{\mathrm{H}}$ 0.096; 11/19} & \makecell{+0.0006 [+0.0002, +0.0013]\\ $p_{\mathrm{H}}$ 0.023; 14/19} & \makecell{+0.009 [+0.003, +0.016]\\ $p_{\mathrm{H}}$ 0.005; 5/19} & 87.7 & 0 \\
(c) fixed references & 19 & 0.8120 & 0.8617 & 0.474 & \makecell{-0.0074 [-0.0095, -0.0053]\\ $p_{\mathrm{H}}$ $<$0.001; 1/19} & \makecell{-0.0207 [-0.0219, -0.0196]\\ $p_{\mathrm{H}}$ $<$0.001; 0/19} & \makecell{-0.154 [-0.164, -0.145]\\ $p_{\mathrm{H}}$ $<$0.001; 19/19} & 88.2 & 0 \\
(c$'$) identity references, $\times1$ & 19 & 0.7336 & 0.7657 & 0.217 & \makecell{-0.0858 [-0.0999, -0.0740]\\ $p_{\mathrm{H}}$ $<$0.001; 0/19} & \makecell{-0.1167 [-0.1247, -0.1099]\\ $p_{\mathrm{H}}$ $<$0.001; 0/19} & \makecell{-0.411 [-0.432, -0.389]\\ $p_{\mathrm{H}}$ $<$0.001; 19/19} & 88.0 & 0 \\
(c$'$) identity references, $\times10^{-2}$ & 19 & 0.8223 & 0.8795 & 0.588 & \makecell{+0.0030 [+0.0015, +0.0043]\\ $p_{\mathrm{H}}$ 0.006; 15/19} & \makecell{-0.0029 [-0.0043, -0.0020]\\ $p_{\mathrm{H}}$ $<$0.001; 0/19} & \makecell{-0.040 [-0.049, -0.032]\\ $p_{\mathrm{H}}$ $<$0.001; 19/19} & 88.2 & 0 \\
(d) diagonal metric & 19 & 0.8229 & 0.8796 & 0.578 & \makecell{+0.0035 [+0.0014, +0.0052]\\ $p_{\mathrm{H}}$ 0.006; 17/19} & \makecell{-0.0028 [-0.0035, -0.0022]\\ $p_{\mathrm{H}}$ $<$0.001; 0/19} & \makecell{-0.049 [-0.055, -0.043]\\ $p_{\mathrm{H}}$ $<$0.001; 19/19} & 88.6 & 0 \\
(d) identity metric & 14 & 0.8053 & 0.8495 & 0.382 & \makecell{-0.0163 [-0.0209, -0.0122]\\ $p_{\mathrm{H}}$ $<$0.001; 0/14} & \makecell{-0.0328 [-0.0363, -0.0296]\\ $p_{\mathrm{H}}$ $<$0.001; 0/14} & \makecell{-0.236 [-0.254, -0.219]\\ $p_{\mathrm{H}}$ $<$0.001; 14/14} & 89.1 & 0 (+5) \\
(e) LM inner solver & 17 & 0.8009 & 0.8461 & 0.384 & \makecell{-0.0174 [-0.0222, -0.0126]\\ $p_{\mathrm{H}}$ $<$0.001; 0/17} & \makecell{-0.0351 [-0.0387, -0.0317]\\ $p_{\mathrm{H}}$ $<$0.001; 0/17} & \makecell{-0.243 [-0.257, -0.229]\\ $p_{\mathrm{H}}$ $<$0.001; 17/17} & 111.0 & 2 \\
\bottomrule
\end{tabular}

%% file: analysis/delta/delta_table.tex
\begin{tabular}{lccccccc}
\toprule
 & \multicolumn{3}{c}{mean} & \multicolumn{4}{c}{difference to $\delta=0$, in units of $10^{-3}$} \\
\cmidrule(lr){2-4}\cmidrule(lr){5-8}
Measure & $\delta=0$ & $\delta=10^{-3}$ & $\delta=10^{-2}$ & $\delta=10^{-3}$ & $p$ & $\delta=10^{-2}$ & $p$ \\
\midrule
seg35 & 0.819 & 0.819 & 0.818 & $-0.27$ [$-0.50$, $-0.03$] & 0.03 & $-1.14$ [$-2.61$, $-0.24$] & 0.03 \\
seg4 & 0.882 & 0.882 & 0.879 & $-0.67$ [$-1.61$, $-0.03$] & 0.11 & $-2.95$ [$-5.26$, $-1.33$] & $<10^{-4}$ \\
HD95 (mm) & 1.82 & 1.83 & 1.84 & $+8.1$ [$-3.9$, $+26.9$] & 0.50 & $+27.4$ [$+5.1$, $+56.7$] & 0.07 \\
L2R SDlogJ & 0.0798 & 0.0791 & 0.0769 & $-0.74$ [$-1.85$, $+0.02$] & 0.13 & $-2.92$ [$-4.94$, $-1.42$] & $<10^{-4}$ \\
in-brain SDlogJ & 0.625 & 0.618 & 0.599 & $-6.8$ [$-16.9$, $-0.1$] & 0.10 & $-26.0$ [$-42.8$, $-13.2$] & $<10^{-4}$ \\
native commits & 27.7 & 26.3 & 25.1 & & & & \\
handoffs at $\rho=0.9$ & 8/19 & 8/19 & 12/19 & & & & \\
min $J$, median & $2.885\cdot10^{-4}$ & $1.385\cdot10^{-3}$ & $1.001\cdot10^{-2}$ & & & & \\
min $J$, smallest & $1.453\cdot10^{-5}$ & $1.000\cdot10^{-3}$ & $1.000\cdot10^{-2}$ & & & & \\
\bottomrule
\end{tabular}

%% file: analysis/cond/cond_table.tex
\begin{tabular}{lrrrrrrrrrrr}
\toprule
 & & \multicolumn{5}{c}{condition number $\kappa$} & \multicolumn{2}{c}{Lipschitz} & \multicolumn{2}{c}{$J$} & \\
\cmidrule(lr){3-7}\cmidrule(lr){8-9}\cmidrule(lr){10-11}
Method & certified & median & p99 & p99.9 & max & $>10$ (\%) & $L(\Phi)$ & $L(\Phi^{-1})$ & p99 & max & $\det F\le0$ \\
\midrule
ours, frozen ($\delta=0$, $\alpha=0$) & 19/19 & $2.56$ & $44.4$ & $153$ & $2032$ & $6.72$ & $10.3$ & $855$ & $4.64$ & $19.2$ & $0$ \\
\quad $\delta=10^{-3}$ & 19/19 & $2.54$ & $43.7$ & $149$ & $1837$ & $6.16$ & $10.0$ & $730$ & $4.52$ & $18.2$ & $0$ \\
\quad $\delta=10^{-2}$ & 19/19 & $2.43$ & $30.6$ & $115$ & $865$ & $4.91$ & $9.16$ & $237$ & $4.14$ & $17.5$ & $0$ \\
\quad $\alpha=10^{-3}$ & 19/19 & $2.28$ & $31.2$ & $127$ & $1317$ & $4.24$ & $5.99$ & $658$ & $3.61$ & $14.9$ & $0$ \\
\quad $\alpha=3\cdot10^{-3}$ & 19/19 & $2.14$ & $20.4$ & $97.4$ & $1345$ & $2.72$ & $4.73$ & $615$ & $3.21$ & $11.5$ & $0$ \\
\quad $\alpha=10^{-2}$ & 19/19 & $1.95$ & $11.1$ & $55.1$ & $967$ & $1.20$ & $3.38$ & $531$ & $2.79$ & $8.28$ & $0$ \\
\midrule
ANTs SyN & 0/19 & $2.16$ & $6.89$ & $12.5$ & $1187$ & $0.22$ & $6.17$ & --- & $3.03$ & $10.5$ & $496$ \\
uniGradICON + IO, eq.\ time & 0/19 & $2.12$ & $8.58$ & $20.4$ & $2.29\cdot10^{4}$ & $0.66$ & $5.36$ & --- & $2.82$ & $10.3$ & $2.95\cdot10^{4}$ \\
FireANTs greedy & 19/19 & $2.13$ & $6.19$ & $10.5$ & $47.2$ & $0.12$ & $4.96$ & $33.8$ & $3.22$ & $12.5$ & $0$ \\
NiftyReg & 0/19 & $2.24$ & $33.2$ & $689$ & $1.57\cdot10^{6}$ & $2.82$ & $3.04$ & --- & $2.78$ & $6.51$ & $1.99\cdot10^{6}$ \\
VPreg & 19/19 & $1.37$ & $2.33$ & $3.20$ & $6.18$ & $0$ & $3.72$ & $3.05$ & $1.94$ & $6.62$ & $0$ \\
TransMorph large$^\dagger$ & 0/19 & $2.67$ & $192$ & $2130$ & $3.67\cdot10^{6}$ & $9.86$ & $6.78$ & --- & $4.70$ & $34.5$ & $1.13\cdot10^{7}$ \\
\midrule
ANTs SyN, variance $1$ & 17/19 & $1.86$ & $4.20$ & $6.06$ & $28.3$ & $0.00$ & $4.28$ & --- & $2.39$ & $6.57$ & $2$ \\
ANTs SyN, variance $0$ & 19/19 & $1.35$ & $1.87$ & $2.14$ & $3.24$ & $0$ & $1.96$ & $2.44$ & $1.54$ & $2.36$ & $0$ \\
FireANTs greedy, $f=0.5$ & 0/19 & $6.21$ & $454$ & $4943$ & $1.07\cdot10^{7}$ & $31.32$ & $13.8$ & --- & $9.47$ & $99.6$ & $4.45\cdot10^{6}$ \\
FireANTs greedy, $f=0.25$ & 0/19 & $5.72$ & $188$ & $2060$ & $4.97\cdot10^{6}$ & $24.32$ & $12.2$ & --- & $9.16$ & $80.2$ & $1.87\cdot10^{6}$ \\
FireANTs greedy, $f=0$ & 0/19 & $5.51$ & $163$ & $1719$ & $3.79\cdot10^{6}$ & $22.87$ & $11.9$ & --- & $9.13$ & $70.4$ & $1.60\cdot10^{6}$ \\
\bottomrule
\end{tabular}

%% file: sections/09_discussion.tex
\section{Discussion and limitations}\label{sec:discussion}

\paragraph{Scope of the guarantee} The certificate and \cref{thm:invertibility} concern the trilinear map
defined by the stored nodal displacement. Anatomical correctness of the correspondence lies outside their scope, as
do transformations obtained by other interpolations of the same field (for instance cubic B-spline
interpolation of the displacement) or by composing fields. The moving image is sampled with a cubic spline in the
data term; this affects the energy, not the map. The energy decrease of each commit is verified with a resolution
guard, not with a rigorous error bound, and in the recommended configuration partly with binary32 energies and an
empirical band (\cref{sec:implementation}); \cref{prop:monotone} is an exact-arithmetic statement checked to that
resolution.

\paragraph{Positivity and a lower bound} The certificate proves $J>0$, not that $J$ is bounded away from zero.
The returned maps contain a few strongly compressed cells: over the $177$ reported final maps the minimum of $J$ is
$1\times10^{-6}$ to $3\times10^{-3}$ (median of order $10^{-4}$), below $10^{-3}$ in about five cells per map. The barrier keeps $J$ away from zero only at its sample points, and
the minimum between them can be much smaller (\cref{fig:toy}(c)). The same tests certify $J>\delta$ for any
$\delta\ge0$. In a post-freeze arm on the validation pairs with the barrier shifted to $\delta$, certifying
$J>10^{-2}$ moved the minimum of $J$ to just above $10^{-2}$ on every map at a cost of $0.001$ seg35 and $0.003$ seg4
and with slightly smoother maps (\cref{sec:results}); a positive $\delta$ therefore buys a quantitative lower bound
cheaply, but its value is a modelling choice, and it was not evaluated on the held-out sets. An upper bound $J<M$ is
decided by the same coefficients, since $\max\beta$ bounds $J_K$ from above by \eqref{eq:bernstein}, so a two-sided
bound on the local volume change is available at the same cost. Invertibility does not imply good conditioning: in
a few percent of the brain cells our maps are more anisotropic than those of the smooth baselines (99th percentile of
the sampled $\kappa=\sigma_1/\sigma_3$ $44$ against $6$ to $9$), and the sampled Lipschitz constant of the inverse is
much larger than for the fold-free baselines (\cref{tab:cond}). A lower bound $\delta$ on $J$ improves both, the
curvature weight mainly the former; certified bounds on the singular values themselves, which would need bounds on
$\lvert\nabla\Phi_u\rvert$ over each cell in addition to those on $J$, are left to future work.

\paragraph{Termination} All reported runs are budget-terminated; none reaches the gradient threshold
(\cref{rem:no-convergence}). The method is designed and evaluated as an anytime method from the certified handoff on: the returned map is the
last committed iterate at the end of the budget, or the initialiser if nothing was committed.

\paragraph{Speed} The method is slower than GPU baselines that do not certify their output by a factor of 20
to 40 ($88$\,s end to end on the held-out pairs against $3.2$\,s for FireANTs greedy, $2.3$\,s for uniGradICON and
$4.0$\,s for VoxelMorph; $36$\,s for uniGradICON with instance optimisation), about six times slower than NiftyReg
and deedsBCV on the CPU ($14$--$15$\,s), about twice as slow as VPreg on the same GPU ($47$\,s for its forward
registration), and faster than ANTs SyN on 8 CPU threads ($188$\,s). Most of the time is spent in binary64 arithmetic on a consumer GPU with low binary64 throughput; on a GPU with
full-rate binary64 the native iterations would be faster, which we have not measured.

\paragraph{Regularity} The deformations are less smooth than those of the baselines by the SDlogJ measure (in-brain SDlogJ $0.62$ on H1
against $0.44$ for ANTs SyN, $0.47$ for uniGradICON at equal time and $0.24$ for VPreg; Learn2Reg SDlogJ $0.079$
against $0.065$ for ANTs). The native level uses no curvature regularisation in the recommended configuration ($\alpha=0$),
so there $\Fout=\Dat+\Bpot$, and by \cref{prop:consistency} a fixed point of the sequential scheme is a stationary
point of the image term plus the barrier: at its fixed points the model has no regulariser. The regularity of a
returned map comes from the path, each step being penalised against the last committed map, and from where the path
is stopped. The time budget acts as the regularisation parameter, as the stopping index does in iterative regularisation, where the iteration must be stopped early because of
semi-convergence \cite[Ch.~7 and 11]{Clason2021notes} \cite[Ch.~6]{EnglHankeNeubauer1996}. The time course on the validation pairs (\cref{fig:valtime}) shows this directly: from $90$ to $179$\,s the
in-brain SDlogJ rises from $0.63$ to $0.68$ while seg35 stays at $0.820$ and seg4 gains $0.003$, and before $90$\,s
the map is smoother and less accurate ($0.39$ and seg35 $0.803$ at $15$\,s). The budget of $90$\,s was fixed before the held-out evaluation.
Because the path is stopped by wall time, the same budget on other hardware stops at a different point of the path,
which is why we report the hardware (\cref{sec:implementation}) and the time course in addition to the end point; on the same machine the sequence of committed maps itself is reproducible bit for bit, and the number of native commits is recorded with every run (\cref{sec:implementation}). Adding curvature regularisation at the native level trades accuracy for smoothness along a frontier
(\cref{fig:frontier}): at $\alpha=10^{-2}$ the in-brain SDlogJ falls from $0.63$ to $0.49$ and seg35 from $0.819$ to
$0.801$, still above every classical baseline of comparable regularity, but no longer significantly so.

\paragraph{Role of the curl term} The curl-change penalty is part of the model because boundary values and the
Jacobian determinant alone do not determine a map (\cref{sec:sequential-model}). At the weights of the frozen
configuration its measured effect is small: removing it changes seg35 and seg4 by less than $0.001$ and moves the two
regularity measures by small amounts in opposite directions (\cref{sec:results}), and the same arm also halves the Poisson
coupling of the metric. The experiments therefore support the combined penalty without isolating a practical role
of the curl term; its effect at larger $\nu$, on inverse consistency and on rotational structures is a question for
further experiments.

\paragraph{What the accuracy measures show} Image similarity and the overlap of tissue classes are weak surrogates
for registration accuracy: a registration that is anatomically meaningless can score well on both
\cite{Rohlfing2012}. We therefore report overlaps of small anatomical structures (35 OASIS structures, 56 LPBA40
structures), HD95 and the time course of these measures, alongside the tissue overlap (seg4) and the MSE ratio; none of
them measures voxelwise correspondence directly.

\paragraph{Where the remaining error is} Label overlap at 1\,mm has a floor set by the labels themselves: shifting a
label map by one voxel already costs a sizeable fraction of its Dice. On the validation pairs the gap $1-\mathrm{seg35}$
splits into this one-voxel floor ($0.164$ on these pairs, a property of the label maps) and an excess of $0.016$ for
the frozen configuration (seg35 $0.819$); $80\%$ of the excess comes from six of the 35 structures, the choroid
plexus and the inferior lateral ventricles of both hemispheres, the third ventricle and the left hippocampus, and
$38\%$ from the six ventricular labels alone, so the remaining error
is concentrated in a few small structures around the ventricles, where the correspondence error exceeds one voxel. A uniform shrinkage of the final
displacement would raise the overlap slightly (a diagnostic on the validation runs: scaling the final displacement by $0.98$ raises seg35 by about $0.002$), which indicates a
systematic over-displacement of deep structures; the returned map is the certified committed iterate, so no such
post-hoc change is applied.

\paragraph{Large local compression} Registering a brain with strongly enlarged lateral ventricles to one with small
ventricles requires the map to compress the ventricles of the fixed image several-fold. On the 20 H4 pairs (volume
ratios $4.1$--$7.6$) the frozen configuration stays ahead of every label-free baseline, by $+0.017$ seg35 and
$+0.019$ seg4 against uniGradICON at equal time and $+0.029$ / $+0.034$ against ANTs, and twice the budget adds
$+0.005$ / $+0.008$ at a rougher map (\cref{sec:results}); the label-supervised TransMorph reaches seg35 $0.87$ there
with about $1.6\times10^5$ certified folded cells per map, and on the CSF class of seg4, where the ventricular
compression is measured, $0.92$ against $0.81$ for the frozen configuration and $0.80$ for ANTs. A pre-freeze check on one
development pair outside H4 (ventricle ratio $11.6$) located a limitation that appears beyond the H4 range. The uncertified pre-optimiser stops well short
of the required compression because of its soft lower bound on $J$; the certified handoff then rescales the whole
displacement; and the certified refinement advances slowly with most ventricle cells in the damping zone of the metric.
Relaxing the pre-optimiser's lower bound let it compress further, but left small values of $J$ elsewhere, so that the
handoff had to rescale the whole field more strongly; a pre-registered selection step on development and tuning pairs
rejected both candidates because a pair that requires expansion lost accuracy, and the configuration was frozen unchanged. The baselines that matched the ventricles of that pair
did so with certified folds, all but one of them in large numbers; one of them, ANTs, folded only in two cells, which
suggests that such compression is nearly compatible with invertibility and that the limitation lies in our solver, not
in the constraint. Two changes follow from this diagnosis: a deformation model for the pre-optimiser that does not approach $J=0$ under
large compression (for instance a log-domain or velocity parametrisation), and a handoff that rescales the displacement
only where the certificate fails instead of everywhere.

\paragraph{Selection and generalisation} Several design choices were made while monitoring the 19 Learn2Reg
validation pairs; we therefore report them as a validation set and base the main comparison on pre-registered held-out pairs (\cref{sec:experiments}). Parameters were tuned for skull-stripped, affinely aligned T1 brain MRI
with a sum-of-squares data term, and the evaluation stays within that setting. The data term is the
limitation that the IBSR18 result exposes. Its 18 images come from two acquisition groups whose white-to-grey
matter intensity ratios differ ($1.30$--$1.50$ against $1.65$--$1.78$), and the per-image percentile scaling of
our pipeline does not remove the difference: the per-structure mean intensities of fixed and moving image differ
four times more than on OASIS, and $29$ of the $36$ pairs lie outside the OASIS range. In a post-hoc analysis the
per-pair deficit against ANTs is graded by this mismatch (Pearson $r=-0.75$ with the per-structure intensity
difference; terciles $-0.007$, $-0.037$ and $-0.075$ Dice), whereas the gains of the baselines, all of which use
correlation metrics invariant to local affine intensity changes, are not ($|r|\le0.15$); on the seven pairs whose
intensity statistics resemble OASIS the two methods are level ($+0.001$, ours higher on five). The roughness of our
maps on IBSR18 is coupled to the same mismatch ($r=0.81$ with the fraction of cells with $J<\tfrac12$) and explains
nothing beyond it: a sum of squares on contrast-mismatched images can only be reduced by moving tissue boundaries.
The post-hoc time curves of \cref{fig:h3tc} separate this from the other candidate explanation, a budget chosen on
OASIS that stops too late on IBSR18. We wrote down beforehand that, if the budget were the cause, the high-mismatch
pairs would peak early and reach the ANTs level at their peak. They do peak at the first read-out and lose $0.022$
Dice to $89$\,s while the map roughens, but their peak remains $0.05$ below ANTs; on the low-mismatch pairs and on
LPBA40 the curves plateau. The stopping point therefore accounts for at most a fifth of the deficit and the data term
for the rest. A contrast-invariant data term is the natural next step; it changes the model and was not evaluated here.

\paragraph{Boundary condition and domain} The displacement is zero on the boundary of the image box. This is
natural for skull-stripped brains with a background margin, and it is what makes \cref{thm:invertibility} global, but
it excludes motion through the field of view.

%% file: sections/A_proofs.tex
\section{Proofs}\label{app:proofs}

\subsection{Proof of \cref{prop:elimination}}\label{app:elimination}
For fixed $\Phi$ the $f$-dependent part of \eqref{eq:joint} is, pointwise,
$\frac{a_J}{2}(J-f)^2+\frac{b_J}{2}(f-f_i)^2$, a strictly convex quadratic in $f$ minimised at
$f^\ast=(a_JJ+b_Jf_i)/(a_J+b_J)$. Then $J-f^\ast=b_J(J-f_i)/(a_J+b_J)$ and $f^\ast-f_i=a_J(J-f_i)/(a_J+b_J)$, so the
minimum value is
\[
  \frac{a_Jb_J^2+b_Ja_J^2}{2(a_J+b_J)^2}(J-f_i)^2=\frac{a_Jb_J}{2(a_J+b_J)}(J-f_i)^2 .
\]
The curl part is identical with $(a_C,b_C,g)$. As $a_J\to\infty$, $a_Jb_J/(a_J+b_J)\to b_J$ and $f^\ast\to J$.
\qed

\subsection{Proof of \cref{lem:edgehull}}\label{app:edgehull}
Take $j=1$; the other axes are analogous. Differentiating \eqref{eq:q1map},
\[
  \partial_{\xi_1}u_r(\xi)=\sum_{b,c\in\{0,1\}}\ell_b(\xi_2)\ell_c(\xi_3)\,\big(u_{K,(1,b,c),r}-u_{K,(0,b,c),r}\big),
\]
and $\partial_{X_1}=h_1^{-1}\partial_{\xi_1}$. The four weights $\ell_b(\xi_2)\ell_c(\xi_3)$ are nonnegative on $[0,1]^3$ and
sum to one, and the four differences are exactly the differences along the four edges parallel to axis $1$.
Therefore $(\nabla\Phi_u(\xi))_{r1}$ lies in $\delta_{r1}$ plus the convex hull of the four edge slopes, i.e.\ in
$[\Gh]_{r1}$. By the inclusion property of interval arithmetic, any interval evaluation of the determinant
over $\Gh$ contains $\det A$ for every real matrix $A\in\Gh$, in particular $\det\nabla\Phi_u(\xi)$
for every $\xi$. \qed

\subsection{Outward rounding}\label{app:soundness}
\begin{lemma}\label{lem:outward}
Let $x,y$ be finite binary64 numbers, $\circ\in\{+,-,\times,\div\}$ (with $y\ne0$ for $\div$), $z=x\circ y$ the
exact real result and $\hat z=\fl(z)$ its round-to-nearest value under gradual underflow. If $\hat z$ is finite,
then $\down\hat z<z<\up\hat z$ or $z=\hat z$.
\end{lemma}
\begin{proof}
If $z$ is representable, $\hat z=z$. Otherwise $\hat z$ is a nearest representable number to $z$, so $z$ lies
strictly between the two representable neighbours of $\hat z$: if $z>\hat z$, the next representable number
above $\hat z$ is $\up\hat z$, and $z<\up\hat z$ because otherwise $\up\hat z$ would be at least as close; the case
$z<\hat z$ is symmetric. This argument does not depend on the spacing of the binary64 numbers and therefore covers
binade boundaries and the subnormal range, including $\hat z=0$.
\end{proof}
Interval operations take, for each endpoint, the extreme of the endpoint operations (four products for
multiplication) and move it one step outward; by \cref{lem:outward} each computed interval contains the exact
range of the operation on the input intervals. Composing operations preserves containment, which proves that every
computed interval in the certificate contains the exact value of the corresponding real expression of the stored
inputs.

\begin{proof}[Proof of \cref{lem:hoisting}]
$d\mapsto\down d$ is nondecreasing on finite binary64 numbers; division by a fixed positive binary64 number followed
by rounding to nearest is nondecreasing (rounding is monotone); and $\down$ is nondecreasing. The composition
$L_h$ is therefore nondecreasing, and so is $U_h$. A nondecreasing function attains its minimum over a finite set
at the argument minimiser and its maximum at the argument maximiser.
\end{proof}

\begin{lemma}[A priori bound for test (ii)]\label{lem:prescreen}
Under the hypotheses of \cref{app:prescreen} on the derivative table and on the finiteness of the computed values,
the floating-point Bernstein coefficients $\hat\beta$ and the bound $\hat{\mathsf B}$ computed by the fixed sequence of
operations of \cref{app:prescreen} satisfy $|\hat\beta_k-\beta_k|\le\hat{\mathsf B}$ for all $k$, where $\beta$ are the exact
Bernstein coefficients of $J_K$ for the stored data; and $\fl(\min_k\hat\beta_k-\hat{\mathsf B})>0$ implies $\min_k\beta_k>0$.
\end{lemma}
The statement with all constants, and the proof, are in \cref{app:prescreen}.

\begin{proof}[Proof of \cref{thm:soundness}]
The certificate first checks that all inputs are finite and that $u_a=0$ on boundary nodes. A cell certified by
test (i) has, by \cref{lem:outward}, an interval matrix $\Gh$ that contains the exact edge-slope hulls, and an
interval determinant whose lower endpoint is positive and is a lower bound of the exact range of the determinant
over $\Gh$; \cref{lem:edgehull} then gives $\min_\xi J_K>0$. By \cref{lem:hoisting} the hoisted evaluation
produces the same endpoints, and the GPU kernel executes the same correctly rounded operations in the same order,
so it produces the same endpoints as well. A cell certified by test (ii) satisfies
$\min\beta>0$ by \cref{lem:prescreen}, and \eqref{eq:bernstein-bound} gives $\min_\xi J_K>0$. A cell
certified by test (iii) has a finite set of leaf boxes covering $[0,1]^3$ (bisection of the reference cube), and on
each leaf interval Bernstein coefficients that contain the exact Bernstein coefficients of $J_K$ restricted to the
leaf (the 27 point values are enclosed by \cref{lem:outward}, and the transformation by $T\otimes T\otimes T$ and the de
Casteljau averages are evaluated in interval arithmetic with exact constants). All lower endpoints are positive, so
\eqref{eq:bernstein-bound} applied to each leaf gives $J_K>0$ on the leaf. A map is certified only if every cell is
certified. A Jacobian witness is a point value computed at an actual point of a box, whose enclosing interval has an
upper endpoint $\le0$; the exact value at that point is then $\le0$.
\end{proof}

\subsection{Proof of \cref{thm:invertibility}}\label{app:invertibility}
Write $\Phi=\Phi_u$, let $\Sigma=\bigcup_K\partial K$ be the union of all cell faces, and let $\Phi_K$ denote the
trilinear polynomial map that coincides with $\Phi$ on $K$, extended to $\R^3$. We use the following standard
properties of the Brouwer degree $\deg(\Phi,U,y)$ for bounded open $U$, $\Phi\in C(\overline U;\R^3)$ and
$y\notin\Phi(\partial U)$ \cite[Ch.~1]{Deimling1985}: (D1)
$\deg(\Id,U,y)=1$ for $y\in U$ and $0$ for $y\notin\overline U$; (D2) homotopy invariance, so $\deg(\Phi,U,y)$ depends
only on $\Phi|_{\partial U}$; (D3) additivity over disjoint open subsets $U_1,U_2\subset U$ when
$y\notin\Phi(\overline U\setminus(U_1\cup U_2))$; (D4) $y\mapsto\deg(\Phi,U,y)$ is constant on each connected component
of $\R^3\setminus\Phi(\partial U)$; (D5) if $\Phi^{-1}(y)\cap\overline U=\{x_1,\dots,x_m\}\subset U$ and $\Phi$ is $C^1$ with
nonsingular derivative in a neighbourhood of each $x_k$, then $\deg(\Phi,U,y)=\sum_k\operatorname{sign}\det\nabla\Phi(x_k)$;
(D6) if $\deg(\Phi,U,y)\ne0$, then $y\in\Phi(U)$.

\emph{Step 1 (local structure).} Since $J_K>0$ on the compact set $K$ and $J_K$ is continuous on $\R^3$, $J_K>0$ on an
open neighbourhood of $K$; by the inverse function theorem $\Phi_K$ is a local diffeomorphism there. Hence for every
$y$ the set $\Phi^{-1}(y)\cap K$ consists of isolated points of the compact set $K$ and is finite, and so is
$\Phi^{-1}(y)\cap\Omb$. The map $\Phi$ is Lipschitz on $\Omb$ (continuous and piecewise polynomial on a convex domain)
and $\Sigma$ has measure zero, so $\Phi(\Sigma)$ has measure zero. Call $y$ \emph{regular} if $y\notin\Phi(\Sigma)$; the
regular points are dense. For regular $y$, every preimage lies in the interior of a cell, where $\Phi$ is smooth with
positive Jacobian determinant.

\emph{Step 2 (global degree).} Because $\Phi=\Id$ on $\partial\Om$, (D1)--(D2) give $\deg(\Phi,\Om,y)=1$ for $y\in\Om$
and $\deg(\Phi,\Om,y)=0$ for $y\notin\Omb$.

\emph{Step 3 (positive local index).} Let $x_0\in\Om$ and $y_0=\Phi(x_0)$. By Step 1 there is an open ball $B\ni x_0$
with $\overline B\subset\Om$ and $\overline B\cap\Phi^{-1}(y_0)=\{x_0\}$, so $y_0\notin\Phi(\partial B)$. Let $W$ be the
component of $\R^3\setminus\Phi(\partial B)$ that contains $y_0$; it is open. Choose a cell $K\ni x_0$. Since
$x_0\in\overline{\operatorname{int}K}$, there is $x\in\operatorname{int}K\cap B$ so close to $x_0$ that a small open
neighbourhood $V\subset\operatorname{int}K\cap B$ of $x$ satisfies $\Phi(V)\subset W$; $\Phi(V)$ is open and nonempty
by Step 1, so it contains a regular point $y$. By (D3) and (D5) applied in $B$, $\deg(\Phi,B,y)$ equals the number of
preimages of $y$ in $B$, which is at least one. By (D4), $\deg(\Phi,B,y_0)=\deg(\Phi,B,y)\ge1$, and the same holds for
every $y'$ in the open set $W$.

\emph{Step 4 ($\Phi(\Om)\subset\Om$).} Suppose $x_0\in\Om$ and $y_0=\Phi(x_0)\notin\Om$. Every neighbourhood of $y_0$
meets $\R^3\setminus\Omb$ (for $y_0\in\partial\Om$ because $\Om$ is a box), so there is a regular $y\in W\setminus\Omb$,
with $W,B$ from Step 3. Then $y\notin\Phi(\partial\Om)=\partial\Om$ and $y\notin\Phi(\partial B)$, and by (D3), (D5)
\[
  \deg(\Phi,\Om,y)=\deg(\Phi,B,y)+\#\big(\Phi^{-1}(y)\cap(\Om\setminus\overline B)\big)\ge1,
\]
contradicting Step 2.

\emph{Step 5 (bijectivity).} Let $y_0\in\Om$. By Step 2 and (D6), $y_0$ has a
preimage; by Step 1 the preimages $x_1,\dots,x_m$ are finitely many, and none lies on $\partial\Om$ because
$\Phi=\Id$ there. With disjoint balls $B_k\ni x_k$ as in Step 3, (D3) gives
$1=\deg(\Phi,\Om,y_0)=\sum_k\deg(\Phi,B_k,y_0)\ge m$, so $m=1$. Together with Step 4 and $\Phi|_{\partial\Om}=\Id$,
$\Phi$ is a bijection of $\Omb$ onto itself that maps $\Om$ onto $\Om$ and $\partial\Om$ onto $\partial\Om$. A continuous
bijection of a compact space onto a Hausdorff space is a homeomorphism.

\emph{Step 6 (Lipschitz inverse).} Let $C=\max_K\max_{\xi\in[0,1]^3}\norm{(\nabla\Phi_K(\xi))^{-1}}$, finite by the
positivity of $J_K$ on compact cells, and fix $\epsilon>0$. For $y,y'\in\Omb$ put $\gamma(s)=y+s(y'-y)$, $s\in[0,1]$
(the segment lies in $\Omb$ by convexity), and $g=\Phi^{-1}\circ\gamma$, which is continuous by Step 5. Fix $s$ and a
cell $K'$ containing $g(s)$. By Step 1, $\Phi_{K'}$ maps a convex neighbourhood $V$ of $g(s)$ diffeomorphically onto a
neighbourhood $W$ of $\gamma(s)$ with $\norm{(\nabla\Phi_{K'})^{-1}}\le C+\epsilon$ on $V$. For $s'$ close to $s$, $g(s')$ lies
in $V$ and in one of the finitely many cells that contain $g(s)$ (the others are at positive distance); if it lies in
$K'$, then $g(s')=(\Phi_{K'}|_V)^{-1}(\gamma(s'))$. Applying this to each cell containing $g(s)$ gives
$\norm{g(s')-g(s)}\le(C+\epsilon)\norm{y'-y}\,|s'-s|$ for all $s'$ near $s$. A continuous function on $[0,1]$ with this
local bound at every point is Lipschitz with the same constant, so
$\norm{\Phi^{-1}(y')-\Phi^{-1}(y)}\le(C+\epsilon)\norm{y'-y}$; let $\epsilon\to0$. \qed

\begin{remark}
The same conclusion follows from Ball's global invertibility theorem \cite[Theorems~1--2]{Ball1981}: the box $\Om$ is bounded, connected and strongly Lipschitz; $\Phi$ is Lipschitz, hence in
$W^{1,p}(\Om)$ for every $p>3$, and $\det\nabla\Phi\ge c>0$ almost everywhere; $\Phi$ agrees on $\partial\Om$ with
$\Phi_0=\Id$, which is one-to-one with $\Phi_0(\Om)=\Om$ satisfying the cone condition and strongly Lipschitz; and
$\int_\Om|\nabla\Phi^{-1}|^q\det\nabla\Phi\,dx<\infty$ for every $q$ because $\nabla\Phi^{-1}$ is bounded. Ball's theorem
then gives a homeomorphism of $\Omb$ onto $\Omb$ with inverse in $W^{1,q}$, but not the Lipschitz inverse of Step 6. We
include the elementary argument above because it uses only the
cellwise structure of $\Phi$ and shows where each hypothesis enters: the boundary condition fixes the degree to
one, which excludes global overlaps (for instance a band of cells bent so that distant parts of the image domain
are mapped onto the same region, which cellwise positivity alone does not exclude); positivity on the closed
cells makes all preimages isolated and the inverse Lipschitz.
\end{remark}

%% file: sections/B_parameters.tex
\section{Parameters of the recommended configuration}\label{app:parameters}

\begin{table}[ht]
\centering\small\setlength{\tabcolsep}{4pt}
\caption{Parameters of the recommended configuration (variant P), read from the configuration record of the frozen
H1 runs. Lengths in mm, intensities in $[0,1]$; $L_\ast$ is the geometric mean of the three side lengths of
the image box in mm.}\label{tab:params}
\begin{tabular}{@{}llr@{}}
\toprule
Stage & Quantity & Value\\
\midrule
Model & proximity weight $\lambda=0.1/L_\ast^2$ & $2.794\times10^{-6}$\\
 & geometric weights $\mu=\nu$ (native) & $0.03$\\
 & curvature weight $\alpha$ (native) & $0$\\
 & barrier weight $\kappa$, activation $J_b$, threshold & $0.002$, $0.05$, $J>0$\\
 & barrier points / geometric quadrature & closed $\{0,\frac12,1\}^3$ / $3^3$ Gauss\\
Inner method & accepted steps per outer iteration $k_{\max}$ & $5$\\
 & BB2 safeguard, first step & $[10^{-8},10]$, $1$\\
 & Armijo $c_1$, halvings & $10^{-4}$, $\le25$\\
 & Jacobian-determinant step prediction safety factor & $0.9$\\
 & metric (native) & \eqref{eq:metric}, $J_{\mathrm{act}}=0.25$\\
Outer iteration & forcing $\eta_0$ & $10^{-2}$\\
 & GF threshold & $10^{-8}+10^{-6}\GF(0)$\\
Certificate & edge hull, then Bernstein; depth / boxes per cell & $48$ / $255$\\
Pre-optimiser & levels / iterations & $8,4,2,1$ / $50,50,100,60$\\
 & $\mu^{\mathrm{pre},(l)}$, $\alpha^{(l)}$ per level & $(1,1,0.01,0.003)$, $(10^{-3},10^{-3},10^{-3},0)$\\
 & $\kappa_{\mathrm{pre}}$, $j_0$ & $10$, $0.2$\\
 & step cap, Adam $(\beta_1,\beta_2)$, learning rate & $0.5$ voxel, $(0.9,0.99)$, $0.5$\\
 & gradient / field smoothing $\sigma$ & $1$ / $1$ voxel\\
 & boundary taper width & $4$ voxels\\
Handoff & scale sequence $\rho_k=\rho_0\,0.9^k$ & $\rho_0=1$, $k=0,\dots,6$\\
Precision & screening band, penalties / barrier & $64$ / $4096$ ($\times2^{-24}$)\\
 & binary64 spot checks & first $3$ per grid, every $50$th\\
Budget & wall clock / export reserve & $90$\,s / $1$\,s\\
\bottomrule
\end{tabular}
\end{table}

\paragraph{Variant M (certified pyramid)} Geometric weights per level $\mu^{(l)}=\nu^{(l)}=(1,1,0.1,0.03)$, curvature
weights $(10^{-3},10^{-3},10^{-3},0)$, metric $P^{-1}=RM^{-1}R$ on the two coarsest levels and \eqref{eq:metric} on the
two finest, cumulative time shares $(0.0208,0.0625,0.1458,1)$ of the solve budget, native handoff scale $\rho_0=0.9$,
wall clock $180$\,s.

%% file: sections/C_prescreen.tex
\section{The a priori error bound of test (ii)}\label{app:prescreen}

This appendix states \cref{lem:prescreen} with all constants and proves it. We first fix the arithmetic model and the
exact sequence of operations, then bound the error stage by stage.

\paragraph{Arithmetic model} Let $\mathsf u=2^{-53}$ and $\gamma_n=n\mathsf u/(1-n\mathsf u)$. For binary64 with
round-to-nearest and gradual underflow, and finite results,
\begin{equation}\label{eq:fp-model}
  \fl(x\pm y)=(x\pm y)(1+\epsilon),\qquad \fl(x\times y)=(x\times y)(1+\epsilon)+\zeta,\qquad
  |\epsilon|\le\mathsf u,\quad|\zeta|\le2^{-1075},
\end{equation}
and multiplication by a power of two is exact unless it underflows \cite[\S2.1--2.2]{Higham2002}. If
$|\theta_i|\le\mathsf u$ and $\rho_i=\pm1$, then $\prod_{i\le n}(1+\theta_i)^{\rho_i}=1+\vartheta_n$ with
$|\vartheta_n|\le\gamma_n$, and $\gamma_m+\gamma_n+\gamma_m\gamma_n\le\gamma_{m+n}$ \cite[Lemma~3.1, 3.3]{Higham2002}. We
write $\hat{\ }$ for computed quantities.

\paragraph{The computation} For one cell, the stored nodal displacements $u_{a,r}$ ($a$ the eight corners, $r$ the
component) and the table $D_{j,p,a}=\fl(1/h_j)\,c_{j,p,a}$, with $c_{j,p,a}\in\{0,\pm\tfrac14,\pm\tfrac12,\pm1\}$ the
reference derivative of the corner shape function at the point $p$ of $\Spts$, are inputs. The exact quantities are
$d_{j,p,a}=c_{j,p,a}/h_j$, $F_{rj}(p)=\delta_{rj}+\sum_au_{a,r}d_{j,p,a}$, $J_p=\det F(p)$ and $\beta=(T\otimes T\otimes T)J$.
The kernel computes, in this order:
\begin{enumerate}[label=(\alph*)]
  \item $\hat t_a=\fl(u_{a,r}D_{j,p,a})$, $\hat s=$ the sequential sum $(((0+\hat t_1)+\hat t_2)+\dots)+\hat t_8$, and
  $\hat F_{rj}=\hat s$ for $r\ne j$, $\hat F_{rr}=\fl(\hat s+1)$;
  \item $\hat\sigma=\max_{p,r,j}\fl(\sum_a|\hat t_a|)$ (same sequential order) and $\hat\varphi=\max_{p,r,j}|\hat F_{rj}|$;
  \item $\hat J_p=\fl\big(\hat F_{00}(\hat F_{11}\hat F_{22}-\hat F_{12}\hat F_{21})+\hat F_{01}(\hat F_{12}\hat F_{20}-\hat F_{10}\hat F_{22})
  +\hat F_{02}(\hat F_{10}\hat F_{21}-\hat F_{11}\hat F_{20})\big)$, evaluated left to right, and $\Jmaxh=\max_p|\hat J_p|$;
  \item three in-place passes (along $x$, $y$, $z$) that replace each middle entry $x_m$ of every line $(x_l,x_m,x_r)$ by
  $\fl\big(2x_m-\fl(\tfrac12\fl(x_l+x_r))\big)$, giving $\hat\beta$;
  \item $\hat e=\fl(\hat\gamma_{12}(\hat\sigma+\hat\varphi)+\uflow)$, $\hat P=\fl(\hat\varphi+\hat e)$,
  $\hat E_J=\fl\big(\mathsf c\,\hat e(3\hat P^2+3\hat P\hat e+\hat e^2)+6\hat\gamma_6\hat P^3\big)$ with $\mathsf c=5.196152422706633\ge3\sqrt3$,
  and $\hat{\mathsf B}=\fl\big(1.01(27\hat E_J+135\hat\gamma_3\Jmaxh)+\uflow\big)$, $\uflow=10^{-290}$;
  \item the cell passes if $\fl(\min_k\hat\beta_k-\hat{\mathsf B})>0$.
\end{enumerate}
The test is applied only if every table entry is finite and every non-zero entry is a normal number. As constructed,
each entry is $(\pm1/h_j)\,f_1f_2$ with $f_i\in\{0,\frac12,1\}$, which equals $\fl(1/h_j)\,c_{j,p,a}$ exactly whenever the entry
is normal (an intermediate can be subnormal only if the entry itself is subnormal or zero). A cell is discarded if
any $\hat\beta_k$, $\hat\sigma$ or $\hat{\mathsf B}$ is not finite. This check excludes overflow anywhere in (a)--(e): in
(a)--(b) every partial sum satisfies $|\text{acc}|\le\text{mag}$ by monotonicity of rounding, so an overflow in a product or
a sum makes $\hat\sigma$ infinite; an overflow or an invalid operation in (c)--(d) produces an infinite or NaN value that
survives every later $+,-,\times$ and reaches $\hat\beta$ (end entries of a line are not changed, middle entries combine
their inputs); and all terms of (e) are non-negative, so an overflow there makes $\hat{\mathsf B}$ infinite.

\begin{lemma}[\cref{lem:prescreen}, full statement]\label{lem:prescreen-full}
Let the table entries be $D_{j,p,a}=\fl(1/h_j)\,c_{j,p,a}$ with $c_{j,p,a}$ zero or a power of two, all finite, and
every non-zero entry a normal binary64 number, and let $\hat\beta$, $\hat\sigma$, $\hat\varphi$, $\Jmaxh$ and $\hat{\mathsf B}$ be
computed by (a)--(e). If every $\hat\beta_k$, $\hat\sigma$ and $\hat{\mathsf B}$ is finite, then $|\hat\beta_k-\beta_k|\le\hat{\mathsf B}$ for
all $k$, and $\fl(\min_k\hat\beta_k-\hat{\mathsf B})>0$ implies $\min_k\beta_k>0$.
\end{lemma}

\paragraph{Step 1: entries of the deformation gradient} Because $c_{j,p,a}$ is zero or a power of two and $D$ has no
subnormal entries, $D_{j,p,a}=d_{j,p,a}(1+\delta_D)$ with $|\delta_D|\le\mathsf u$. By \eqref{eq:fp-model},
$\hat t_a=u_{a,r}d_{j,p,a}(1+\vartheta_2)+\zeta_a$, and the sequential sum with seven roundings gives
\[
  |\hat s-s|\le\gamma_9\sum_a|u_{a,r}d_{j,p,a}|+8(1+\gamma_7)2^{-1075},\qquad s=\sum_au_{a,r}d_{j,p,a}.
\]
Reading the same model backwards, $\sum_a|u_{a,r}d_{j,p,a}|\le(1+\gamma_{10})\hat\sigma+10^{-321}$, so
$\gamma_9\sum_a|u_ad_a|\le\gamma_{10}\hat\sigma+10^{-321}$. The diagonal entries carry one more rounding,
$|\fl(\hat s+1)-(\hat s+1)|\le\gamma_1\hat\varphi$. Hence, for all $p,r,j$,
\begin{equation}\label{eq:entry-error}
  |\hat F_{rj}-F_{rj}|\le\gamma_{10}\hat\sigma+\gamma_1\hat\varphi+10^{-319}\le\hat e ,
\end{equation}
where the last inequality uses $\hat e\ge(1-6\mathsf u)\gamma_{12}(\hat\sigma+\hat\varphi)+(1-\mathsf u)\uflow$ (the
constant $\hat\gamma_{12}$ and the three operations of (e) contribute at most six relative roundings),
$\gamma_{12}(1-6\mathsf u)\ge\gamma_{10}+\gamma_1$ and $(1-\mathsf u)\uflow\ge10^{-319}$. Consequently
$|F_{rj}|\le\hat\varphi+\hat e\le(1+\gamma_1)\hat P$.

\paragraph{Step 2: determinant} Write $\hat F=F+E$ with $|E_{rj}|\le\hat e$ (Step 1) and $|F_{rj}|\le P_\ast:=(1+\gamma_1)\hat P$.
The determinant is linear in each row, so $\det\hat F-\det F$ is the sum over the seven non-empty sets $\mathcal R$ of rows of
the determinants in which the rows in $\mathcal R$ are taken from $E$ and the others from $F$. By Hadamard's inequality each
such determinant is at most the product of the Euclidean row norms, which are at most $\sqrt3\hat e$ for rows of $E$
and $\sqrt3P_\ast$ for rows of $F$. Summing over $|\mathcal R|=1,2,3$,
\begin{equation}\label{eq:det-perturb}
  |\det\hat F-\det F|\le3\sqrt3\,\hat e\,(3P_\ast^2+3P_\ast\hat e+\hat e^2).
\end{equation}
The evaluated formula (c) is a signed sum of six products of three entries; along any path from the inputs to the
result there are at most five roundings (one product, one difference, one product, at most two sums), so
\begin{equation}\label{eq:det-eval}
  |\hat J_p-\det\hat F|\le\gamma_5\sum_{\text{6 monomials}}|\hat F\hat F\hat F|+A_{\det}\le6\gamma_5\hat\varphi^3+A_{\det},
\end{equation}
where $A_{\det}\le6(1+2\hat\varphi)\,2^{-1075}$ collects the absolute errors of underflowing products. Together,
$|\hat J_p-J_p|\le(1+4\mathsf u)E_J^\ast+A_{\det}$ with $E_J^\ast=3\sqrt3\,\hat e(3\hat P^2+3\hat P\hat e+\hat e^2)+6\gamma_6\hat P^3$,
since $\gamma_5<\gamma_6$ and $\hat\varphi\le\hat P$.

\paragraph{Step 3: Bernstein transform} Let $\tau$ denote the exact map $T\otimes T\otimes T$ and $\hat\tau$ its
floating-point evaluation (d). Then $\hat\beta-\beta=\big(\hat\tau(\hat J)-\tau(\hat J)\big)+\tau(\hat J-J)$. Every row of
$\tau$ has absolute row sum at most $27$ ($T$ has absolute row sums $1,3,1$), so
$|\tau(\hat J-J)|\le27\max_p|\hat J_p-J_p|$. For the rounding of the passes, one replacement of a middle entry commits
at most $\gamma_2(2|x_m|+\tfrac12|x_l|+\tfrac12|x_r|)\le3\gamma_2X$ for inputs bounded by $X$ (the doubling is exact
because an overflow would be caught by the finiteness check, and the halving is exact unless it underflows, which is
accounted for in $A_\tau$), and the two end entries of each line are not changed. The inputs of the three passes are bounded by $\Jmaxh$,
$3\Jmaxh(1+3\gamma_2)$ and $9\Jmaxh(1+3\gamma_2)^2$, and an error committed in one pass is amplified by at most $3$ in each
later pass. Hence
\[
  |\hat\tau(\hat J)-\tau(\hat J)|\le\big(9\cdot3+3\cdot9+27\big)\gamma_2\Jmaxh(1+3\gamma_2)^2+A_{\tau}
  =81\gamma_2\Jmaxh(1+3\gamma_2)^2+A_\tau ,
\]
where $A_\tau$ collects underflow in the passes (only the halving can underflow; $A_\tau\le27\cdot9\cdot2^{-1075}$).

\paragraph{Step 4: assembling the bound} Combining Steps 2 and 3,
\[
  |\hat\beta_k-\beta_k|\le27(1+4\mathsf u)E_J^\ast+81\gamma_2(1+3\gamma_2)^2\Jmaxh+27A_{\det}+A_\tau .
\]
We compare with $\hat{\mathsf B}$. The evaluation (e) of $\hat E_J$ and $\hat{\mathsf B}$, including the rounded constants
$\hat\gamma_n$ and $1.01$, involves at most twenty relative roundings in any term, so
$\hat{\mathsf B}\ge1.01(1-\gamma_{20})\big(27E_J^\ast+135\gamma_3\Jmaxh\big)+(1-\mathsf u)\uflow$, where we used
$\mathsf c\ge3\sqrt3$ and $\hat\gamma_n\ge(1-2\mathsf u)\gamma_n$. Since $135\gamma_3\ge81\gamma_2(1+3\gamma_2)^2$ and
$1.01(1-\gamma_{20})\ge1+4\mathsf u+10^{-3}$, it remains to absorb the absolute terms $A=27A_{\det}+A_\tau$, together with
the absolute error (at most $2^{-1075}$ per product) of products in (e) that underflow, for instance
$\fl(135\hat\gamma_3)\Jmaxh$ when $\Jmaxh$ is tiny; these are covered by $\uflow$ in the same way. If
$\hat\varphi<1$, then $A\le27\cdot18\cdot2^{-1075}+243\cdot2^{-1075}<10^{-320}<(1-\mathsf u)\uflow$. If $\hat\varphi\ge1$, then
$\hat P\ge1$ and the slack $10^{-3}\cdot27\cdot6\gamma_6\hat P^3\ge10^{-16}\hat P^3$ exceeds $A\le10^{-320}(1+\hat P)$. In both cases
\begin{equation}\label{eq:prescreen-final}
  |\hat\beta_k-\beta_k|\le\hat{\mathsf B}\qquad\text{for all }k.
\end{equation}

\paragraph{Step 5: the decision} For finite binary64 numbers $x,y$ under gradual underflow, $\fl(x-y)>0$ if and only if
$x>y$: every binary64 number is an integer multiple of $2^{-1074}$, so $x>y$ implies $x-y\ge2^{-1074}$, which is
representable, and monotone rounding gives $\fl(x-y)\ge2^{-1074}>0$; if $x\le y$, then $\fl(x-y)\le0$. Overflow is
impossible here because $\hat{\mathsf B}\ge0$ and $\min_k\hat\beta_k\le\hat\beta_{\text{corner}}=\hat J_{\text{corner}}$. Therefore the test (f) passes exactly when $\min_k\hat\beta_k>\hat{\mathsf B}$, and by
\eqref{eq:prescreen-final} then $\min_k\beta_k\ge\min_k\hat\beta_k-\hat{\mathsf B}>0$; \eqref{eq:bernstein-bound} gives $J_K>0$ on
the whole cell. \qed

\paragraph{Remarks} (i) The constants used by the implementation are conservative: the propagated transform error is
$81\gamma_2\Jmaxh$ to first order, against $135\gamma_3\Jmaxh$ in $\hat{\mathsf B}$, and the entry error uses $\gamma_{12}$ where
$\gamma_{10}+\gamma_1$ suffices. (ii) The entry bound of Step 1 is nearly attained: on 300 adversarial cells evaluated with exact rational
arithmetic, the largest ratio of the entry error to $\gamma_{10}\hat\sigma+\gamma_1\hat\varphi$ was $0.974$, while the transform
error stayed below $0.05$ of $81\gamma_2\Jmaxh(1+3\gamma_2)^2$. On 280 cells---random cells with displacement scales from
$10^{-3}$ to $10^8$ and
spacings from $0.013$ to $7.7$, cells collapsed to within $10^{-14}$ of a degenerate cell, large common translations and
displacements of size $10^{-300}$---the largest ratio of the actual error, computed with exact rational arithmetic
from the same stored data, to $\hat{\mathsf B}$ was $4.5\times10^{-3}$, and the bound was never violated
(script in the ancillary files). (iii) The positivity of the maps reported in this paper does not rely on
this lemma once the audit counts are in place: the post-hoc audit re-certifies every committed iterate with tests (i)
and (iii) only (\cref{sec:rounding}). (iv) The recorded lower bound $\fl(\min\hat\beta-\hat{\mathsf B})$ of a cell certified by
test (ii) is itself a rigorous lower bound of $\min_\xi J_K$: it exceeds $\min\hat\beta-\hat{\mathsf B}$ by at most
$\mathsf u\,|\min\hat\beta-\hat{\mathsf B}|\le\mathsf u\Jmaxh$ (the minimum coefficient is at most a corner value), while
$\hat{\mathsf B}$ exceeds the error bound of Step 4 by at least $(135\gamma_3-81\gamma_2(1+3\gamma_2)^2)\Jmaxh>200\,\mathsf u\Jmaxh$.
(v) Since $\hat\sigma+\hat\varphi\ge\frac12$ for every cell (either a diagonal entry is at least $\frac12$, or its dot
product has magnitude above $\frac12$), $\hat e\ge6\times10^{-16}$, and the terms involving $\uflow$ are never the active
ones; the case analysis above covers them for completeness. (vi) The analysis assumes that the kernel is executed
without fused multiply--add contraction or reassociation. It is compiled without fast-math, and an inspection of the
generated code on our platform (no fast-math flags or FMA intrinsics in the intermediate representation, no fused
multiply--add instructions in the assembly; re-checked on 2026-10-18) found none; unlike the interval kernels, this
kernel has no regression test of its compiled code.